\pdfoutput=1
\documentclass[aps,prx,reprint,superscriptaddress,longbibliography,floatfix]{revtex4-2}
\usepackage[T1]{fontenc}
\usepackage{textcomp}
\usepackage{xspace}
\usepackage{amsmath,amssymb}
\usepackage{graphicx}
\usepackage{booktabs}
\usepackage{array}
\usepackage{adjustbox}
\usepackage{multirow}
\usepackage{xcolor}
\usepackage{pifont}
\usepackage{tikz}
\usepackage{newfloat}
\DeclareFloatingEnvironment[name=Algorithm]{algorithm}
\makeatletter
\let\ftype@algorithm\ftype@figure
\makeatother
\usepackage{algorithmic}

\usetikzlibrary{arrows.meta,calc,positioning,shapes.geometric,decorations.pathreplacing}
\usepackage[hidelinks]{hyperref}
\usepackage{orcidlink}
\usepackage{url}
\usepackage{microtype}
\usepackage{placeins}
\usepackage{needspace}
\makeatletter
\let\cd@section\section
\renewcommand{\section}{\needspace{5\baselineskip}\cd@section}
\let\cd@subsection\subsection
\renewcommand{\subsection}{\needspace{4\baselineskip}\cd@subsection}
\makeatother
\definecolor{clv}{HTML}{D55E00}
\definecolor{fbb}{HTML}{0072B2}
\definecolor{bpg}{HTML}{009E73}
\definecolor{isp}{HTML}{CC79A7}
\definecolor{fdg}{HTML}{4D4D4D}

\newcommand{\us}{\,\textmu s}
\newcommand{\ms}{\,ms}
\newcommand{\pct}{\,\%}
\newcommand{\ci}[1]{\ding{\numexpr181+#1\relax}}
\newcommand{\head}[2][3pt]{\par\vspace{#1}\noindent\textbf{#2}}
\newcommand{\Tc}{T_{\mathrm{comp}}}
\newcommand{\Tn}{T_{\mathrm{net}}}
\newcommand{\Tf}{T_{\mathrm{fb}}}
\newcommand{\Th}{T_{\mathrm{hand}}}

\newcommand{\ie}{\textit{i.e.}}

\newcommand{\sys}{CoDecoder\xspace}

\begin{document}

\title{\sys: Make Quantum Decoders Up~to~100\texorpdfstring{$\times$}{×} Faster and Even More Accurate}

\author{Yicheng Guang\,\orcidlink{0009-0006-6180-9497}}
\email[Contact author: ]{yicheng.guang@colorado.edu}
\affiliation{Department of Computer Science, University of Colorado Boulder, Boulder, Colorado, USA}
\author{Yuliang Xiong\,\orcidlink{0009-0000-7258-1770}}
\affiliation{Department of Computer Science, University of Colorado Boulder, Boulder, Colorado, USA}
\author{Xun Gao\,\orcidlink{0009-0005-7278-4005}}
\email[Contact author: ]{xun.gao@colorado.edu}
\affiliation{Department of Physics and JILA, University of Colorado Boulder, Boulder, Colorado, USA}
\author{Yueqi Chen\,\orcidlink{0009-0007-4205-3273}}
\email[Contact author: ]{yueqi.chen@colorado.edu}
\affiliation{Department of Computer Science, University of Colorado Boulder, Boulder, Colorado, USA}

\begin{abstract}
Fault-tolerant quantum computers need fast and accurate error decoding. We present \sys, a small neural frontend that composes with an existing decoder to improve accuracy and reduce mean decoding time. It predicts each shot's logical outcome, retains confident predictions and routes the remaining shots to the unchanged fallback. A logical-operator-aware output head scores shifted logical operators, combining equivalent logical lines on surface codes and deciding logical qubits jointly on bivariate-bicycle (BB) codes. Fused GPU inference and efficient routing reduce the overhead paid on every shot. In registered tests under circuit-level noise, composition reduces failures by 6.4--31.4\pct{} relative to correlated matching on surface codes of distance 7--17 and by 22--50\pct{} relative to full-DEM Relay-BP on two BB codes, with 2.3--124-fold speed-ups across these comparisons. With our re-implementations of the neural decoder Cascade, composition reduces decoding time in all four settings. A joint decision of both decoders on the routed shots keeps this speed and makes the composition more accurate than Cascade in all four, with 14--18\pct{} fewer failures on the $[[72,12,6]]$ code. On the $[[72,12,6]]$ BB code, a conventional head with independent logical outputs fails 3.8--5.0 times as often as our joint class score under matched training conditions. Per-shot analysis shows how composition gains accuracy. On retained shots, the frontend corrects more fallback errors than it introduces, even when it is less accurate overall.
\end{abstract}

\maketitle

\section{Introduction}
\label{sec:intro}

A quantum computer protects its logical qubits by measuring parity checks, called stabilizers, in every
round~\cite{dennis,fowler2012}. A decoder reads these outcomes and predicts whether the logical qubits flipped. The
decoder has to keep pace with the rounds, about one per microsecond on superconducting processors~\cite{google_below}.
Otherwise, a backlog grows, and every logical operation that depends on a decoded result waits~\cite{battistel}. The
decoder also has to fail rarely, because every wrong prediction is a logical error. Decoders are therefore compared on
memory experiments, which keep a logical state for some rounds and then measure it. Each run is a shot, and a comparison measures each decoder's failure rate and mean decoding time.

The decoders that fail least are slow. Consider the $[[72,12,6]]$ bivariate-bicycle (BB) code~\cite{bravyi} at a
physical error rate of 0.3\pct{}. In our tests, the belief-propagation decoder Relay-BP~\cite{relaybp} fails on
1.38\pct{} of shots when it decodes one type of check alone and on 0.55\pct{} when it decodes both types together. The
second setting, however, takes 26 times as long, 85.4\ms{} per shot on one CPU core. Neural decoders such as
AlphaQubit~2~\cite{alphaqubit2}, Cascade~\cite{cascade} and the NTU decoder~\cite{ntu} push accuracy further, and larger
models take longer. We call an accurate but slow decoder of this kind a strong decoder.

A natural approach is to put a fast decoder in front of a strong one. The frontend answers shots on which it is confident and routes the others to the strong decoder, which serves as a fallback. Decoder Switching~\cite{decoder_switching} and Chongder~\cite{chongder} use confidence to route difficult shots to matching-based decoders. Decoder Switching matched or improved its fallback's accuracy with simulated times at distances 5 to 11. These results raise a broader question. Under what conditions can decoder composition improve both accuracy and measured decoding time, across code families and fallback algorithms? \looseness=-1 A small neural frontend makes this question particularly challenging because it can fail more often than the fallback when used alone.

Composition does not require the frontend to outperform the fallback on every shot. The composition uses the fallback's decision on every routed shot, so its accuracy differs from the fallback's only on the shots retained by the frontend. The composition is more accurate exactly when the frontend makes fewer errors than the fallback on this retained subset (\textbf{C1}). It is faster when the fallback work saved exceeds the frontend inference and handoff costs (\textbf{C2}). Prediction accuracy and confidence ranking determine the first condition. Inference cost, handoff and the work required by routed shots determine the second. We derive these relations in Section~\ref{sec:overview} and use them to guide the design.

\begin{figure}[tbp]
\centering
\begin{tikzpicture}[x=1cm,y=1cm,font=\footnotesize,
  box/.style={draw,rounded corners=1.5pt,align=center,inner sep=2pt},
  flow/.style={-{Stealth[length=4pt]},line width=0.6pt},
  lab/.style={font=\scriptsize,inner sep=1pt}]
\fill[black!6] (0,-0.05) rectangle (6.85,3.15);
\fill[black!3] (6.97,-0.05) rectangle (8.6,3.15);
\node[anchor=north west,font=\scriptsize\bfseries,text=black!70] at (0.06,3.12) {GPU};
\node[anchor=north west,font=\scriptsize\bfseries,text=black!70] at (7.03,3.12) {CPU};
\node[box,draw=clv,fill=clv!13,line width=0.7pt,minimum width=1.7cm,minimum height=0.8cm] (net) at (1.08,2.3)
  {\textbf{\sys}\\[-1pt]{\scriptsize decision $y^*$,}\\[-2pt]{\scriptsize confidence}};
\node[draw,diamond,aspect=1.9,inner sep=0.5pt,fill=white,line width=0.6pt] (dec) at (3.95,2.3) {$\geq\tau$?};
\node[box,fill=white,minimum width=1.0cm,minimum height=0.5cm,line width=0.6pt] (ans) at (6.1,2.3) {answer};
\node[box,fill=white,minimum width=1.6cm,minimum height=0.5cm] (buf) at (1.08,0.95) {};
\foreach \i/\b in {0/0,1/1,2/0,3/0,4/1,5/0,6/1,7/0}
  \node[font=\scriptsize,text=black!75] at (0.45+0.18*\i,0.95) {\b};
\node[lab,below=2pt of buf,align=center] {detector bits\\from the readout};
\node[box,draw=fbb,fill=fbb!12,line width=0.7pt,minimum width=2.5cm,minimum height=0.62cm] (fb) at (4.3,0.95)
  {\textbf{strong decoder}\\[-1pt]{\scriptsize a black box}};
\node[box,draw=fbb,dashed,fill=white,line width=0.6pt,minimum height=0.62cm] (fbc) at (7.785,0.95)
  {\scriptsize strong\\[-2pt]\scriptsize decoder};
\draw[dashed,line width=0.5pt,draw=fbb] (fb.south) -- (fb.south |- 0,0.3) -- node[lab,above,pos=0.75,text=fbb] {or}
  (fbc.south |- 0,0.3) -- (fbc.south);
\draw[flow] (buf.north) -- node[lab,left] {\ci{1}} (net.south);
\draw[flow] (net.east) -- node[lab,above] {confidence} node[lab,below] {\ci{2}} (dec.west);
\draw[flow] (dec.east) -- node[lab,above] {$1-q$} node[lab,below] {\ci{3}\ $y^*$} (ans.west);
\draw[flow] (dec.south) -- node[lab,right] {$q$} (dec.south |- fb.north);
\draw[flow,fbb] (buf.east) -- node[lab,above] {\ci{4}} (fb.west);
\draw[flow] (fb.east) -- (ans.south |- fb.east) -- (ans.south);
\draw[flow,dashed,clv] (net.north) -- (net.north |- 0,2.84) -- node[lab,above,pos=0.5,text=clv] {class scores \ci{5}}
  (ans.north |- 0,2.84) -- (ans.north);
\end{tikzpicture}
\caption{\sys and a strong decoder as a composition, shown for one shot. \protect\ci{1}~\sys reads the shot's
detector bits and returns its decision $y^*$ and a confidence. \protect\ci{2}~The confidence is compared with the
threshold $\tau$. \protect\ci{3}~If it reaches $\tau$, \sys's decision is the answer. \protect\ci{4}~Otherwise, the
strong decoder decodes the shot from its original detector bits, which happens for a fraction $q$ of all shots. The
strong decoder is a black box. It can be a neural decoder such as Cascade on the GPU, or correlated matching or
Relay-BP on the CPU. \protect\ci{5}~A joint decision can combine \sys's class scores with the strong decoder's logits on the
routed shots where they differ (dashed).}
\label{fig:overview}
\end{figure}
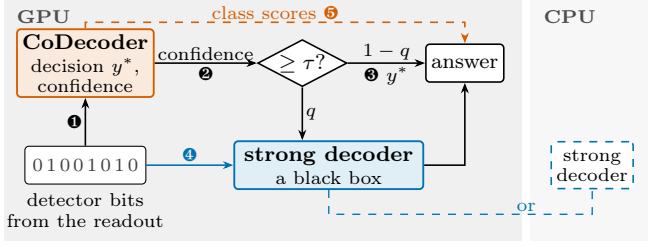

We present \sys, a small neural frontend that composes with an existing decoder through a confidence threshold. It predicts a shot's logical outcome and keeps that prediction when its confidence reaches the threshold. Otherwise, the fallback decodes the original detector bits. The fallback's algorithm is unchanged and may run on a CPU or GPU. When the fallback also reports logits, as a neural decoder does, both decoders can decide together the routed shots on which they differ. We write \sys for the frontend alone and \sys + X for its composition with decoder X. Figure~\ref{fig:overview} shows the interface.

To meet C1, \sys uses a logical-operator-aware output head, the module that turns convolutional features into a prediction and confidence. Instead of pooling features without logical-operator geometry, it scores the flips of multiple shifted logical operators. On the surface code, equivalent logical lines contribute to one observable prediction. On BB codes, a translate generally acts on different logical qubits, so the head combines their scores to decide all logical qubits jointly. The gap between the best and next-best outcomes supplies the confidence for routing. Earlier neural decoders predicted joint logical outcomes directly on the toric code~\cite{wagner_hld,end_decoder}. Our head derives them from shifted operators (Section~\ref{sec:neural-decoding}).

To meet C2, we combine selective routing with efficient execution (Section~\ref{sec:threshold}). Fused GPU kernels and CUDA graphs reduce the overhead of running a small network on every shot. A resident kernel supports successive surface-code shots, while a compiled router reduces the handoff cost to matching. For throughput, a multi-shot BB engine shares weight loads across shots, and GPU fallbacks process routed shots in batches. We choose the network and routing threshold on development and calibration data and freeze them before testing. Single-shot latency and batched cost are evaluated separately.

We evaluate \sys with four strong decoders under circuit-level noise, namely correlated matching on surface codes of distance 7 to 17, full-DEM Relay-BP on two BB codes, and our re-implementations of Cascade on both code families (Section~\ref{sec:rq1}). Cascade's code and trained networks are not public, so we implement its published architecture and train it on our circuits. Composition reduces failures by 6.4--31.4\pct{} relative to correlated matching and by 22--50\pct{} relative to full-DEM Relay-BP, with 2.3--124-fold speed-ups across these comparisons. With our Cascades, composition reduces decoding time on both GPU models. A joint decision on the routed shots, which also reads Cascade's logits, keeps this speed and makes the composition more accurate than Cascade in all four settings, with 14--18\pct{} fewer failures on BB72. In batches, \sys + correlated matching decodes the surface code at $p=0.6\pct$ up to distance 11 in less than 1\us{} per syndrome round, about the round period of current superconducting processors~\cite{google_below}, against 3.2--10.8\us{} for correlated matching alone on one core. \sys + Cascade on the multi-shot engine decodes BB72 at $p=0.2\pct$ in 0.733 and 0.891\us{} per round on an RTX PRO 6000 and an H100, against 12.00 and 7.34\us{} for Cascade alone. To our knowledge, \sys is the first open-source design that brings a strong decoder below 1\us{} per round at a physical error rate as high as 0.6\pct{} while reducing its failures. The tests compare each composition and its fallback on the same fresh shots, with configurations fixed in advance. Every speed-up compares the slowest run of the composition with the fastest run of the decoder it is compared against.

\looseness=-1 Matched output-head comparisons isolate the effect of the head while holding the convolutional trunk, training data and budget fixed (Section~\ref{sec:rq2}). On BB72, independent logical outputs fail 3.8--5.0 times as often as our joint class score. With 2\pct{} of shots routed to full-DEM Relay-BP, only the joint class score's composition beats Relay-BP alone. On the surface code, the logical-line vote outperforms a pooled head and a single line, all trained on the final answer alone. Adding line supervision makes the vote worse. Per-shot analysis explains the composition's accuracy gain (Section~\ref{sec:why}). On retained shots, the frontend corrects more fallback errors than it introduces. Confidence concentrates frontend failures among routed shots, but routing helps only if the fallback is more accurate on those shots. Runtime measurements then examine C2 by separating frontend, handoff and fallback costs, and by measuring the throughput of batched execution (Section~\ref{sec:time}).

This paper makes four contributions.
\begin{itemize}
\item We state the accuracy and cost conditions for useful decoder composition, the first following Decoder Switching~\cite{decoder_switching}, and use them to guide a frontend that leaves the fallback unchanged.
\item We design a logical-operator-aware neural output head that produces decoding predictions and confidence from scores of shifted logical operators, deciding all logical qubits together on BB codes.
\item We build GPU engines and routing support for single-shot and batched decoding, including a resident surface-code kernel and a multi-shot BB engine.
\item We evaluate composition, and a joint decision with neural fallbacks, across two code families and four strong decoders, measure the output head's contribution, and explain the sources and limits of accuracy gains and speed-ups. We release the training and decoding code, weights and commands for the matching and Relay-BP tests~\cite{convlines_code}.
\end{itemize}

\section{Background}
\label{sec:background}

\subsection{Quantum Error Correction}
\label{sec:qec-background}

\head[0pt]{Stabilizer codes and logical operators.}
A stabilizer code encodes logical qubits in physical data qubits.
Its stabilizers are Pauli products whose measurements provide information about errors without revealing the stored logical information.
A nontrivial logical Pauli operator commutes with every stabilizer but implements a nonidentity operation on the encoded logical qubits.
The code distance is the minimum number of data qubits needed to support such an operator.
Multiplying a logical operator by a stabilizer changes its physical representation but preserves its action on the encoded state.
Thus, different physical operators can implement the same logical action.

\head{Memory experiments and decoding.}
A memory experiment tests whether a code preserves a logical state.
It prepares the state, repeatedly measures the stabilizers, and finally measures the data qubits.
Each run is a shot.
A detector checks a parity of measurement outcomes against its expected value in the absence of faults.
The resulting detector bits form the decoder's input.
A logical observable is measured by multiplying the final data-qubit outcomes on the support of a chosen logical Pauli operator.
The decoder predicts which logical observables have flipped relative to their expected values.
The simulator records the actual flips as ground truth.
A shot fails if any predicted flip is wrong, so failure on a code with several logical observables is a block failure.
An X-basis memory experiment prepares and reads out logical X eigenstates.\looseness=-1

A detector error model (DEM) represents circuit noise as independent error mechanisms.
Each mechanism has a probability and specifies which detector bits and logical observables it flips.
Stim~\cite{stim} derives the DEM from the circuit and samples the test shots.
The decoder infers the logical effect of the faults from their detector bits.

\head{Code families.}
We focus on rotated surface codes and bivariate-bicycle (BB) codes.
A rotated surface code of distance $d$ uses $d^2$ data qubits and $d^2-1$ stabilizers~\cite{google_below}.
The BB codes $[[72,12,6]]$ (\ie, BB72), and $[[144,12,12]]$ (\ie, the gross code), are quantum low-density parity-check (qLDPC) codes~\cite{bravyi}.
In the notation $[[n,k,d]]$, $n$ is the number of data qubits, $k$ the number of logical qubits, and $d$ the code distance.
Both BB codes encode 12 logical qubits, with checks on an $l\times m$ torus of size $6\times6$ for BB72 and $12\times6$ for the gross code.

\head{Decoder families.}
Matching, belief propagation, and neural networks provide three different ways to infer logical flips from detector bits.
\looseness=-1 PyMatching (PM) finds a minimum-weight perfect matching on the graph of a graphlike DEM~\cite{dennis,pymatching1,sparseblossom}.
Its correlated mode uses an additional reweighting pass to exploit correlations between X and Z errors~\cite{fowler_corr}.
NVIDIA's Ising pre-decoders are convolutional networks that remove local error patterns before matching~\cite{ising,ising_code}.
BP-OSD runs belief propagation (BP) and falls back to ordered-statistics decoding (OSD) when BP does not converge~\cite{bposd,roffe}.
Relay-BP chains BP runs, each initialized from the previous run's final marginals with randomly drawn memory strengths, to find several valid corrections~\cite{relaybp}.
Its sector setting decodes the X detectors, whereas its full-DEM setting decodes all detectors.
Cascade is a convolutional neural decoder that follows the geometric structure of the code~\cite{cascade}.
Matching and BP decoders return a correction, whereas a neural decoder such as Cascade also reports a logit for each logical observable.
These methods differ in accuracy, computational cost, and the shots on which they fail.

\subsection{Decoder Composition}
\label{sec:overview}

\head[0pt]{Routing by confidence.}
A fast decoder can answer the shots on which it is confident and pass the others to a stronger decoder, as in Decoder Switching~\cite{decoder_switching} and Chongder~\cite{chongder}.
We call the fast decoder the frontend and the stronger decoder the fallback.
Here the frontend is a neural network $N$, and the fallback $B$ is treated as a black box.
Both predict the same logical observables from the original detector bits.
The composition keeps the network's answer when its confidence reaches a threshold $\tau$ and otherwise uses the fallback's answer.
The routed shots form a fraction $q$ of all shots.

\head{Accuracy depends on the selected shots.}
Let $K$ and $R$ be the kept and routed shots, and let $F_{X,S}$ count the failures of decoder $X$ on a set of shots $S$.
The composition uses $N$ on $K$ and $B$ on $R$, giving
\begin{equation}
F_{\mathrm{composition}}-F_B=F_{N,K}-F_{B,K},
\label{eq:identity}
\end{equation}
where $F_B$ counts the fallback's failures over all shots.
The composition therefore fails less often than the fallback exactly when the network fails less often on the kept shots.
This identity is the finite-sample form of Eq.~(3) of Decoder Switching~\cite{decoder_switching}.
Likewise, the composition fails less often than the network exactly when $F_{B,R}<F_{N,R}$.
These relations require the fallback to produce the same decision for a routed shot in the composition as it would for that shot alone.
Appendix~\ref{app:reg} describes how our tests preserve this property.

A network can therefore improve a fallback even if it is less accurate over all shots.
Its advantage must hold on the subset that its confidence selects, while the fallback must be better on the routed subset for the composition to improve on both components.
Different error patterns alone are insufficient.
The confidence must separate shots on which the network should answer from those on which the fallback should answer.

An optimal decoder selects the logical class with the highest posterior probability given the detector bits, where a class groups faults with the same logical effect.
A routing rule based on those same bits cannot reduce its expected failure probability.
Composition gains therefore correct suboptimal fallback decisions.
A joint decision $J$ of both decoders on the routed shots (Section~\ref{sec:joint}) adds $F_{J,R}-F_{B,R}$ to
Eq.~(\ref{eq:identity}), so it can also correct the fallback on the routed shots.

\head{Latency and throughput.}
A decoder serving a long computation can process independent syndrome windows~\cite{skoric_window,tan_window} or separate logical blocks in batches.
We call this the throughput mode and report its amortized cost per syndrome round, which must remain below the round period for the decoder to keep pace.
At a feed-forward point, the next quantum operation depends on a decoded result, as a correction after a T gate does.
The decoder must then answer each shot as it arrives.
We call this the reaction mode and measure it at batch size one.

\head{The cost of composition.}
Every shot pays for the network and for the handoff that selects and dispatches the answer.
Only routed shots pay for the fallback.
The mean composite time per shot is
\begin{equation}
\Tc=\Tn+T_{\mathrm{fb}}+\Th,
\label{eq:comp}
\end{equation}
where $\Tn$ is the network time, $T_{\mathrm{fb}}$ is the fallback work averaged over all shots, and $\Th$ is the handoff overhead.
The handoff includes launch, result transfer, and routing work as applicable to the execution path.
The fallback may run on either a CPU or a GPU.

Let $T_B$ be the fallback's mean time per shot when run alone, and let $\kappa$ be its mean time on routed shots divided by $T_B$.
Then $T_{\mathrm{fb}}=q\kappa T_B$, and the speed-up is
\begin{equation}
\frac{T_B}{\Tc}=\frac{1}{q\kappa+(\Tn+\Th)/T_B}.
\label{eq:speedup}
\end{equation}
The composition is faster when the fallback work saved exceeds the cost of the network and the handoff.
Routed shots can be more expensive than average, so a small routed fraction does not by itself imply a large speed-up.
For the classical fallbacks in our measurements, $\kappa>1$ (Section~\ref{sec:where}).
For $q\kappa>0$, the speed-up cannot exceed $1/(q\kappa)$.
The same accounting applies to amortized costs in batches when all terms use a consistent timing basis.

Together, these relations identify \textbf{two necessary conditions for a frontend to improve both the accuracy and speed of a fallback decoder}.
\looseness=-1 First, the frontend must fail less often than the fallback on the shots it keeps (\textbf{C1}).
Meeting C1 requires both accurate predictions and confidence scores that select the shots on which the frontend has an advantage.
Second, network inference, handoff and routed fallback work together must cost less than the fallback alone (\textbf{C2}).

\section{Design of \sys}
\label{sec:frontend}

\looseness=1 \sys meets the two conditions for improving both the accuracy and speed of a fallback decoder through two parts of its design.
To meet \textbf{C1}, a logical-operator-aware output head combines scores for multiple representations of logical operators into a prediction and a confidence, allowing the frontend to keep shots on which it fails less often than the fallback.
When the fallback reports logits, a joint decision extends this advantage to the routed shots.
To meet \textbf{C2}, confidence-based composition limits the fallback work, while specialized GPU engines and a compiled router reduce the cost of neural inference and handoff paid on every shot.
We jointly choose the network size and routing threshold to preserve the accuracy advantage while saving more fallback work than this fixed cost.
Section~\ref{sec:neural-decoding} describes the output head, and Section~\ref{sec:threshold} the composition, the joint decision and the GPU engines.

\subsection{Logical-Operator-Aware Neural Decoding}
\label{sec:neural-decoding}

The output head connects the network's features to the code's logical operators. On the surface code, parallel logical lines represent the same observable up to stabilizers, so their scores can be aligned and combined into one vote. On BB codes, translating a logical operator can change its logical class. The head therefore scores joint observable-flip patterns instead of separate votes.

\begin{figure}[tbp]
\centering
\begin{adjustbox}{max width=\columnwidth}\begin{tikzpicture}[x=1cm,y=1cm,font=\footnotesize]
\begin{scope}
\def\s{0.42}
\node[font=\small,anchor=north west] at (-0.74,2.46) {(a)};
\foreach \r in {0,...,3}{\foreach \c in {0,...,3}{
  \pgfmathtruncatemacro{\t}{mod(\r+\c,2)}
  \ifnum\t=0
    \ifnum\r<2 \fill[black!45] (\c*\s,\r*\s) rectangle ++(\s,\s);
    \else \fill[black!15] (\c*\s,\r*\s) rectangle ++(\s,\s); \fi
  \fi
  \draw[black!45,line width=0.3pt] (\c*\s,\r*\s) rectangle ++(\s,\s);}}
\foreach \r/\f in {1/black!45,3/black!15}
  \filldraw[fill=\f,draw=black!45,line width=0.3pt] (0,\r*\s) arc[start angle=270,end angle=90,radius=\s/2] -- cycle;
\foreach \r/\f in {0/black!45,2/black!15}
  \filldraw[fill=\f,draw=black!45,line width=0.3pt] (4*\s,\r*\s) arc[start angle=-90,end angle=90,radius=\s/2] -- cycle;
\foreach \c in {0,2}
  \filldraw[fill=white,draw=black!45,line width=0.3pt] (\c*\s,0) arc[start angle=180,end angle=360,radius=\s/2] -- cycle;
\foreach \c in {1,3}
  \filldraw[fill=white,draw=black!45,line width=0.3pt] (\c*\s,4*\s) arc[start angle=180,end angle=0,radius=\s/2] -- cycle;
\draw[black,line width=1.2pt] (-0.06,0) -- (4*\s+0.06,0);
\draw[clv,line width=1.2pt] (-0.06,2*\s) -- (4*\s+0.06,2*\s);
\foreach \r in {0,...,4}{\foreach \c in {0,...,4}{\filldraw[fill=white,draw=black,line width=0.35pt] (\c*\s,\r*\s) circle (0.045);}}
\node[anchor=west] at (4*\s+0.24,0) {obs.};
\node[anchor=west,text=clv] at (4*\s+0.24,2*\s) {$j$};
\draw[decorate,decoration={brace,amplitude=2.5pt},line width=0.5pt] (-0.3,0) -- (-0.3,2*\s) node[midway,left=2pt] {$\delta_j$};
\node[anchor=north] at (2*\s,-0.27) {X checks shaded};
\end{scope}
\begin{scope}[shift={(2.87,0.05)}]
\def\L{1.36}\def\dx{0.5}\def\dy{0.34}\def\u{0.22667}\def\ta{0.25}\def\tb{0.75}
\draw[black!35,line width=0.4pt] (\dx,\dy) rectangle ++(\L,\L);
\foreach \xx/\yy in {0/0,1/0,0/1,1/1} \draw[black!35,line width=0.4pt] (\xx*\L,\yy*\L) -- ++(\dx,\dy);
\fill[clv!22] ($(0.5*\u,0.5*\u)+\tb*(\dx,\dy)$) rectangle ++(5*\u,5*\u);
\draw[clv,line width=0.5pt] ($(0.5*\u,0.5*\u)+\tb*(\dx,\dy)$) rectangle ++(5*\u,5*\u);
\foreach \xx/\yy in {0.5/5.5,5.5/5.5,5.5/0.5} \draw[clv,line width=0.5pt] ($(\xx*\u,\yy*\u)+\ta*(\dx,\dy)$) -- ($(\xx*\u,\yy*\u)+\tb*(\dx,\dy)$);
\fill[clv!35,opacity=0.8] ($(0.5*\u,0.5*\u)+\ta*(\dx,\dy)$) rectangle ++(5*\u,5*\u);
\draw[clv,line width=0.6pt] ($(0.5*\u,0.5*\u)+\ta*(\dx,\dy)$) rectangle ++(5*\u,5*\u);
\foreach \i in {1,...,5}{\draw[black!18,line width=0.25pt] (\i*\u,0) -- (\i*\u,\L); \draw[black!18,line width=0.25pt] (0,\i*\u) -- (\L,\i*\u);}
\draw[black!70,line width=0.5pt] (0,0) rectangle (\L,\L);
\node[anchor=north] at (\L/2,-0.01) {$W$};
\node[anchor=east] at (-0.01,\L/2) {$H$};
\node[anchor=north west] at ($(\L,0)+0.45*(\dx,\dy)$) {$T$};
\node[anchor=south,text=clv] at (0.9,\L+0.36) {stem $3{\times}5{\times}5$};
\node[anchor=north] at (0.85,-0.27) {grid $T{\times}H{\times}W$};
\end{scope}
\begin{scope}[shift={(5.4,-0.06)}]
\def\c{0.4}
\node[font=\small,anchor=north west] at (-0.66,2.52) {(b)};
\foreach \i in {0,...,6}{\draw[black!25,line width=0.3pt] (\i*\c,0) -- (\i*\c,6*\c); \draw[black!25,line width=0.3pt] (0,\i*\c) -- (6*\c,\i*\c);}
\draw[black!70,line width=0.5pt] (0,0) rectangle (6*\c,6*\c);
\foreach \xx in {-0.1,6*\c+0.1} \draw[-{Stealth[length=3pt]},black!70,line width=0.5pt] (\xx,2.6*\c) -- (\xx,3.4*\c);
\foreach \yy in {-0.1,6*\c+0.1} \draw[-{Stealth[length=3pt]Stealth[length=3pt]},black!70,line width=0.5pt] (2.6*\c,\yy) -- (3.4*\c,\yy);
\node[anchor=west,inner sep=1pt] at (3.4*\c+0.04,-0.1) {$x$};
\node[anchor=south,inner sep=1pt] at (-0.1,3.4*\c+0.04) {$y$};
\foreach \a in {0,...,5}{\foreach \b in {0,...,5}{
  \filldraw[fill=white,draw=black!50,line width=0.3pt] (\a*\c+0.12,\b*\c+0.2) circle (0.045);
  \filldraw[fill=white,draw=black!50,line width=0.3pt] (\a*\c+0.28,\b*\c+0.2) circle (0.045);}}
\foreach \a/\b in {2/3,2/5,5/0,5/1,5/4,5/5} \fill[black] (\a*\c+0.28,\b*\c+0.2) circle (0.07);
\foreach \a/\b in {4/4,4/0,1/1,1/2,1/5,1/0} \fill[clv] (\a*\c+0.28,\b*\c+0.2) circle (0.07);
\fill[black] (0.32,-0.43) circle (0.07);
\draw[-{Stealth[length=4pt]},line width=0.7pt] (0.46,-0.43) -- (1.0,-0.43);
\fill[clv] (1.14,-0.43) circle (0.07);
\node[anchor=west,inner sep=1pt] at (1.25,-0.43) {$g=x^2y$};
\end{scope}
\end{tikzpicture}\end{adjustbox}
\caption{Output-head geometry. (a) At $d=5$, surface-code line $j$ and the observable's line (obs.) differ by the dark X stabilizers, whose detectors give $\delta_j$. The cube shows the detector grid and stem footprint. (b) On the BB72 torus, translating a logical operator $\ell_a$ (black) by $g=x^2y$ (vermillion) wraps around the torus and changes its logical class. Arrows identify periodic edges.}
\label{fig:head}
\end{figure}
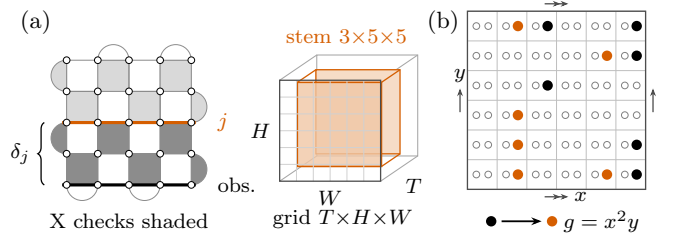

\vspace{3pt}
\head{Surface-code output head.} The head asks one query per logical line, a row $j$ of data qubits whose X operators form a logical operator equivalent to the measured observable (Figure~\ref{fig:head}(a)). Each query uses sigmoid gates to weight and sum site features. A multilayer perceptron (MLP) turns the sum into a score $z_j$. The flip of a line is the parity of the error chains that cross it. In the main surface-code tests, the network learns from the final answer alone, so $z_j$ is a learned score rather than a calibrated prediction of the line's flip.

Line $j$ and the observable's line differ by the X stabilizers between them. Their flips therefore differ by $\delta_j$, the parity of the X detectors of these stabilizers over all rounds. The head aligns the scores and sums them,
\begin{equation}
c=\sum_j w_j(-1)^{\delta_j}z_j+b_0,
\label{eq:vote}
\end{equation}
with learned weights $w_j$ and bias $b_0$. It predicts an observable flip when $c>0$.

\vspace{3pt}
\head{BB-code output head.}\label{sec:bb-net} We call the BB-code network \sys-BB. BB logical operators have no row structure, and equivalent translates alone give too few members for a vote (Appendix~\ref{app:train}). The head instead asks one query per measured logical operator $a$ and torus shift $g\in G$. Each logit $z_{ag}$ predicts the translate's flip.

Let $k=12$ be the number of measured observables and $y\in\{0,1\}^k$ their joint flip pattern. Each translated logical can be expressed as a product of measured logical operators and X stabilizers. Its flip under a candidate class $y$ is
\begin{equation}
t_{ag}(y)=\phi_{ag}(y)\oplus\delta_{ag},
\label{eq:translate-flip}
\end{equation}
where $\phi_{ag}(y)$ is the parity of the observable flips in that product, and $\delta_{ag}$ is the detector parity of the corresponding stabilizers over all rounds and the final layer. We precompute the operator decomposition once per code and compute $\delta_{ag}$ from each shot's detector bits.

The head scores class agreement with these logits,
\begin{equation}
\begin{split}
S(y)={}&\sum_{a,g}w_a\bigl(t_{ag}(y)-\tfrac12\bigr)z_{ag}\\
&+\sum_j\beta_j\bigl(y_j-\tfrac12\bigr),
\end{split}
\label{eq:bbvote}
\end{equation}
with learned weights $w_a$ and biases $\beta_j$, and returns $y^*=\arg\max_y S(y)$. The $2^k=4{,}096$ classes are the observable flip patterns measured by the X-basis memory experiment. For one logical whose translates are equivalent, $S(1)-S(0)$ reduces to Eq.~(\ref{eq:vote}) with a shared vote weight. Class enumeration grows exponentially with the number of measured logicals, a limitation discussed in Section~\ref{sec:discussion}. Appendix~\ref{app:train} gives the query construction, scoring algorithm and computational cost.

\vspace{3pt}
\head{Convolutional networks.} Both heads read features from a small convolutional trunk with a stem and residual blocks. The surface-code trunk places detectors on a dense space-time grid, with zeros at sites without detectors and separate convolution weights for X- and Z-centred sites. The BB trunk places X and Z detectors on a periodic space-time grid. Circular padding and shared weights exploit torus translation symmetry, while convolution offsets follow fault correlations in the DEM and check adjacency in the Tanner graph. Appendix~\ref{app:train} gives each network's architecture, size and symmetry properties.

\vspace{3pt}
\head{Training objectives.} We train on fresh Stim shots with exact observable-flip labels. In the main surface-code tests, the loss is the binary cross-entropy $\operatorname{BCE}(c,y)$ of the final logit. We omit individual line supervision because it makes the final vote fail more often (Section~\ref{sec:rq2}). The surface-code comparisons with Cascade use an earlier network trained with line supervision, as specified in Section~\ref{sec:rq1n}.
On BB codes, the loss supervises both the joint class and every translated logical flip,
\begin{equation}
\begin{split}
\mathcal{L}_{\mathrm{BB}}={}&\operatorname{CE}\bigl(\operatorname{softmax}S,y\bigr)\\
&+\frac{1}{k|G|}\sum_{a,g}\operatorname{BCE}\bigl(z_{ag},t_{ag}(y)\bigr).
\end{split}
\label{eq:bb-loss}
\end{equation}
BCE takes a logit as its first argument. Both terms are averaged over the batch, and the auxiliary term has weight 1. Training derives translate labels from the true class, while decoding uses only detectors and the learned network. Larger networks grow from trained smaller ones (See Appendix~\ref{app:train} for optimizer and budgets).

\subsection{Confidence-Based Composition and Efficient Execution}
\label{sec:threshold}

To meet C2, we reduce both the fallback work and the fixed cost of neural inference and handoff. The resulting composition must satisfy $\Tn+\Th+q\kappa T_B<T_B$ (Section~\ref{sec:overview}).

\vspace{3pt}
\head{Confidence and routing.} On the surface code, the confidence is $|c|$, the gap between the scores of the two observable outcomes. On BB codes, it is the gap between the two highest joint class scores,
\begin{equation}
m=S(y^*)-\max_{y\ne y^*}S(y).
\label{eq:class-margin}
\end{equation}
The BB margin compares complete patterns of the 12 observable flips. Neither score is a calibrated probability. \sys uses the confidence only to rank shots.

As Figure~\ref{fig:overview} shows, the network computes a prediction and confidence on the GPU. The composition compares the confidence with a threshold $\tau$. A shot at or above $\tau$ keeps the neural prediction. Otherwise, the fallback supplies the answer from the original detector bits. The fallback needs no change to its decoding algorithm and can run on a CPU or GPU.

\vspace{3pt}
\head{Joint decision on routed shots.}\label{sec:joint} A routed shot discards the network's prediction, which is
already computed. When the fallback also reports a logit $\ell_k$ for each observable $k$, as a neural decoder does,
both decoders decide together the routed shots where they differ. On BB codes, the joint decision is
$\hat y=\arg\max_y[S(y)+w\sum_k y_k\ell_k]$, whose second term is the fallback's log-likelihood of class $y$ up to a
constant, so $\hat y$ can be a class that neither decoder chose. On the surface code, it is the sign of $c+w\,\ell$.
We fit $w$ on separate fresh shots and freeze it.

\vspace{3pt}
\head{Choosing the composition.} Network size and routing jointly affect C1 and C2. A larger frontend may improve predictions but costs more on every shot. Routing more shots adds fallback work and can replace correct neural predictions with fallback errors. We choose configurations on development and calibration data, then freeze each composition before its independent test. For a chosen routed fraction $q$, we set $\tau$ to a calibration confidence quantile, measured through the deployed GPU engine. The fixed threshold can yield a slightly different routed fraction on test shots.

For the classical comparisons, the calibration rule balances an accuracy target against a frozen speed limit. Appendix~\ref{app:train} gives the target, search grids and fallback rule when no candidate qualifies. The Cascade configurations and their follow-up tests are in Appendix~\ref{app:cascade}.

\vspace{3pt}
\head{Efficient execution.} A small network can still incur substantial runtime overhead. The three-block surface-code network launches about 40 kernels per shot in PyTorch. Our fused Triton engine combines the computation into $L+2$ kernels for an $L$-block network and replays them with a CUDA graph. At $d=7$, a resident kernel keeps weights on the GPU and accepts successive shots without relaunching the network. The BB engine similarly fuses inference, class scoring and top-two selection. A compiled C++ router joins the graph-based surface-code frontend to correlated PM and returns either the neural or fallback answer, reducing handoff overhead.

In throughput mode, the multi-shot BB engine loads each weight tile once for a group of shots. GPU fallbacks process routed shots in batches. Batching amortizes execution cost but adds queueing delay. We distinguish mean cost from tail latency and assess C2 within a consistent execution mode and timing basis (Appendix~\ref{app:timing}).

\section{Experiment I: Accuracy and Speed of Decoder Composition}
\label{sec:rq1}

This experiment asks whether placing \sys before a strong decoder can reduce both its logical failure rate and its mean decoding time (\textbf{RQ1}). We test four strong decoders from two code families, covering correlated matching on surface codes, Relay-BP on BB codes, and our re-implementations of Cascade on both families. Each composition is compared with its fallback alone on the same fresh shots, with the network and routing threshold fixed before the test. We describe the setup, then compare matching, BP and neural decoders.

\begin{figure*}[tp]
\centering
\includegraphics[width=\textwidth]{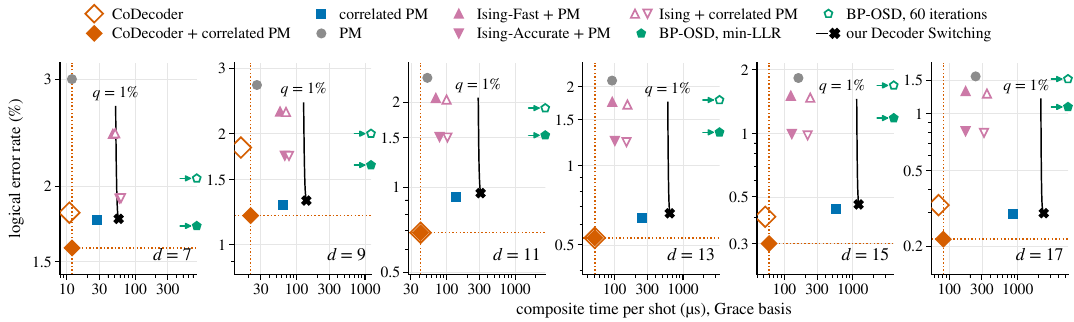}
\caption{Surface-code accuracy and decoding time at $p=0.6\pct$.
Each panel plots logical error rate against composite time per shot
on the Grace basis. Dotted lines intersect at \sys + correlated PM.
Open diamonds show \sys alone ($q=0$), coinciding with the composition
at $d=11$ and 13. Crosses mark Decoder Switching at its frozen routed
fraction, and its curves vary this fraction from 1\pct{} to the frozen value.
BP-OSD points exceeding the time range are placed at the right edge.
The 95\pct{} Wilson intervals are smaller than the markers.}
\label{fig:surface}
\end{figure*}

\subsection{Setup}
\label{sec:setup}

\head[0pt]{Fallback choices.} In the fast mode, $q=0$ and the frontend decides every shot. In the balanced mode, the fallback is correlated PM on surface codes and sector Relay-BP on BB codes. In the accurate mode, full-DEM Relay-BP receives 0.25--2\pct{} of the BB shots. We also compose the frontend with our re-implementations of Cascade. These choices leave the confidence-based interface unchanged, and each composition keeps confident neural predictions and routes the rest to its fallback.

\head{Test sets and statistical comparisons.} Each test, except the held-out joint decision, was registered before its
test shots were drawn and timed in a registered session on shots drawn after its rules were public
(Appendix~\ref{app:reg}). The surface-code test holds
2,000,000 shots at $d=7$ and 1,000,000 at $d=9$--17, and Table~\ref{tab:bb} gives the sizes of the BB tests. Each
accuracy comparison applies the exact McNemar test to the shots on which exactly one of the two decoders fails. A
claim needs all of its accuracy comparisons and a lower time (Appendix~\ref{app:stats}). Every reduction in failures
is relative to the compared decoder. Our claim rests on each composition's comparison with its strong decoder
alone, and the comparisons with other decoders, such as BP-OSD on the GPU, give context.

\head{Cascade baselines.} Cascade's code and trained networks are not public. We therefore re-implemented Cascade from its
paper~\cite{cascade}. On our BB72 circuits, our re-implementation has 14 blocks of width 512 (19,687,937 parameters).
Our $d=13$ surface-code implementation has 13 blocks of width 256 (2.53 million parameters). Below, Cascade means these re-implementations, not the published networks. On BB72,
Table~\ref{tab:cascade} reports the copy trained under a fitted circuit-level noise model that reproduces the baseline failure rates
of Cascade's paper (Appendix~\ref{app:cascade}). Each comparison timed Cascade alone in the same job as its compositions, on
each card. Our surface-code Cascade takes 385\us{} per 13-round shot on an RTX PRO 6000 and 430\us{} on an H100, or 29.6 and 33.1\us{} per round, below the roughly 40\us{} per cycle reported on an H200~\cite{cascade}. Its speed therefore rules out a slow baseline as the source of our speed-ups.

Table~\ref{tab:cascade} reports failures from Cascade's compiled single and batched paths on the same fresh shots. Each BB72 shot contains 6 syndrome rounds and each $d=13$ surface-code shot 13 rounds.

\head{Hardware and timing.} \looseness=1We call the tests whose fallback is correlated PM or Relay-BP the classical tests. We
timed their GPU parts on NVIDIA H100 80GB GPUs and their CPU parts on a Linux workstation with an AMD
Ryzen 9 9950X CPU, and their composite times use the Grace basis of Appendix~\ref{app:timing}. Each of our times is
the slowest of our processes, and each baseline decoder's time is the fastest of its three, except Ising's
(Appendix~\ref{app:timing}). 

We also timed every Cascade comparison end to end on two cards, an NVIDIA RTX PRO
6000 Blackwell Server Edition GPU with an AMD EPYC 9555 CPU and an H100 with an Intel Xeon Platinum 8480+ CPU. There
too, each composition's time is the slowest of its three processes and Cascade alone's the fastest of its three, and each card
is compared on its own (Appendix~\ref{app:timing}). This rule counts against our compositions in every comparison. Mean times
over the processes would only raise every speed-up that we report (Table~\ref{tab:means}). We
use both cards because the comparisons were trained, validated and registered on the RTX PRO 6000, and the H100 is the
card of our other tests. Cascade's own paper used an H200~\cite{cascade}.

\head{\sys alone.} Beside each composition, we also report \sys alone ($q=0$). Its failures count the network's
own wrong decisions on the same test shots, and its time is the network term plus the handoff.

\begin{figure*}[tp]
\centering
\includegraphics[width=\textwidth]{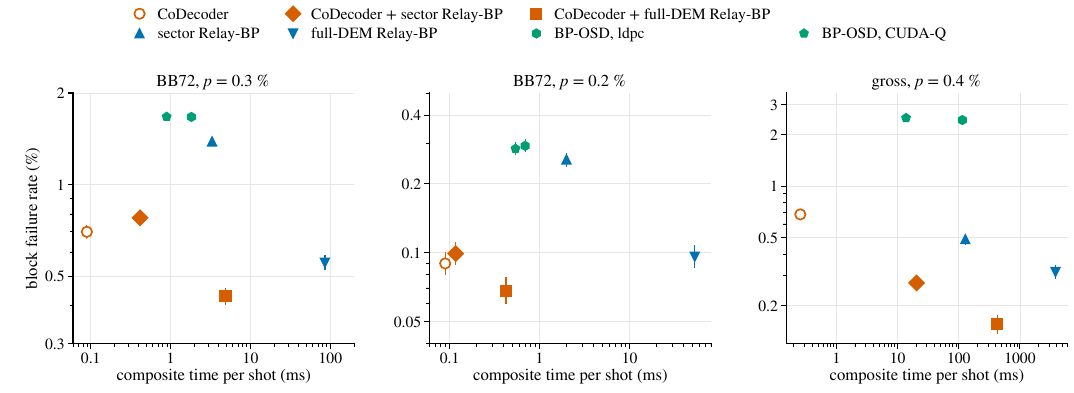}
\caption{Block failure rate against composite time per shot on the BB codes, with 95\pct{} Wilson intervals. \sys
and its compositions are vermillion, with hollow circles for \sys alone, diamonds for \sys + sector Relay-BP and
squares for \sys + full-DEM Relay-BP. The two Relay-BP decoders alone are blue, as every fallback is in the
figures of this paper. CPU work is on the Grace basis and GPU work on the H100, timed in a registered timing
session. Table~\ref{tab:bb} lists every count and time.}
\label{fig:bb}
\end{figure*}

\begin{table*}[tp]
\centering
\caption{Accuracy and speed of \sys composed with our Cascade implementations, with and without the joint decision.}
\label{tab:cascade}
\footnotesize
\setlength{\tabcolsep}{3.7pt}
\begin{adjustbox}{max width=\linewidth}
\begin{tabular}{@{}l r r r l l r r r r@{}}
\toprule
& & \multicolumn{2}{c}{failures} & \multicolumn{2}{c}{accuracy} & \multicolumn{2}{c}{RTX PRO 6000} &
\multicolumn{2}{c}{H100} \\
\cmidrule(lr){3-4}\cmidrule(lr){5-6}\cmidrule(lr){7-8}\cmidrule(lr){9-10}
decoder & routed (\%) & single & batched & single & batched & single & batched & single & batched \\[-2.5pt]
\scriptsize & & & & & & \scriptsize\textmu s/shot & \scriptsize\textmu s/round & \scriptsize\textmu s/shot & \scriptsize\textmu s/round \\
\midrule
\multicolumn{10}{@{}l}{\emph{BB72, $p=0.2\pct$, 11,000,000 shots}} \\
Cascade alone & & 8,074 & 8,073 & & & 1,719.5 & 12.00 & 1,477.1 & 7.34 \\
\sys & 0 & 10,398 & 10,398 & & & & 0.444 & & 0.688 \\
\sys + Cascade & 2 & 7,986 & 7,984 & better & better & 168.1 & 1.355 & 189.4 & 1.702 \\
speed-up & & & & & & 10.23 & 8.85 & 7.80 & 4.31 \\
\sys + Cascade, joint decision$^{a}$ & 2 & 6,597 & 6,588 & better & better & 174.9 & 1.356 & 195.8 & 1.707 \\
speed-up$^{a}$ & & & & & & 10.08 & 8.85 & 9.68 & 4.03 \\
\addlinespace[1pt]
\multicolumn{10}{@{}l}{\emph{BB72, $p=0.3\pct$, 17,000,000 shots}} \\
Cascade alone & & 75,979 & 75,973 & & & 1,724.1 & 12.00 & 1,482.8 & 7.34 \\
\sys & 0 & 112,101 & 112,101 & & & & 0.444 & & 0.688 \\
\sys + Cascade & 10 & 75,740 & 75,735 & better & better & 296.6 & 2.331 & 316.3 & 2.315 \\
speed-up & & & & & & 5.81 & 5.15 & 4.69 & 3.17 \\
\sys + Cascade, joint decision$^{a}$ & 10 & 65,189 & 65,185 & better & better & 315.7 & 2.334 & 351.8 & 2.252 \\
speed-up$^{a}$ & & & & & & 5.61 & 5.14 & 5.37 & 3.05 \\
\addlinespace[1pt]
\multicolumn{10}{@{}l}{\emph{Surface code, $d=13$, $p=0.6\pct$, 1,000,000 shots}} \\
Cascade alone & & 4,936 & 4,947 & & & 384.7 & 10.47 & 430.1 & 9.16 \\
\sys & 0 & 8,114 & 8,114 & & & & & & \\
\sys + Cascade & 10 & 4,917 & 4,927 & non-inferior & non-inferior & 106.0 & 2.34 & 129.1 &
2.10 \\
speed-up & & & & & & 3.63 & 4.47 & 3.33 & 4.35 \\
\sys + Cascade, joint decision$^{a}$ & 10 & 4,886 & 4,884 & better & better & 108.1 & 2.34 & 127.2 & 2.15 \\
speed-up$^{a}$ & & & & & & 3.56 & 4.45 & 3.30 & 4.23 \\
\addlinespace[1pt]
\multicolumn{10}{@{}l}{\emph{Surface code, $d=13$, $p=0.35\pct$, 30,000,000 shots}} \\
Cascade alone & & 2,354 & 2,352 & & & 384.8 & 10.46 & 429.3 & 9.12 \\
\sys & 0 & 3,943 & 3,943 & & & & & & \\
\sys + Cascade & 1 & 2,325 & 2,322 & better & better & 69.5 & 1.36 & 83.0 & 1.27 \\
speed-up & & & & & & 5.53 & 7.66 & 5.17 & 7.18 \\
\sys + Cascade, joint decision$^{a}$ & 1 & 2,306 & 2,308 & better & better & 69.4 & 1.36 & 91.3 & 1.29 \\
speed-up$^{a}$ & & & & & & 5.54 & 7.65 & 4.64 & 7.06 \\
\bottomrule
\end{tabular}
\end{adjustbox}

\vspace{1pt}
\parbox{\textwidth}{\scriptsize Times are in \textmu s per shot (single) or per syndrome round (batched). Accuracy compares each composition with Cascade alone in the same mode. Better denotes significantly fewer failures. Non-inferior uses a margin of 10\pct{} of Cascade's development failure rate (Appendix~\ref{app:stats}). Blank timing entries were not registered. $^{a}$Both decoders decide together the routed shots where they differ (Section~\ref{sec:joint}). These rows were timed in a later session with Cascade alone, on fresh shots and another H100, and their speed-ups compare with Cascade alone there (Appendix~\ref{app:cascade}).}
\end{table*}

\subsection{Composing matching decoders}
\label{sec:rq1m}

We first evaluate \sys alone against plain and correlated PM on the surface code, then measure how composition with correlated PM changes accuracy and decoding time. We compare the composition with the other baseline decoders and our reproduction of Decoder Switching on the same test shots.

\head{\sys alone beats plain PM.} \sys alone beats plain PM in accuracy and time at every distance. At $d=7$, it takes 10.90\us{} against PM's 11.89\us{} and fails on 36,156 shots against 60,016. At $d=9$, it takes 15.31\us{} against 26.30\us{} and fails on 18,305 shots against 27,028. It also beats correlated PM at $d=11$ and 13, where the frozen routing rule sends no shots to the fallback. There it fails on 6,887 and 5,318 of 1,000,000 shots, against 9,252 and 6,344 for correlated PM, in 3.3 and 4.9 times less time. Network depth sets its time at small distances. With one block at $d=7$ and 9 and five at $d=11$--17, its GPU time is 8.37, 14.21, 40.4, 50.1, 51.2 and 68.5\us{}, against PM's 11.89, 26.30, 52.2, 91.7, 160.1 and 242.5\us{} (Tables~\ref{tab:surface} and~\ref{tab:time}).

\head{Composition significantly improves the accuracy--speed tradeoff.} As Figure~\ref{fig:surface} shows, \sys + correlated PM fails 6.4--31.4\pct{} less often than correlated PM alone at $d=7$--17, with a 2.3--10.3-fold speed-up on the Grace basis. At $d=15$, routing 1\pct{} of the shots cuts the network's failures from 4,030 to 3,012. At $d=7$, 9 and 17, routing 2, 6 and 1\pct{} reverses the network's accuracy deficit against correlated PM. It also fails less often than PM, the four pipelines combining Ising-Fast or Ising-Accurate with PM or correlated PM, and BP-OSD in two settings. From $d=9$ on, its composite time is 0.48--0.81 of the fastest of these baselines. At $d=7$, plain PM is slightly faster, 11.89 against 11.94\us{}, but fails 1.9 times as often. Ising's failures use NVIDIA's official PyTorch numerics and its times use its TensorRT plans. Table~\ref{tab:surface} in Appendix~\ref{app:stats} reports all failure counts and times.

\head{\sys + correlated PM beats Decoder Switching.} We re-implemented Decoder Switching~\cite{decoder_switching} from this
test's decoders, with PM as the frontend that routes by its complementary gap and correlated PM as the fallback,
whereas the published system falls back to belief matching (Appendix~\ref{app:stats}). \sys + correlated PM failed
less often than this re-implementation at all six distances, in 4.8 to 29 times less time
(Appendix~\ref{app:stats}).

\subsection{Composing belief-propagation decoders}
\label{sec:rq1b}

We compose \sys with full-DEM and sector Relay-BP on BB72 and the gross code. We compare standalone decoding and each composition with the Relay-BP and BP-OSD baselines (Figure~\ref{fig:bb}).

\head{\sys alone is fast but does not replace the strongest baseline.} On BB72, \sys alone beats sector Relay-BP and both BP-OSD decoders in accuracy and time. However, the BB72 network fails more often than full-DEM Relay-BP at $p=0.3\pct$ (1,398 against 1,107) and shows no significant difference at $p=0.2\pct$ (269 against 288). On the gross code, the network alone takes 0.019 of CUDA-Q BP-OSD's time (Figure~\ref{fig:bb}), but fails on 1,024 shots against full-DEM Relay-BP's 472.

\head{Composition significantly improves the accuracy--speed tradeoff.} \sys + full-DEM Relay-BP fails 22--50\pct{} less often than full-DEM Relay-BP alone, with an 8.9--124-fold speed-up. On BB72 at $p=0.3\pct$, routing 1.99\pct{} of the shots reduces failures to 857 from 1,107, with a 17-fold speed-up. At $p=0.2\pct$, routing 0.245\pct{} reduces failures to 204 from 288, with a 124-fold speed-up, the largest in our tests. On the gross code, routing 2\pct{} reduces failures to 234 from 472, with an 8.9-fold speed-up, timed on 25 routed shots and 75 full-DEM Relay-BP shots per process. These gains require composition rather than a standalone network advantage over the fallback.

Composing with sector Relay-BP also beats it and both BP-OSD decoders on both axes on BB72. On the gross code, this composition fails less often than all three and is faster than sector Relay-BP and ldpc BP-OSD, but takes 1.49 times as long as CUDA-Q BP-OSD, 20.35 against 13.70\ms{} per shot. Without the Grace conversion, it would take 13.57\ms{} and be faster (Appendix~\ref{app:stats}).

\begin{table*}[tp]
\centering
\caption{The output-head comparison. On BB72, independent outputs fail 3.8 to 5.0 times as often as the joint class
score at matched supervision. On the surface code, the line vote trained on the final answer fails least.}
\label{tab:head}
\footnotesize
\setlength{\tabcolsep}{4.5pt}
\begin{adjustbox}{max width=\linewidth}
\begin{tabular}{@{}l l l r r r r l@{}}
\toprule
& & & \multicolumn{2}{c}{alone} & \multicolumn{2}{c}{with fallback at $q$} & \\
\cmidrule(lr){4-5}\cmidrule(lr){6-7}
arm & output head & supervision & seed 1 & seed 2 & seed 1 & seed 2 & comparison \\
\midrule
\multicolumn{8}{@{}l}{\emph{Surface code, $d=9$, $p=0.6\pct$, $q=1\pct$, correlated PM alone 12,989}} \\
a & line vote & with lines & 15,018 & 13,981 & 13,015 & 12,202 & \\
b & line vote & final answer$^{*}$ & 10,661 & 10,822 & 9,921 & 9,919 & better than a \\
c & centre line & final answer & 16,798 & 15,088 & 14,470 & 13,056 & worse than a and b \\
d & pooled head & final answer & 478,245\rlap{$^{\dagger}$} & 13,684 & 473,494\rlap{$^{\dagger}$} & 11,931 & worse than b,
inconclusive against a \\
e & pooled head & with lines & 14,924 & 15,859 & 12,891 & 13,667 & inconclusive against a and d \\
\multicolumn{8}{@{}l}{\emph{BB72, $p=0.3\pct$, $q=2\pct$, full-DEM Relay-BP alone 5,228}} \\
f & joint class score & with translates$^{*}$ & 6,766 & 6,714 & 4,246 & 4,256 & \\
g & joint class score & final answer & 6,974 & 6,954 & 4,210 & 4,254 & worse than f at $q=0$ \\
h & independent outputs & final answer & 33,748 & 34,702 & 18,280 & 19,416 & worse than f and g \\
i & independent outputs & with translates & 27,788 & 25,840 & 12,504 & 10,892 & worse than f, better than h \\
\bottomrule
\end{tabular}
\end{adjustbox}

\vspace{2pt}
\parbox{\textwidth}{\scriptsize Failures per million shots. Comparisons apply both alone and with fallback unless specified. Section~\ref{sec:headdesign} gives the statistical criteria. $^{*}$Designs used in Section~\ref{sec:rq1}. $^{\dagger}$Training failed.}
\end{table*}

\subsection{Composing large neural decoders}
\label{sec:rq1n}

We first evaluate \sys alone against our reproduction of Cascade on BB72 and the surface code at $d=13$, then measure how composition, with and without a joint decision on the routed shots, changes accuracy, single-shot latency and batched cost (Table~\ref{tab:cascade}). 
We also repeat the timing with TensorRT engines for Cascade.

\head{\sys alone is fast but less accurate.} \looseness=1On BB72, the six-block \sys-BB network has 841,053 parameters and takes 0.444 and 0.688\us{} per round in batches on the RTX PRO 6000 and H100, 26.8 and 10.6 times less than Cascade's network. However, it fails 29\pct{} and 48\pct{} more often at $p=0.2$ and 0.3\pct{}. At $d=13$, the surface-code network fails 64\pct{} and 68\pct{} more often than Cascade at $p=0.6$ and 0.35\pct{}. The BB72 standalone batched times use the multi-shot engine, whereas the compositions in Table~\ref{tab:cascade} use the BB engine.

\head{Composition significantly improves the accuracy--speed tradeoff.} \looseness=1Composing \sys with Cascade improves accuracy in three of the four settings and is non-inferior in the fourth, while reducing mean decoding time on both cards. On BB72, the six-block network routes 2\pct{} of the shots at $p=0.2\pct$ and 10\pct{} at 0.3\pct{}. Tests on 11,000,000 and 17,000,000 fresh shots show 1.1\pct{} and 0.31\pct{} fewer failures on both of Cascade's compiled paths. Across the BB and multi-shot engines, mean single-shot latency is 5.8--12.5 times lower on the RTX PRO 6000 and 4.6--9.1 times lower on the H100. Batched cost per shot is 5.1--16.3 and 3.1--8.2 times lower, respectively (Table~\ref{tab:cascade}, Section~\ref{sec:batched-throughput} and Appendix~\ref{app:timing}).

On the surface code at $d=13$, we use a network trained with line supervision and thresholds fitted on development shots. Routing 10\pct{} at $p=0.6\pct$ gives 4,917 failures against Cascade's 4,936 in 1,000,000 single-mode shots, a non-inferior result. Routing 1\pct{} at $p=0.35\pct$ gives 2,325 against 2,354 in 30,000,000 single-mode shots, with 1.2--1.3\pct{} fewer failures across the single and batched modes. At these two error rates, single-shot latency is 3.63 and 5.53 times lower on the RTX PRO 6000 and 3.33 and 5.17 times lower on the H100. Batched cost is 4.47 and 7.66 times lower on the RTX PRO 6000 and 4.35 and 7.18 times lower on the H100. The frontend decides 90\pct{} and 99\pct{} of the shots, although tail latency increases on both cards (Section~\ref{sec:discussion}).

The BB72 gains are small because 96--97\pct{} of Cascade's failures occur on routed shots (Section~\ref{sec:cascade-why}).

\head{A joint decision recovers routed failures.} With the joint decision of Section~\ref{sec:joint} on the routed
shots, the composition fails 18\pct{} and 14\pct{} less often than Cascade alone on the test shots of BB72 and
1.0--2.0\pct{} less often at $d=13$. It thus beats Cascade's accuracy in all four settings. The gain is smaller at
$d=13$, where the network has one logical observable. There, the joint decision can only choose between the two answers, and
routing has already sent it the shots on which the network's logit is weakest. Timed later, the composition took at most
2.4\pct{} longer with it than without it in 15 of the 16 timings, and a separate run of the 16th at equal GPU clock
found 0.2\pct{} (Appendix~\ref{app:cascade}).

\head{The speed advantage persists with faster Cascade engines.} TensorRT runs our Cascades 3.1--4.7 times faster per single shot on BB72 and 1.3--1.4 times faster at $d=13$. Even with Cascade alone on its faster engine, every composition takes 2.3--16.3 times less time per shot on the two cards (Appendix~\ref{app:trt}).

\section{Experiment II: Contribution of the Output Head}
\label{sec:rq2}

This experiment asks whether the logical-operator-aware output head improves decoding accuracy, both alone and in composition, relative to a conventional head trained on the same trunk, data and budget (\textbf{RQ2}). Because the head produces both the prediction and its confidence, we count failures alone and in composition. We compare the heads on fresh surface-code and BB72 shots under matched supervision. We describe the setup before comparing the two code families.

\subsection{Setup}
\label{sec:headdesign}

\head[0pt]{Two heads, two kinds of supervision.} To isolate the output head, each comparison holds the trunk, its initial
weights, the training shots and their order, and the budget fixed. The trunk is a three-block network of
width 128, at $d=9$ and $p=0.6\pct$ on the surface code and at $p=0.3\pct$ on BB72. Every arm (one combination of head
and supervision) trains from scratch on $2\times10^8$ shots in two seeds with the optimizer and schedule that trained
the networks of Section~\ref{sec:rq1}. The first head is \sys's own. The second is a conventional head that we call
a pooled head. It keeps the number of queries, the gated sum, the MLP and about the same multiply-adds, but its queries
are learned and carry no line or translate geometry. On BB72, it outputs 12 independent logical logits, and its
confidence is the smallest of their magnitudes. A comparison of the two heads thus changes the query geometry, the
scoring and the confidence together. The supervision is the final answer alone or with every line or translate flip.
The surface code adds a head that reads the centre line alone (Appendix~\ref{app:head}).

\head{Test sets and statistical comparisons.} We compare the arms on 1,000,000 fresh test shots at $d=9$ and 500,000 on BB72, alone ($q=0$) and in
compositions at the $q$ fixed before the test, 1\pct{} at $d=9$ and 2\pct{} on BB72. Each comparison tests the two seeds
separately. It is better or worse only when both seeds reject in the same direction after Holm's correction, and
inconclusive otherwise (Appendix~\ref{app:stats}).

\subsection{Comparing output heads on BB72}
\label{sec:head-bb}

We first compare the joint class score with independent logical outputs when each network decodes alone, then compose each with full-DEM Relay-BP at the same fixed routing fraction. Both comparisons use the matched supervision settings in Table~\ref{tab:head}.

\head{The joint class score improves standalone accuracy.} As Table~\ref{tab:head} shows, at matched supervision a pooled
head with independent logical outputs fails 3.8--5.0 times as often as the joint class score in both seeds, with and
without the translate supervision. At $q=2\pct$, it still fails 2.5--4.6 times as often.

\head{Only the joint class score improves accuracy in composition.} The joint class score's compositions
fail on 4,210--4,256 test shots per million at $q=2\pct$, about 19\pct{} below the 5,228 of full-DEM Relay-BP alone.
The compositions with independent outputs fail on 10,892--19,416, above Relay-BP alone (Figure~\ref{fig:headcurves} in
Appendix~\ref{app:head}).

\subsection{Comparing output heads on the surface code}
\label{sec:head-surface}

We first compare the line vote with a single centre line and a pooled head when trained on the final answer alone, then compose each with correlated PM. We next examine how adding line supervision changes the vote and why.

\head{The line vote improves standalone accuracy.} Without the line supervision, the line vote fails
on 10,661 and 10,822 test shots per million, and the centre line alone on 16,798 and 15,088. Seed 1's pooled head never
trained and failed on 478,245 shots per million, close to always guessing the same answer. \looseness=-1 Seed 2's pooled head trained and still failed on 2,862 more shots per million than the vote.

\head{The line vote retains its advantage in composition.} At $q=1\pct$, the line vote fails on 9,921 and 9,919 shots per million, compared with 12,989 for correlated PM alone. Seed 2's pooled head fails on 2,012 more shots per million than the vote. Both the centre-line and pooled-head comparisons favour the vote at both $q$.

\head{The line supervision hurts the vote.} With it, the vote fails 41\pct{} (seed 1) and 29\pct{} (seed 2) more often,
and the comparison favours the vote without it in both seeds at both $q$ (Appendix~\ref{app:head}). Against the pooled
head, the vote with the line supervision is inconclusive, since only seed 2 favours it.

\head{Why the line supervision hurts.} Each line target repeats the observable's flip up to a known parity, so the line
supervision asks every line for the same answer. Exploratory diagnostics of both seeds on 200,000 validation shots
measure this redundancy. With the line supervision, the noise of the nine lines' scores correlates at 0.84--0.87, so
the lines act as near-copies of one decoder. At the end of training, the line loss's gradient on the trunk is also
2.4--3.6 times that of the final loss. Without it, the correlation falls to 0.23--0.24, and the vote fails 3.7 and
3.8 times less often than its best line alone.

As such, we conclude that on BB72, of the two heads tested, only the joint class score lets the tiny network help
full-DEM Relay-BP at the routing fixed before the test. On the surface code at $d=9$, the vote over equivalent logical
lines beats a pooled head and a single line when the heads learn from the final answer alone. With the line
supervision, the comparison is inconclusive, and the vote itself fails more often.

\section{Experiment III: Sources and Limits of Accuracy Gains}
\label{sec:why}

This experiment asks which shots produce the composition's accuracy gain and under what conditions composition increases failures (\textbf{RQ3}). We decompose the gain shot by shot using the saved predictions of both decoders, relating their errors to the frontend's confidence and the kept and routed subsets. On the surface code, we analyze two networks trained with line supervision at $d=13$ and 17, routing 2 and 8\pct{} of the shots. On BB72, we analyze the comparison with our Cascade. This analysis is exploratory (Appendix~\ref{app:why}). We first explain how complementary errors and confidence-based routing improve accuracy, then identify when routing increases failures.

\subsection{Why composition improves accuracy}
\label{sec:accuracy-gains}

\head[0pt]{The decoders make complementary errors.}
\label{sec:differ}
\looseness=1 A frontend can be less accurate overall and still correct some of the fallback's errors. Recall from Eq.~(\ref{eq:identity}) that only the shots on which exactly one decoder fails separate the composition from its
fallback. These discordant shots split between the two decoders. At $d=13$, only the network fails on 5,416 test shots,
only correlated PM on 3,834, and both on 2,715. At $d=17$, the counts are 7,380, 1,811 and 1,132. The two kinds of
discordant shot look alike in every syndrome measure that we used, and both are near-ties for matching
(Appendix~\ref{app:why}).

\begin{figure}[tbp]
\centering
\includegraphics[width=\columnwidth]{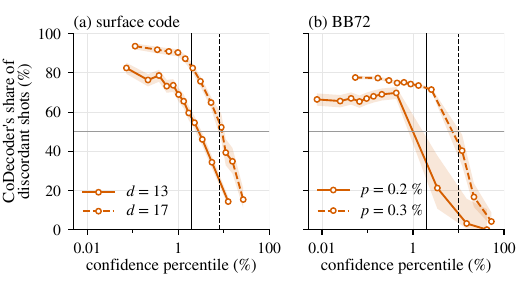}
\caption{Error complementarity versus \sys's confidence percentile (exploratory test shots). The ordinate is \sys's failure share among shots on which exactly one decoder fails, with Wilson 95\pct{} intervals. Vertical lines mark routing thresholds and the grey line marks 50\pct{}. Fallbacks are (a) correlated PM at $d=13$ (solid) and 17 (dashed), and (b) our Cascade on BB72 at $p=0.2\pct$ (solid) and $0.3\pct$ (dashed).}
\label{fig:why}
\end{figure}

\head{Confidence identifies shots that benefit from fallback decoding.}
\label{sec:rank}
\looseness=1 The composition keeps the network's most confident shots, so the confidence has to rank the network's failures. On the test
shots, the network's least confident 1, 2 and 5\pct{} of shots hold 46, 67 and 89\pct{} of its failures at $d=13$ and
43, 65 and 88\pct{} at $d=17$. Random routing would capture a share equal to the routed fraction. Calibration and
development shots show the same concentration on every code, the joint class score's margin on BB codes included
(Appendix~\ref{app:why}).

\looseness=1 These routed shots are hard for both decoders. At the thresholds of these tests, correlated PM fails on 2,882 of the
19,405 routed shots at $d=13$ (14.9\pct), against 0.65\pct{} of all shots, and on 2,229 of 79,652 at $d=17$
(2.8\pct). The network fails on these shots two ($d=13$) to four ($d=17$) times as often, so handing them over removes
more failures than it adds. Calibration preserves this order except for sector Relay-BP on BB72 (Appendix~\ref{app:why}).

\head{The kept shots determine the accuracy gain.}
\label{sec:gain}
\looseness=-1 Figure~\ref{fig:why}(a) joins the two findings. Among the discordant shots, it plots the share that the network fails
against the network's confidence. Below the threshold, the network is the decoder that fails on 72\pct{} ($d=13$) and
85\pct{} ($d=17$) of the discordant shots. Above it, the network fails on 37\pct{} and 35\pct. The threshold therefore
hands correlated PM the shots on which correlated PM is usually the one that is right, and it keeps the shots on which
the network usually is. As a result, the kept shots hold 2,278 shots that only correlated PM fails against 1,356 that
only the network fails at $d=13$, and 520 against 285 at $d=17$. By Eq.~(\ref{eq:identity}), the differences of 922 and 235 failures per million shots give the entire gain over correlated PM.

\label{sec:cascade-why}
\looseness=-1 The same ordering holds for Cascade on BB72 (Figure~\ref{fig:why}(b)).
The frontend is more accurate on the kept shots, while Cascade is more accurate on the routed shots. 
The accuracy gain is smaller because the routed shots contain 96--97\pct{} of Cascade's failures, leaving few errors for the frontend to correct on the kept shots. Most corrections repair a small-margin mistake on one logical qubit using the joint class score over all twelve (Appendix~\ref{app:why}). With the joint decision of Section~\ref{sec:joint}, the network also corrects Cascade on the routed shots, and the composition fails 14--18\pct{} less often than Cascade on BB72 (Section~\ref{sec:rq1n}). Routing hands Cascade the whole 12-bit answer, including bits that the network gets right. On BB72, the joint decision repairs 14 to 49 shots for each one it breaks, and about a third of its repairs are shots that both decoders fail, which it fixes by combining the bits of their answers. At $d=13$, the single logit leaves nothing to combine, and in single mode at $p=0.6\pct$ the joint decision repairs 249 shots while breaking 218.

\subsection{When composition fails}
\label{sec:lose}

\looseness=-1 Recall that a composition beats both of its decoders when the network beats the fallback on the kept shots (C1) and the
fallback beats the network on the routed ones (Section~\ref{sec:overview}). Our tests include fallbacks that miss the
second part. Figure~\ref{fig:frontier} (Appendix~\ref{app:train}) follows each composition's failures on calibration shots as
$q$ grows.

\head{A weaker fallback adds failures.} \looseness=-1 Sector Relay-BP is the weaker decoder on the BB72 network's least confident
shots once $q$ passes about 1\pct. At $p=0.3\pct$, routing 0.5\pct{} to it lowers the network's 1,396 failures to
1,230, and routing 2\pct{} raises them to 1,560. At $p=0.2\pct$, the network alone already beat sector Relay-BP, so the rule kept the fast
mode. In an exploratory run on BB72 at $p=0.1\pct$, full-DEM Relay-BP lost as well. The least confident 0.29\pct{} of
10,000,000 shots held 344 of the network's 348 failures, but full-DEM Relay-BP failed on 469 of these shots, so routing
raised the failures to 473. As such, we conclude that concentration alone is not enough, and the fallback must also
beat the network on the routed shots.

\section{Experiment IV: Runtime Costs and Throughput}
\label{sec:time}
\label{sec:where}

This experiment asks how neural inference, handoff and routed fallback work contribute to the composition's time, and how batching changes its throughput (\textbf{RQ4}). These measurements examine C2 by separating the cost paid on every shot from the fallback work paid only on routed shots (Eq.~(\ref{eq:speedup})). We decompose reaction-mode time, then compare amortized per-round engine costs in batches.

\subsection{Single-shot runtime decomposition}
\label{sec:runtime-breakdown}

We separate frontend inference, handoff and routed fallback work to identify which component dominates reaction-mode time (Table~\ref{tab:time} in Appendix~\ref{app:timing}).

\head{Frontend inference.} On an H100, the resident kernel takes 8.37\us{} per shot at $d=7$. The BB engine takes 88.3\us{} per shot for the BB72 network and 0.254\ms{} for the gross-code network. These GPU inference times form the network term. The composite time adds the handoff and the routed fallback work.

\head{The dominant cost depends on the code family.} The network term takes 66--98\pct{} of the composite time on the surface code, and the routed CPU term at most
28\pct{} ($d=9$). The routed CPU term takes
78--99.9\pct{} in every BB composition but one, the balanced mode on BB72 at $p=0.2\pct$
(23\pct).

\head{Routed shots take longer to decode.} Relative to each fallback's mean decoding time, routed shots cost 1.3--1.9 times as much for correlated PM, 5--17 for sector Relay-BP and 2.5--5.7 for full-DEM Relay-BP.

\subsection{Batched decoding throughput}
\label{sec:batched-throughput}

We decode 4,096 independent memory shots per batch on each card, first comparing frontend engines and then measuring the full composition with Cascade. Costs are amortized per shot or per syndrome round.

\head{The multi-shot engine raises the frontend's throughput.} On the multi-shot engine, a three-block BB72 network takes 1.870\us{} per shot
host to host on the RTX PRO 6000, or 0.312\us{} per round, and 2.82\us{} per shot, or
0.470\us{} per round, on the H100. This cost is 2.20 and 1.98 times lower than that of
the BB engine on the same batches in the same job. The six-block network takes 0.444 and
0.688\us{} per round on the two cards, and Cascade's network 11.94 and 7.29\us{} per
round. Eq.~(\ref{eq:speedup}) predicts the speed-ups of the compositions from these costs (Appendix~\ref{app:timing}).

\head{The compositions cost a fraction of Cascade's time.} \looseness=-1 On the multi-shot engine and the RTX PRO 6000,
the BB72 compositions took 0.733\us{} per round at $p=0.2\pct$ and 1.709\us{} at 0.3\pct, against
12.00\us{} for Cascade alone at both rates in the same job, 16.3 and 7.0
times less. On the H100, they took 0.891 and 1.482\us{} against 7.34\us{} for Cascade alone at both
rates, 8.2 and 4.9 times less. The costs at
$p=0.2\pct$ lie below the round period of current superconducting processors, about 1\us~\cite{google_below}.

These costs measure the capacity of one GPU for one BB72 block on independent memory shots in batches of 4,096, without the
waits of batch formation and of Cascade's queue (Appendix~\ref{app:timing}). In a registered batched timing, \sys +
correlated PM also stays below 1\us{} per round on the surface code up to $d=11$ (Appendix~\ref{app:timing}).

\section{Discussion and Future Work}
\label{sec:discussion}

Sections~\ref{sec:rq1} to~\ref{sec:time} tested trained networks on memory experiments, timed on two GPU models with
CPU times on the Grace basis. In this section, we discuss how far these results reach and give our plan for each
limit.

\looseness=-1 \head{The CPU basis decides the narrowest margins.} Every Grace value scales a time measured on one core of our
workstation by a factor fitted over NVIDIA's 18 published CPU baseline timings. On this basis, PM is faster than
\sys + correlated PM at $d=7$. Without the conversion, PM is also faster at $d=9$ and 11, and the balanced mode on the gross code
becomes faster than BP-OSD from CUDA-Q (Appendix~\ref{app:timing}). 

The speed claims of the classical tests are
therefore claims about composite benchmark costs on the Grace basis. NVIDIA's own accounting of its Ising pre-decoders
also times PM on a Grace CPU and leaves out the transfer of syndrome data between the GPU and the CPU, which its
NVQLink estimates find negligible~\cite{ising}.
We compared speeds on the GPU and CPU hosts that we measured, not on FPGAs, where the accurate mode would still fail
less often than full-DEM Relay-BP and cut a hardware fallback's load to 0.245--2\pct{} of the shots.

\looseness=-1 \head{Codes with many logical qubits.} The cost of the joint class score doubles with every added logical qubit
(Section~\ref{sec:bb-net}), so the score does not extend directly to codes with many more logical qubits. Such codes
need another way to find the best class, such as a search over classes. We will build one and measure the time and
memory of the score apart from those of the trunk.

\head{Scope of the head comparison.} Section~\ref{sec:rq2} compared each of \sys's heads with one pooled head on
one trunk, at one noise level and in two seeds, which give only an observed range of training runs. This comparison
does not tell which part of the joint class score causes its advantage. Our surface-code networks grew through a
curriculum that the comparison did not test, in one training run per distance. We will add seeds and larger distances
to the comparison of heads, repeat the training of the surface-code networks, and test the redundancy of
Section~\ref{sec:head-surface} directly with smaller line weights.

\head{Long computations.} The final measurement of every data qubit defines the observables and the parities $\delta$
that the vote and the score use. A long computation, however, is decoded in windows that end without it.
Parallel window decoders~\cite{skoric_window,tan_window} let these windows be decoded in parallel, as the throughput mode
needs.
We will run the frontend on parallel windows and build the multi-shot engine for the surface-code networks.

\head{Feed-forward points and tails.} At a feed-forward point~\cite{battistel}, the tail matters. Google's real-time
decoder for a distance-5 surface code, for example, had an average latency of 63\us{} over a 1.1\us{}
cycle~\cite{google_below}. In front of our Cascades, every routed decoder had a higher
99.9th-percentile single-shot latency than Cascade alone on both cards, because a routed shot passes through both
networks. 

On BB72, the compositions were 7.2--10.3\pct{} higher on the RTX PRO 6000 and
9.7--19.1\pct{} higher on the H100. At $d=13$ and $p=0.6$ and 0.35\pct, the composition took
486 and 499\us{} against 397 and 402\us{} for Cascade alone
on the RTX PRO 6000, and 626 and 574\us{} against 546 and
542\us{} on the H100. We will bound the tail by capping the fallback's time per routed shot. A cap changes the decision policy, so its accuracy must be re-evaluated.

\head{Limits of routing.} Routing helps only where the fallback beats the network on the routed shots and the network
beats the fallback on the kept ones (Section~\ref{sec:lose}). The thresholds also assume that the decoder knows $p$. We
will choose the fallback, the depth and $q$ per noise level from calibration shots, as the rule already does for $q$.

\head{Better scores for the joint decision.} \looseness=-1 The joint decision gains only where both decoders' class scores tell
which one is right, and correlated PM and Relay-BP report no such score. A matching decoder can supply the complementary gap, the weight difference between the best correction in the
other logical class and its own, which approximates the log-likelihood ratio of the two
classes~\cite{meister_soft,smith_gap}. In an exploratory look at our surface-code test shots, we took this gap from a
second correlated decode forced into the other class. With 2\pct{} of the shots routed, a joint decision with this gap
failed 5--18\pct{} less often than \sys + correlated PM in Table~\ref{tab:surface} at $d=11$--17. The second decode
runs only on the routed shots where the decisions differ, 17--27\pct{} of them. A BP decoder's posterior marginals give
approximate log-likelihood ratios of the observables. When Relay-BP's valid corrections fall in
different logical classes, their weight gap needs no extra decoding. At $d=13$, Cascade reports a logit, but the network's
single logit is weakest on the shots that it routes (Section~\ref{sec:rq1n}). There, the votes of the individual
logical lines, which the head sums into that logit, may still tell which decoder is right. We will test these scores
on fresh shots, with their timing.

\head{Training the network for its fallback.} \looseness=-1 We train the network alone on the final answer, so it never learns
which shots its fallback gets wrong. In an exploratory look at $d=13$, a perfect choice between the two answers on the
routed shots would remove 30\pct{} of the composition's failures. We will train the network to complement its fallback, weighting the loss toward
the fallback's failures or training it with its deferral, as learning to defer does~\cite{mozannar_defer}, and measure how much of the 30\pct{} it recovers on fresh shots.

\head{Scope of the Cascade results.} Our Cascade comparisons cover two re-implementations, each trained once per
setting and timed on two GPU models. Neither is the authors' implementation. On BB72, the circuit-level noise model is fitted to the
failure rates of Cascade's paper. At $d=13$, the width is inferred, so a wider network may be more accurate and slower
(Appendix~\ref{app:cascade}). A second training seed or a mismatched noise level with the routing policy held fixed
would test whether both results hold beyond these conditions.

\section{Related Work}
\label{sec:related}

\head[0pt]{Pre-decoders and hierarchical decoders.} \looseness=-1 Hierarchical decoders~\cite{delfosse_hier} and pre-decoders~\cite{chamberland_pre,clique,promatch} let a cheap first stage handle easy syndromes and pass the rest, or the residual of every shot, to a global decoder.
NVIDIA's Ising pre-decoders~\cite{ising} follow this pattern on a GPU, and a larger unreleased one reportedly beats correlated matching up to distance 13.
LATTE~\cite{latte} runs a convolutional local decoder of about 3,000 parameters on an FPGA, applies the local corrections it is confident about, and sends the remaining syndrome to sparse blossom on the CPU.
At $p=0.1\pct$, it matches the base decoder's accuracy with a 6.4 times average speed-up per block.
In these designs, the global decoder still sees most shots, or the composition aims at its accuracy.
\sys predicts the whole shot's logical outcome and routes at most 10\pct{} to the fallback.

\head{Decoders that route by confidence.} \looseness=-1 Decoder Switching~\cite{decoder_switching} hands a decoding window to belief-matching when the gap of a soft-output matching or union-find decoder is small.
Chongder~\cite{chongder} escalates to PyMatching the 3.3--6.2\pct{} of shots with the lowest maximum softmax probability of a small CPU network.
Other decoders use confidence as soft output~\cite{meister_soft}, for post-selection and abstention~\cite{smith_gap,bombin_ps}, for decoding an outer code~\cite{yoked}, and as the agreement of an ensemble~\cite{harmony}.
\sys also routes by confidence but differs in its output head, GPU engines, joint decision with a neural fallback and tests, which compare each composition with its fallback alone up to distance 17, on two BB codes and in front of two neural decoders.

\head{Learning to defer.} \looseness=-1 A model that may pass a decision to a downstream expert learns to defer~\cite{madras_defer}, and Mozannar and Sontag~\cite{mozannar_defer} train the classifier and the deferral rule together with a consistent surrogate loss.
\sys with a strong decoder is such a system, whose rule thresholds the network's confidence alone.
This simple rule sufficed in front of matching, and in front of our Cascade on BB72, a joint decision on the routed shots gave 14--18\pct{} fewer failures than Cascade alone (Section~\ref{sec:rq1n}).

\head{Neural decoders.} \looseness=-1 They range from early networks for topological codes~\cite{torlai,meinerz} to decoders for experimental data~\cite{varbanov} and AlphaQubit's recurrent transformer~\cite{alphaqubit}.
AlphaQubit~2~\cite{alphaqubit2} reaches near-optimal logical error rates on surface and color codes and decodes the surface code in real time up to distance 11.
Among recent BB-code networks~\cite{blue_bb,cascade,ntu}, Cascade convolves along the geometric structure of the code with up to 14 blocks of width 512, and its group also built an attention-based decoder for logical circuits~\cite{bonilla_universal}.
Yan \emph{et al.}~\cite{ntu} train transformer foundation decoders that surpass correlation-aware matching on a distance-19 planar surface code and Relay-BP on BB72 at low physical error rates.
These networks generally read out each logical observable with a separate logit, as does the pooled head in Section~\ref{sec:rq2}, and can serve as \sys's fallbacks, as our Cascade re-implementations do.

\enlargethispage{2\baselineskip}%
\head{Joint logical predictions.} \looseness=-1 Wagner \emph{et al.}~\cite{wagner_hld} post-correct an underlying toric-code decoder with a network whose softmax output names one of the 16 logical classes of the two encoded qubits, and map each syndrome to a canonical translate.
END~\cite{end_decoder} scores each of the 16 classes at every lattice position and averages the scores over torus translations after relabelling them by syndrome parities, which makes it equivariant.
\sys-BB shares the joint class and the parity correction with END, but gives one flip logit per translated logical operator and maps the logits to classes through the GF(2) map $\phi_{ag}$ of Eq.~(\ref{eq:translate-flip}), which BB codes need because their translates are other logical operators.
Its joint class margin is \sys-BB's routing confidence.

\head{Fast decoding systems.} \looseness=-1 Dennis \emph{et al.} proposed minimum-weight perfect matching for the surface code~\cite{dennis}, and sparse blossom made it fast~\cite{sparseblossom}.
Union-find trades accuracy for almost-linear time~\cite{unionfind}, Micro Blossom accelerates matching in hardware~\cite{microblossom}, localized statistics decoding speeds up OSD~\cite{lsd}, and Tesseract searches for the most probable error with A*~\cite{tesseract}.
For BB codes, IBM's FPGA decoder runs sector Relay-BP~\cite{relaybp} on synthetic gross-code syndrome streams at 24\,ns per BP iteration, enough for an average decoding time under 1\us{} per cycle below $p=0.3\pct$~\cite{relaybp_fpga}.
A study of FPGA-tailored qLDPC decoders~\cite{fpga_qldpc} finds that Relay-BP's tail latencies exceed the real-time budget without problematic backlog in its examples, and Vegapunk~\cite{vegapunk} decodes BB codes up to $[[784,24,24]]$ in real time.
Each of these decoders can serve as \sys's fallback.

\section{Conclusion}
\label{sec:conclusion}

We presented \sys, a small neural frontend that composes with an existing decoder through confidence-based routing. Its logical-operator-aware output head produces predictions and confidence, while fused GPU inference and efficient handoff reduce execution costs. Under circuit-level noise, composition reduced failures by up to 50\pct{} relative to correlated matching and full-DEM Relay-BP, with 2.3--124-fold speed-ups. With our re-implementations of Cascade, composition reduced decoding time in all four settings. A joint decision on the routed shots kept this speed and made the composition more accurate than Cascade in all four, with 14--18\pct{} fewer failures on BB72. Matched head comparisons support the output design. On BB72, independent logical outputs failed 3.8--5.0 times as often as the joint class score. On the surface code, the line vote outperformed the simpler heads when trained on the final answer alone.

Per-shot analysis explains how a frontend can improve a fallback's accuracy even when it is less accurate overall. On retained shots, it corrects more fallback errors than it introduces. To also improve on the frontend alone, the fallback must be more accurate on routed shots. A joint decision with a fallback that reports logits also corrects the fallback on routed shots. Speed gains require the saved fallback work to exceed frontend inference and handoff costs. These conditions connect complementary predictions and confidence ranking to efficient execution, and define when composition can improve accuracy and speed without changing the fallback.

\section*{Data Availability}
The decoder implementations comprise about 7,700 lines of Python, Triton, CUDA and C++. We release these implementations, the training code and the trained weights under the Apache License 2.0, together with the circuits and records supporting this study~\cite{convlines_code}. Appendix~\ref{app:names} describes the release contents.

\begin{acknowledgments}
We thank Yidong Zhou (Rutgers University) for checking and reproducing our artifacts.
We used Anthropic's Claude Opus 5.5 for implementation and for polishing the text, and OpenAI's GPT-6 Astra for review.
We checked every result and take full responsibility. X.G. acknowledges funding from the U.S. Department of Energy, Quantum Systems Accelerator.
\end{acknowledgments}

\clearpage
\appendix
\section{Training and Routing Details}
\label{app:train}

This appendix gives the circuits and grids, the architecture details and sizes of the networks, the BB-code head,
the losses and growth, the optimizer, the budgets and the cost of training, and the rule that chose each routed
fraction, with the routing frontiers it read.

\head{Circuits and grids.} NVIDIA's single-parameter circuit-level noise sets two-qubit-gate and idle errors at $p$ and
preparation and measurement errors at $2p/3$. NVIDIA's circuits leave the last round and the final readout of the data qubits
noiseless. A surface-code shot has 336 detectors at $d=7$ and 4,896 at $d=17$, and its
detector grid has $8\times8\times8$ sites at $d=7$ and $18\times18\times18$ at $d=17$. The checks of both BB codes
are built from the two polynomials $A=x^3+y+y^2$ and $B=y^3+x+x^2$, whose monomials shift the torus. The Relay-BP circuits set
two-qubit gate, idle, reset and measurement errors at $p$ in every round and leave only the final readout of the data
qubits noiseless. A BB72 shot has
432 detectors on a $7\times6\times6$ grid, and a gross-code shot has 1,728 on a $13\times12\times6$ grid.

\head{Networks.} We chose the $3\times5\times5$ stem kernel for its reach. Every matching edge and every X-Z pair of one
fault mechanism lies within one layer and two sites, so one stem step sees every mechanism that touches a site. A
$3\times3\times3$ box would cover 95.5\pct{} of the edge probability mass. The surface-code stem embeds the detector
bits with four constant channels, the site's type and whether it lies in the first or the last layer. Its weights are
separate for X- and Z-centred sites, since the same offset means a different edge for the two types. Each block applies
LayerNorm, projects $C$ channels down to $C/4$, applies a type-aware $3\times3\times3$ convolution with the SiLU
activation (the input times its sigmoid), and projects back up through a zero-initialized layer. LayerNorm replaces
batch normalization. The reason is that batch statistics would not follow the averaged weights that the network ships
with. The logical-line head's values are linear projections of the site features. Its gate takes the query-key
product, the dot product of the line's query with the site's key (another linear projection), plus a bias for the
site's signed distance to the line, its time layer and its type. The six surface-code networks differ in depth and
width. They have one block of width 64 at $d=7$ and 9 (90,257 and 90,467 parameters) and five blocks of width 128 at
$d=11$--17 (517,669--518,683). The main test's $d=9$ network had three blocks of width 128 (389,395 parameters). The BB stem's
taps, the offsets that a convolution reads, are the DEM's co-firing offsets, the shifts between two detectors that one fault flips. One stem step thus sees
every fault that touches a site, including those of the far-reaching monomials $x^3$ and $y^3$. Each BB block's taps
are the Tanner offsets between checks that share a data qubit, at three time offsets. The BB stem has 96 co-firing taps
on BB72 and 131 on the gross code, and each BB block has 26 and 29 Tanner offsets. The BB72 networks have three blocks
of width 128 (483,357 parameters, 118 million multiply-adds per shot), and the gross-code network has 12 (2,723,613
parameters).

\head{The BB-code head.} A vote over the translates equivalent to one logical would have a single member for 10 of the
12 BB72 logicals and two members on the gross code, too few to vote. This is why \sys-BB asks one query per
logical $a$ and translate $g\in G$, the $lm$ shifts of the torus, and scores whole classes by
Eq.~(\ref{eq:bbvote}). The $2^k$ classes are the flip patterns of the observables that an X-basis memory experiment
measures, not the $4^k$ classes of logical Pauli errors. Algorithm~\ref{alg:score} gives the head with the sizes of BB72. Shifting a shot by $u$ moves $z_{ag}$ to $z_{a,g+u}$, so the trunk and the translated logits are equivariant. The complete score need not be,
because a shift also maps the shot's class to another class by a map that depends on the detector parities, and the
bias term of Eq.~(\ref{eq:bbvote}) is not invariant under this map. Training forms the label $t_{ag}$ from the true class, and
decoding knows only $\delta_{ag}$. A translate whose weighted logit $w_az_{ag}$ is positive raises the score of every
class under which it flips and lowers the score of every other class.

\begin{algorithm}[tbp]
\caption{The joint class score of \sys-BB, with the sizes of BB72}
\label{alg:score}
\footnotesize
\begin{algorithmic}[1]
\REQUIRE detector bits $\xi\in\{0,1\}^{432}$, trunk features $f_s\in\mathbb{R}^{128}$ at the 252 sites $s$ of the
$7\times6\times6$ grid
\ENSURE decided class $y^*$ and its margin
\STATE \emph{Once per code}
\FORALL{logicals $a$ and shifts $g$ of the torus}
\STATE solve $\pi_g\ell_a=\sum_j\mu_{ag,j}\ell_j+\sum_{r\in\mathcal{C}_{ag}}H_r$ over GF(2)
\STATE $\phi_{ag}(y)\gets\sum_j\mu_{ag,j}\,y_j\bmod 2$
\STATE $P_{i,ag}\gets1$ for every X detector $i$ of a check in $\mathcal{C}_{ag}$
\ENDFOR
\STATE \emph{Per shot}
\STATE $\delta\gets\xi P\bmod 2$
\FORALL{sites $s$ and heads $h$}
\STATE $n_s\gets\mathrm{LN}(f_s)$ and $v_{sh}\gets(W_Vn_s)_h$
\STATE $\gamma_{sh}\gets\sigma\bigl((W_Kn_s)_h\cdot \zeta_h/\sqrt{32}+b_h\bigr)$
\ENDFOR
\FORALL{432 queries $(a,g)$}
\STATE $\bar v_{ag,h}\gets\frac{1}{64}\sum_s\sigma(\rho_{a,h,t(s),s-g})\,\gamma_{sh}v_{sh}$ for each head $h$
\STATE $z_{ag}\gets\mathrm{MLP}(\zeta+e_a+W_O\bar v_{ag})$
\ENDFOR
\FORALL{classes $y\in\{0,1\}^{12}$, in the order of the index $\sum_j2^jy_j$}
\STATE $S(y)\gets\sum_{a,g}w_a\bigl(\phi_{ag}(y)\oplus\delta_{ag}-\tfrac12\bigr)z_{ag}+\sum_j\beta_j\bigl(y_j-\tfrac12\bigr)$
\ENDFOR
\RETURN $y^*=\arg\max_yS(y)$ and its margin $S(y^*)-\max_{y\neq y^*}S(y)$
\end{algorithmic}
\end{algorithm}

\head{Queries and sizes.} All $12lm$ queries share one query vector $\zeta$, the key, value and output projections $W_K$,
$W_V$ and $W_O$ of size $128\times128$ (the key without bias), a gate bias $b_h$ per head, and the MLP, a LayerNorm followed by linear layers of
$128\to256$ and $256\to1$ with SiLU between them. Each logical $a$ adds an embedding $e_a$ and a table $\rho_a$ of position
biases over the 4 heads of 32 channels, the $T$ time layers and the $lm$ offsets $s-g$ of a site from the translate.
The table starts at $-1$ on the X checks that share a data qubit with $\ell_a$ and at $-3$ elsewhere, the weights at $w_a=1/lm$ and
the biases at $\beta_j=0$. A site's gate is thus a content gate shared by all queries times a position gate, and each
query divides its sum over the sites by 64. On BB72 the head takes 45.5 million multiply-adds per shot, 8.3 million for
the projections and the content gates, 13.9 million for the sums, 21.3 million for the output projections and the MLPs
and 2.0 million for the parities and the scores.

\head{Logicals, parities and class bits.} The X logicals $\ell_j$ are the circuit's observables $j=0,\dots,k-1$, and
the X stabilizers $H_r$ are read off its final detectors. Elimination over GF(2) writes each translate $\pi_g\ell_a$, the shift of $\ell_a$ by $x^{g_1}y^{g_2}$ with
$g=g_1m+g_2$, as a product of logicals, which gives $\phi_{ag}$, times a set $\mathcal{C}_{ag}$ of X stabilizers. The parity $\delta_{ag}$ sums every X detector of a check in $\mathcal{C}_{ag}$, in every round and in the final
layer. This sum is the final data parity on the product of these checks, which the translate fixes, so any valid
choice of $\mathcal{C}_{ag}$ gives the same $\delta_{ag}$. Class $y$ has the index $\sum_j2^jy_j$, with bit $j$ recording observable flip $y_j$.

\head{The joint score for one logical.} With one logical whose translates $j$ all have $\phi_j(y)=y$,
$S(1)-S(0)=\sum_j w(-1)^{\delta_j}z_j+\beta$. This difference is Eq.~(\ref{eq:vote}) with
$w_j=w$ and $b_0=\beta$, and the margin is $|c|$. The decided class changes under a shift, since a shift maps logical
classes to other classes.

\head{Losses and growth.} On surface codes the loss is the binary cross-entropy of $c$ alone. The networks of our
earlier surface-code tests added the binary cross-entropy of every line score. On BB codes the loss is the
cross-entropy of the score's softmax over the 4,096 classes plus the binary cross-entropy of the translate logits
against their flips $t_{ag}$, averaged over the $12lm$ translates and the batch and added with weight 1. A $d=13$
network trained from scratch stayed at the constant-guess loss. We therefore grow the surface-code networks from the
first training stage of the main test's three-block $d=9$ network in steps of two distances, to $d=11$, 13, 15 and 17. Each step copies the stem, the blocks and the projections and
interpolates the line embeddings and the vote weights, and a new block starts as the identity through its
zero-initialized up-projection. The networks at $d=7$ and 9 train from scratch. The gross-code network also stayed at
chance when trained from scratch, so its first three blocks trained while $p$ rose from 0.1 to 0.4\pct{} before it
grew to 6 and 12 blocks.

{\widowpenalty=10000 %
\head{Optimizer.} We use AdamW (momentum 0.9 and 0.99, weight decay 0.01) on batches of 1,024 shots. The learning rate
peaks at $10^{-3}$ after 1,000 warm-up steps (500 on BB codes) and decays along a cosine to 2\pct{} of the peak. At
$d=17$ and in the stages that continue a network at its own distance, it peaks at $2\times10^{-4}$ after 500 steps. We
clip gradients at 1 and compute in bf16. We deploy an exponential moving average of the weights (0.999).\par}

\head{Training budgets.} Over their chains, the surface-code networks saw $2\times10^8$ shots at $d=7$, $4\times10^8$
at $d=9$, $8\times10^8$ at $d=11$, $6\times10^8$ at $d=13$, $9\times10^8$ at $d=15$ and $10^9$ at $d=17$, as many as
the networks of our earlier tests. Each stage saw $2\times10^8$ shots, except the continuations at $d=11$
($4\times10^8$) and $d=15$ ($10^8$). The gross-code network saw $2.4\times10^8$ shots over its three stages of 3, 6
and 12 blocks. Both BB72 networks trained from scratch at $p=0.3\pct$ on $2\times10^8$ independent shots~each. The
one-block network at $d=9$ saw the three-block network's $4\times10^8$ shots in the same order, in the same two stages,
and took 1.46\,h on one H100.

\head{Training cost.} The surface-code stages ran on an RTX PRO 6000 Blackwell GPU at $d=7$ and for the continuation
at $d=9$, and on one H100 for the first stage at $d=11$. From there on they ran on two or four H100s, where each step
averages the gradient of one batch of 1,024 shots over the GPUs. The chain from $d=11$ to 17 took 11\,h. The first stage of
the three-block $d=9$ network is the line vote of the output-head comparison trained on the final answer in seed~1
(Appendix~\ref{app:head}).
Every stage passed the check for runs that do not train. Each BB72 network took 3.0\,h on an H200 slice, and the
gross-code network took 5.8\,h on an H200 slice and 7.7\,h on an H100. Two BB72 networks trained independently at
$p=0.3\pct$ fail on 1,376 and 1,398 of the same 200,000 test shots. Two surface-code networks of the recipe of the
first stage at $d=9$ fail on 10,661 and 10,822 of the same 1,000,000.

\head{The networks of the Cascade comparison.} We trained a three-block BB72 network from scratch in each noise setting
of the Cascade comparison, at $p=0.55\pct$ on about $2\times10^8$ shots. The six-block network grew from the final EMA
weights of the fitted three-block network. Its first three blocks are copies, and three new blocks start as the
identity through their zero-initialized up-projections. On 100,000 monitoring shots, the grown network decided every
shot as the three-block network did before training began. It then trained with a fresh optimizer, schedule and EMA on
another $2\times10^8$ shots, about $4\times10^8$ in all. Each of these trainings took 3.9--6.1\,h on one RTX PRO 6000,
the six-block stage the longest.

\head{Choosing the routed fraction.} Recall that a rule fixed before each test chooses $q$ on calibration shots alone
(Section~\ref{sec:threshold}). On surface codes, ranking by $|c|$ equals ranking by the larger sigmoid probability,
since $\max\{\sigma(c),1-\sigma(c)\}=\sigma(|c|)$ for the logistic sigmoid $\sigma$. The rule searches $q$ from 0 to 16\pct{} on surface codes (30\pct{} at $d=17$) and from 0 to
20\pct{} on BB codes. Among the $q$ within the speed limit, it takes the smallest $q$ with at most $M-2\sqrt{2M}$
calibration failures. The target $M$ is a planning count near the most accurate baseline decoder's on surface codes and
90\pct{} of sector Relay-BP's on BB codes. The margin $2\sqrt{2M}$ is two standard deviations of the difference of two
independent counts near $M$. It guides the choice of $q$, and the paired tests of Section~\ref{sec:setup} decide each
claim. If no $q$ qualifies, the rule falls back to the $q$ with the fewest calibration failures, a choice also fixed
before the test. No distance of the main test needed this fallback, but the one-block network at $d=9$ did.

\head{The chosen fractions.} On 200,000 fresh calibration shots per distance, the rule chose $q=2$, 0, 0, 0, 1 and
1\pct{} for the main test's networks at $d=7$ to 17, and the test sets routed 2.00, 0, 0, 0, 0.96 and 1.00\pct. The
rule takes the smallest qualifying $q$, and at $d=9$--13 these networks alone already qualified. For the one-block network at $d=9$, the rule chose
6\pct{} on its own 200,000 calibration shots, on which the network alone failed 3,784 times, the composition at 6\pct{} failed 2,508 times and correlated PM 2,631. On BB72 the rule chose the fast mode, since the network alone
already failed far less often than sector Relay-BP (146 against 523 calibration failures at $p=0.2\pct$). The accurate
mode routes 0.25\pct{} ($p=0.2\pct$) and 2\pct{} ($p=0.3\pct$) on BB72 and 2\pct{} on the gross code, whose balanced
mode routes 1\pct. Section~\ref{sec:rq1n} and Appendix~\ref{app:cascade} detail Cascade configuration selection, and Appendix~\ref{app:reg} the gross-code modes.

\begin{figure}[tbp]
\centering
\includegraphics[width=\columnwidth]{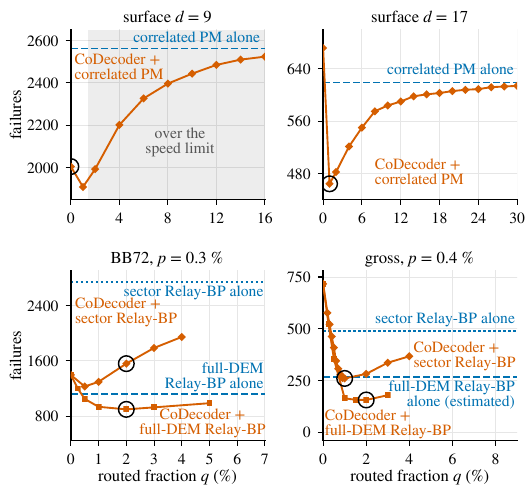}
\caption{Routing frontiers on calibration shots. Failures of \sys with each fallback against the routed fraction $q$, from
\sys alone at $q=0$, with each fallback alone as a blue line. Rings mark the registered $q$ (CPU terms in Table~\ref{tab:time}). Grey marks $q$ exceeding the frozen speed limit. The gross full-DEM
count of 268 combines its least confident 3\pct{} with a 4,100-shot sample (95\pct{} interval 199--415).}
\label{fig:frontier}
\end{figure}

\head{Routing frontiers.} Figure~\ref{fig:frontier} follows the failures of the composition on calibration shots as $q$
grows. At $d=9$ the main test's three-block network alone fails on 2,005 of 200,000 calibration shots and correlated
PM on 2,562. Routing
1\pct{} leaves 1,909 failures, and the speed limit that the rule was frozen with admits no larger $q$. At $d=17$ the network alone fails on 672 and
correlated PM on 619. Routing 1\pct{} leaves 464 failures, and routing more raises them again, to 614 at 30\pct.
Routing to full-DEM Relay-BP on BB72 behaves the same way. At $p=0.3\pct$ it cuts the network's 1,396 failures to 900
at $q=2\pct$. The failures rise again to 989 at 5\pct, where most extra routed shots are ones that the network decides
correctly.

\head{The routed fraction and the network's deficit.} The main test's networks fall behind correlated PM on the
calibration shots only at $d=7$ and 17, by factors of 1.01 and 1.09, so little routing suffices. Their fewest
calibration failures come at $q=2\pct$ at $d=7$ and at 1\pct{} at $d=9$--17. The one-block network at $d=9$ falls
behind by a factor of 1.44, and its fewest calibration failures come at 6\pct. In our earlier tests, the network at $d=17$ alone failed 2.9
times as often as correlated PM, and its composition routed 8\pct{} of the shots.

\section{Engine Details}
\label{app:engines}

This appendix gives the kernel details of the engines of Section~\ref{sec:threshold} and the check of the multi-shot
engine against the BB engine.

\head{Fused engine.} Before deployment, we fold the constant input channels through the stem into a per-site table
and fold the head's query-key product into one matrix. Each block kernel computes one matrix product per tap on a
zero-padded dense grid. The 27 gathers
of a $3\times3\times3$ kernel are thus independent loads at constant offsets. Each block's epilogue prepares the next
block's input, and the final block also computes the head's partial sums. An atomic counter finds the last head
program to finish, which forms $c$. Matrix inputs are fp16 with fp32 accumulation. Each kernel after the first starts
early through programmatic dependent launch and waits only where it reads its predecessor's output.

\head{Resident kernel.} The resident $d=7$ kernel runs as one cluster of seven thread blocks with the weights in
shared memory. Block 0 polls a request line in mapped host memory, forwards each shot to the other blocks over
distributed shared memory, and publishes the logit with one system-scope release store.
The kernel multiplies on tensor cores in fp16 with fp32 accumulation. It uses 7 of 132 H100 multiprocessors and one spinning host core.

\head{Compiled router.} The C++ router copies each shot's detector bits into the graph's input, launches the graph
and spins on a word in mapped host memory. It then returns the sign of $c$ or calls correlated PM, compiled from
PyMatching 2.4.0 into the same library. A one-thread tail kernel writes the word on which the router spins, with a
launch counter and the logit.

\head{BB engine.} The BB engine runs \sys-BB in $L+4$ Triton kernels, the stem, the blocks, pooling, the vote and
the top two. Without atomics, its outputs are deterministic. On the gross code it runs with launch settings tuned
for the H100.

\head{Screening.} At $d=7$ the fused engine is slower than PM, 14.5 against 11.89\us{} per shot in the speed re-test,
so we wrote the resident kernel of
Section~\ref{sec:threshold}. On the gross code, the tuned BB engine takes
0.254\ms{} on an H100 against 3.43\ms{} for the fp32 PyTorch CUDA graph of the earlier test, both timed in the speed
re-test.

\head{Multi-shot engine.} A program of the multi-shot engine loads each weight tile once for its group of shots, and
each shot keeps the arithmetic of the BB engine. The launch settings depend on the batch size, in six tiers. In
88,750,000 shot evaluations of five networks at 14 batch sizes from 1 to 4,100 on an RTX PRO 6000, no difference from
the BB engine was observed in any decision or in the bit pattern of any FP32 margin. These are repeated evaluations of
the same shot sets, not independent~shots.

\section{Statistics and Detailed Comparisons}
\label{app:stats}

This appendix reports statistics for the classical, Decoder Switching, Cascade and output-head comparisons.

\head{Classical tests.} Each comparison applies the exact two-sided McNemar test and Newcombe's paired interval of the
difference in failure rate, which must lie below zero. The level is $\alpha=0.025$ on surface codes and $0.05/3$ on BB
codes. A claim needs all of its comparisons and its speed condition, a direct comparison of times. It is thus an
intersection-union test, whose error rate stays at most the per-comparison $\alpha$. At $d=15$ and 17, BP-OSD read the
first 200,000 test shots, paired with the predictions of \sys + correlated PM on the same subset.

\head{Comparisons on the surface code.} Table~\ref{tab:surface} gives every decoder's failures and composite time. In
the main test, at each distance, the claim needs eight accuracy and eight speed comparisons, with PM, correlated PM,
the four Ising pipelines and BP-OSD in two settings. All 48 accuracy comparisons of the main test hold, each with a
97.5\pct{} Newcombe interval below zero. The weakest is the comparison with correlated PM at $d=13$, 5,318 against 6,344 failures (McNemar
$p=4.8\times10^{-35}$). Every other p-value lies below $10^{-58}$, and 41 of the 48 lie below $10^{-300}$. On the first
200,000 test shots at $d=15$, \sys + correlated PM fails 586 times against 2,361 and 3,368 for BP-OSD in its two settings, and at
$d=17$ 439 times against 2,171 and 3,046. Of the main test's 48 speed comparisons, 46 hold. The two exceptions are the
comparisons with PM at $d=7$ and 9, where PM takes 11.89 and 26.30\us{} against 11.94\us{} for \sys + correlated PM and 28.93\us{}
for its composition with the three-block network. The main test's claim against every baseline decoder therefore holds at $d=11$--17 and
fails at $d=7$ and 9. Each speed comparison is a direct inequality of composite times on the Grace basis, without a
statistical~test. At $d=9$, the second registered test compared a one-block network's composition with PM and correlated PM on
other shots, and both comparisons hold (Appendix~\ref{app:reg}). The surface-code results of this paper thus rest on 42
paired accuracy comparisons, the main test's 40 at the other distances and the second test's two, and all of them
hold. At $d=9$, the one-block composition is also faster than all eight baselines in the speed re-test.

\begin{table*}[tp]
\centering
\caption{The surface-code tests at $p=0.6\pct$. At every distance but $d=9$, \sys + correlated PM fails less often than every
baseline decoder and our Decoder Switching. At $d=9$, the second test shows this against PM and correlated PM, and the
other entries are context from the main test. \sys + correlated PM takes less time than each decoder except PM
at $d=7$.}
\label{tab:surface}
\footnotesize
\setlength{\tabcolsep}{3.3pt}
\begin{adjustbox}{max width=\linewidth}
\begin{tabular}{@{}l rrrrrr rrrrrr@{}}
\toprule
& \multicolumn{6}{c}{failures} & \multicolumn{6}{c}{composite time per shot on the Grace basis (\textmu s)} \\
\cmidrule(lr){2-7}\cmidrule(l){8-13}
decoder & $d=7$ & 9 & 11 & 13 & 15 & 17 & $d=7$ & 9 & 11 & 13 & 15 & 17 \\
\midrule
\sys + correlated PM & 31,629 & 11,978\rlap{$^{b}$} & 6,887 & 5,318 & 3,012 & 2,180 & 11.94 & 21.2\rlap{$^{b}$} & 41.5 & 51.2 & 59.6 &
81.6 \\
\sys & 36,156 & 18,305\rlap{$^{b}$} & 6,887 & 5,318 & 4,030 & 3,296 & 10.90 & 15.3\rlap{$^{b}$} & 41.5 & 51.2 &
52.3 & 69.6 \\
\midrule
PM & 60,016 & 27,028\rlap{$^{b}$} & 24,365 & 21,086 & 18,259 & 15,726 & 11.89 & 26.3 & 52.2 & 91.7 & 160.1 & 242.5 \\
correlated PM & 35,135 & 12,802\rlap{$^{b}$} & 9,252 & 6,344 & 4,396 & 2,973 & 27.60 & 63.9 & 137.9 & 253.1 & 562.7 &
846.0 \\
Ising-Fast + PM & 48,752 & 22,908 & 20,650 & 17,491 & 15,043 & 13,149 & 47.14 & 57.2 & 69.3 & 92.3 & 123.8 & 173.1 \\
Ising-Accurate + PM & 38,157 & 17,400 & 15,040 & 12,440 & 9,932 & 8,083 & 58.99 & 67.2 & 78.6 & 100.8 & 128.2 & 173.0 \\
Ising-Fast + correlated PM & 48,873 & 22,856 & 20,405 & 17,101 & 14,733 & 12,764 & 51.07 & 69.8 & 99.7 & 153.9 & 238.7 &
354.4 \\
Ising-Accurate + correlated PM & 38,204 & 17,379 & 14,977 & 12,298 & 9,769 & 7,918 & 61.91 & 77.3 & 102.5 & 149.0 & 218.4
& 321.3 \\
BP-OSD, min-LLR & 34,385 & 16,409 & 15,250 & 13,405 & 2,361\rlap{$^{a}$} & 2,171\rlap{$^{a}$} & 2,184 & 8,655 & 31,365 &
87,743 & 193,252 & 478,767 \\
BP-OSD, 60 iterations & 41,186 & 19,959 & 19,033 & 17,791 & 3,368\rlap{$^{a}$} & 3,046\rlap{$^{a}$} & 5,951 & 19,034 &
52,051 & 130,649 & 287,949 & 647,847 \\
\midrule
our Decoder Switching & 35,322 & 13,159 & 9,512 & 6,594 & 4,609 & 2,990 & 57.41 & 139.0 & 313.3 & 639.7 & 1,217.2 &
2,402.3 \\
\midrule
\multicolumn{13}{@{}l}{\emph{Main test at $d=9$, three-block network}} \\
\sys + correlated PM & & 10,417 & & & & & & 28.9 & & & & \\
PM & & 27,481 & & & & & & 26.3 & & & & \\
correlated PM & & 12,967 & & & & & & 63.9 & & & & \\
\bottomrule
\end{tabular}
\end{adjustbox}

\vspace{2pt}
\parbox{\textwidth}{\scriptsize Failures on 2,000,000 test shots at $d=7$ and 1,000,000 at $d=9$--17. \sys + correlated PM
routes 2.00, 5.97, 0, 0, 0.96 and 1.00\pct{} of the test shots, so at $d=11$ and 13 it is
\sys alone. The rows of \sys alone are descriptive, and their time is the network term plus the handoff. Every
other time comes from the registered speed re-test, with GPU parts on H100 GPUs and CPU parts on one core of a Linux
workstation, converted to the Grace basis. Ising's times come from TensorRT plans, whereas Ising's failures come from
NVIDIA's official PyTorch numerics. Our Decoder Switching routes 16\pct{} of the shots at $d=7$--15 and 30\pct{} at
$d=17$. Its time is that of its frontend and its routed correlated PM. $^{a}$On the first 200,000 test shots.
$^{b}$The second registered test at $d=9$, with a one-block network, on 1,000,000 other shots. Its composition's time comes
from that test's own timing (Appendix~\ref{app:reg}). The other $d=9$ entries, and the last block, come from the main
test.}
\end{table*}

\head{Our Decoder Switching.} Decoder Switching~\cite{decoder_switching} routes a shot to its strong decoder when the
complementary gap of its weak decoder is small. The gap is the least weight of a correction in the other logical class
minus the weight of the weak decoder's correction. Our re-implementation decodes every shot with PM, computes the gap by
a second matching on PM's graph with the boundary split by the logical class, and routes the shots with the smallest
gaps to correlated PM. A secondary registration, frozen with the surface-code test, fixed its routed fraction by the
networks' rule on the same calibration shots. Within the speed limit, the rule admits no routed fraction at any
distance, because the frontend alone takes 4.0 to 8.3 times PM's time. The registration therefore drops the speed
limit. No $q$ then meets the failure target, so the rule takes the $q$ with the fewest calibration failures, the
largest of the grid at every distance (16\pct{} at $d=7$--15 and 30\pct{} at $d=17$). In the main test, with the
three-block network at $d=9$, our composition failed less often than this re-implementation at every distance, by
10--35\pct{} (exact McNemar tests with Holm's procedure over the six
distances at $\alpha=0.05$, largest adjusted $p=7.2\times10^{-53}$). At every distance, it also failed on more test
shots than correlated PM alone, its limit as $q$ grows.

\begin{table}[tbp]
\centering
\caption{BB-code results. Composing \sys with full-DEM Relay-BP reduces failures and decoding time.}
\label{tab:bb}
\footnotesize
\setlength{\tabcolsep}{3.4pt}
\begin{adjustbox}{max width=\linewidth}
\begin{tabular}{@{}l rr rr rr@{}}
\toprule
& \multicolumn{2}{c}{BB72, 0.3\pct} & \multicolumn{2}{c}{BB72, 0.2\pct} & \multicolumn{2}{c}{gross, 0.4\pct} \\
\cmidrule(lr){2-3} \cmidrule(lr){4-5} \cmidrule(lr){6-7}
test shots & \multicolumn{2}{c}{200,000} & \multicolumn{2}{c}{300,000} & \multicolumn{2}{c}{150,000} \\
decoder & failures & ms & failures & ms & failures & ms \\
\midrule
\sys, BB72 network 1 & 1,376 & 0.091 & & & & \\
\sys, BB72 network 2 & 1,398 & 0.089 & 269 & 0.089 & & \\
\sys, gross network & & & & & 1,024 & 0.255 \\
\sys + sector Relay-BP & 1,554 & 0.415 & 297 & 0.117 & 407 & 20.35 \\
\sys + full-DEM Relay-BP & 857 & 4.86 & 204 & 0.422 & 234 & 428 \\
\midrule
sector Relay-BP & 2,769 & 3.30 & 767 & 1.99 & 733 & 127.6 \\
BP-OSD, ldpc & 3,337 & 1.82 & 880 & 0.691 & 3,646 & 114.0 \\
BP-OSD, CUDA-Q & 3,342 & 0.891 & 856 & 0.540 & 3,754 & 13.7 \\
\midrule
full-DEM Relay-BP & 1,107 & 85.4 & 288 & 52.8 & 472 & 3,811 \\
\bottomrule
\end{tabular}
\end{adjustbox}

\vspace{2pt}
\parbox{\columnwidth}{\scriptsize Failures on the test shots and composite time per shot in ms. All BB72 compositions use network 2. The two BB72 networks were trained independently.}
\end{table}

\head{The accurate mode on BB codes.} The McNemar test of the second BB72 network alone against full-DEM Relay-BP at
$p=0.2\pct$ gives $p=0.33$. At $p=0.3\pct$ the accurate mode takes longer than the 0.891\ms{} of BP-OSD on the GPU,
and at 0.2\pct{} it also beats the three faster decoders on both axes. On the gross code, the accurate mode and
full-DEM Relay-BP disagree on 264 shots, and the composition fails on 13 of them.

\head{The balanced mode on BB codes.} On the gross code, routing the least confident 1\pct{} to sector Relay-BP leaves
407 failures in 150,000 shots, against 733 for sector Relay-BP alone and 3,646 and 3,754 for the two BP-OSD decoders.
This balanced mode takes 20.35\ms, against 127.6\ms{} for sector Relay-BP and 114.0\ms{} for BP-OSD from ldpc. BP-OSD
from CUDA-Q, the fastest decoder, takes 13.70\ms. The balanced mode takes 1.49 times as long. Without the Grace
conversion, the balanced mode takes 13.57\ms{} and is the faster of the two. On BB72, the balanced mode sends the accurate mode's routed shots to sector Relay-BP. It fails less often than sector
Relay-BP alone, 297 against 767 at $p=0.2\pct$ and 1,554 against 2,769 at 0.3\pct, and it beats the three faster
decoders on both axes. However, it fails more often than the second BB72 network alone (269 and 1,398).

{\clubpenalty=10000 %
\head{The fast mode on BB codes.} The fast mode beats sector Relay-BP and both BP-OSD decoders on both axes on BB72. It
fails 2.0--3.3 times less often than these decoders, and against the fastest of them it takes a sixth of the time at
$p=0.2\pct$ and a ninth at 0.3\pct. At $p=0.3\pct$ the first BB72 network fails on 1,376 of 200,000 shots, against
2,769--3,342 for sector Relay-BP and the two BP-OSD decoders. It takes 0.091\ms, against 0.891\ms{} for the fastest of
them. At $p=0.2\pct$ the second network, trained at $p=0.3\pct$, fails on 269 of 300,000 shots against 767--880, in
0.089\ms{} against 0.540\ms{} for BP-OSD from CUDA-Q. On the gross code the network alone takes a fiftieth of the
fastest baseline decoder's time and fails more often than sector Relay-BP (1,024 against 733).\par}

\head{Cascade comparison.} The first registered test on BB72 judged each accuracy comparison as one of 132 comparisons
under one Holm procedure at $\alpha=0.05$, with the exact McNemar test and Newcombe's interval. ``Better'' and
``worse'' mean a significant difference in block failure rate, and ``n.s.'' means that the test detected no significant
difference. Its secondary statement asks whether the upper end of Newcombe's interval at the one-sided level $0.05/32$
lies below a margin frozen before the test, 10\pct{} of Cascade's block failure rate on the development shots. In the
follow-up, each compiled comparison joins the single and the batched comparison as an intersection-union test, whose
p-value is the larger of the two. Holm's procedure jointly tested all four comparisons at $\alpha=0.05$.

\head{Cascade on the surface code.} One Holm procedure at $\alpha=0.05$ judged 40 hypotheses, ten per error rate and
mode, with the exact McNemar test on the shots where exactly one of two decoders fails. A row is better or worse than
Cascade alone when its test is rejected. A row is non-inferior when the test against a margin of 10\pct{} of Cascade
alone's development failure rate is rejected and the row is not worse. The two modes read the same confirmation shots,
so they are two analyses of one sample, not independent replications. The planned power of the test for ``better'' was
0.342 for the composition at $p=0.6\pct$ and 0.999 at 0.35\pct, and the test was rejected at 0.35\pct{} only. In all, 34 of
the 40 hypotheses were rejected. Appendix~\ref{app:cascade} gives the tests of the joint decision.

\head{Output-head comparison.} Each comparison tests each seed on its own, with the exact two-sided McNemar test on
that seed's discordant shots and Newcombe's 95\pct{} interval of the paired difference. Discordant counts are never
summed over seeds, because both seeds read the same test shots. The comparison's p-value is the larger of the two
seeds' p-values when their differences have the same sign and 1 otherwise, an intersection-union test. Holm's procedure
at $\alpha=0.05$ runs over these p-values, 14 on the surface code and 10 on BB72. The tiny p-values concern the
trained networks, and the intervals reflect shot noise, not the spread between training runs. Without a fixed margin, we claim no equivalence.

\section{Timing and CPU Bases}
\label{app:timing}

This appendix gives the composite time and the timing of the classical tests, their CPU basis and their speed margins,
the timing of the Cascade comparisons on both cards, the clock record of each card and the batched timing of the
classical compositions. Times come from the registered speed re-test, except those of the second $d=9$ test, the TensorRT comparison and the joint decision (Appendices~\ref{app:reg}, \ref{app:trt} and~\ref{app:cascade}).

\head{The composite time.} The reaction mode is timed at batch size one, where a GPU cannot spread its launch costs
over a batch. The throughput mode's cost per syndrome round is the host-to-host time per shot over a long batched run
divided by the rounds per shot, or, in the third stage of the speed re-test, the GPU term plus the CPU term per shot (below). The network term $\Tn$ of Eq.~(\ref{eq:comp}) starts from detector bits in GPU
memory. The composite time of a baseline decoder adds the same kinds of parts, so a composite time is a benchmark cost
that sums separately measured parts. We reserve latency for complete measurements from host-memory detector bits to a host-memory decision.

\head{Timing in the reaction mode.} The speed re-test timed every GPU part of the classical tests on an H100 GPU and
every CPU part on one core of a Linux workstation, ours and the baseline decoders' alike, with every input in memory.
The one exception is BP-OSD from CUDA-Q, which the re-test timed host to host on the H100 machines. Its host work
therefore ran on one core of their CPUs, an Intel Xeon Platinum 8470 for the BB codes and the surface code at $d=7$--11 and an
AMD EPYC 9554 at $d=13$--17.
$\Tn$ is the mean device time over 200,000 timing shots from 100 back-to-back CUDA-graph replays. $\Tf$ is the routed
fraction times the fallback's mean time per routed shot, decoded one at a time on one core. At $d=7$, $\Th$ is the
resident kernel's host round trip minus its device time, which includes reading the shot from mapped host memory. At
$d=9$--17, $\Th$ is the router's mean time from the graph launch to the decision, minus the device time and the routed
decodes, and at least 1.1\us. At $d=7$, the handoff thus includes a transfer from host memory that
NVIDIA's setting neglects, which only lengthens our times. On BB codes, $\Th$ is the host-side allowance of 1.1\us{}
per shot for the routing decision, packing and indexing. Elsewhere the composite excludes the CPU--GPU transfer of the shot, as does NVIDIA's NVQLink-based setting. On BB codes, this pinned-memory round trip takes at most 19.5\us.
A GPU process counts only if its clock record shows no outside interference and its outputs equalled the deployed
ones. For \sys and its compositions, we report the slowest of five GPU processes (three for the gross-code network) and the slowest of
three CPU processes. For each baseline decoder, we report the fastest of three processes. Ising's GPU time is
instead the smallest accepted mean among all repetitions of its validated TensorRT plans. The re-test ran each
plan in five repetitions of 100 CUDA-graph replays in each of two processes and in five runs of NVIDIA's trtexec. A minimum over so
many repetitions favours Ising. Its PM share is the fastest of three processes. The composite
time leaves out the copy of the shot into GPU memory at $d=9$--17, averages the CPU term over shots, and uses the
1.1\us{} allowance as the handoff at $d=9$--17.

\head{The CPU host.} The CPU parts ran on a Linux workstation with an AMD Ryzen 9 9950X CPU (16 cores) under Ubuntu
24.04. For the timing, the frequency governor was set to performance with turbo boost on. At most eight timed
processes ran at once, each on its own physical core of one core complex die, whose SMT siblings were taken offline.
Every other process was confined to the other die. Everything the timed processes read sat in a tmpfs, and no timed
process read the disk in any window. Each window began and ended with a reference check on every timed core. The
check requires every reference value to stay within 1.10 times the session's reference and each segment's spread to
stay at most 6\pct. All 46 windows and all 290 processes passed at their first attempt.

\head{Shot counts.} The slowest CPU items decoded fewer shots, so that the CPU session fits in one evening. Each subset
is the first $N$ shots in shot order. The frontend of our Decoder Switching decoded 50,000 shots per distance. On the gross code, full-DEM Relay-BP decoded 75 shots and ldpc BP-OSD 500. The balanced and accurate routed terms used 300 and 25 shots.

\head{CPU times and the Grace basis.} We state every CPU time of a composite on one core of NVIDIA's Grace CPU, on the
surface code and on both BB codes, because NVIDIA's own accounting of its Ising pre-decoders sums the same kinds of
parts on this CPU. This accounting times the pre-decoders on a GB300 GPU with host-device transfers disabled and PM on
a Grace Neoverse-V2 CPU. It states no core or thread count, and our Grace basis assumes one core. It ignores the
transfer of syndrome data between the two, because its estimates for
NVIDIA's NVQLink architecture find this transfer negligible next to both runtimes~\cite{ising}. We use the same
setting, and our host-side handoff $\Th$ stays in every composite. We have no GH200 or GB300 machine. The speed
re-test therefore converted every CPU time from the workstation to a Grace core by a factor
fitted over all 18 of NVIDIA's published CPU baseline timings. These are PM at $d=13$, 21 and 31 and correlated PM at $d=5$, 9,
13, 17, 21 and 31, each at two error rates~\cite{ising}. The re-test timed NVIDIA's circuits for these 18 points on
the same workstation and took the slowest of three processes per point. The slowest time gives the smaller factor,
which makes the CPU baselines faster. At each point $i$, the ratio $r_i$ is NVIDIA's time per shot, its time per
round times $d$, over ours. A decoder's factor is the geometric mean of its $n$ ratios,
$f=\exp\bigl(\frac{1}{n}\sum_i\ln r_i\bigr)$, and its spread $s$ is the standard deviation of $\ln r_i$. PM's factor is
1.4543 ($s=0.400$, $n=6$) and correlated PM's is 1.5383 ($s=0.146$, $n=12$). A rule fixed before the data uses one
common factor when $|\ln f_{\mathrm{PM}}-\ln f_{\mathrm{corr}}|\le\max(s_{\mathrm{PM}},s_{\mathrm{corr}})$, and
otherwise gives each PM decoder its own factor and every other CPU decoder the geometric mean of the two. Here
$0.056\le0.400$, so one common factor converts every CPU time from the workstation, on the surface code and on both
BB codes. It is the geometric mean of the ratios at all 18 points, 1.5098. Correlated PM has 12 of these points
and PM six, so this mean weights correlated PM's factor twice as much as PM's and lies above the geometric mean of
the two factors, 1.4957. The host work of BP-OSD from CUDA-Q is not converted.
Table~\ref{tab:config} lists every baseline decoder's settings, taken from each decoder's published examples, with the
OSD order of the BB-code BP-OSD decoders raised from 10 to 60. Every output counts, even if BP does not converge.

\head{The raw basis.} Table~\ref{tab:raw} gives every composite time with its CPU parts as timed on the workstation
core, beside its Grace-basis value. On this raw basis, PM takes 7.88, 17.4 and 34.6\us{} at $d=7$, 9 and 11 and is
faster than \sys + correlated PM, which takes 11.59, 19.15 and 41.5\us{} there, with the one-block composition at $d=9$. Thus 45 of the
48 surface speed comparisons hold on the raw basis, against 47 on the Grace basis. With the main test's three-block
network at $d=9$ (28.9\us{} on both bases), the counts are 45 and 46. On the gross code, the balanced mode takes 13.57\ms{} on the raw
basis and is faster than BP-OSD from CUDA-Q, which takes 13.70\ms{} on both bases. All other comparisons agree across bases.

\begin{table*}[tp]
\centering
\caption{Composite times per shot before and after the Grace conversion. Each cell gives the time with every CPU part
as timed on one core of the workstation (raw), then its Grace-basis value.}
\label{tab:raw}
\footnotesize
\setlength{\tabcolsep}{4pt}
\begin{adjustbox}{max width=\linewidth}
\begin{tabular}{@{}l cccccc@{}}
\toprule
\multicolumn{7}{@{}l}{\emph{Surface code, \textmu s, raw / Grace}} \\
decoder & $d=7$ & 9 & 11 & 13 & 15 & 17 \\
\midrule
\sys + correlated PM & 11.59 / 11.94 & 19.15 / 21.2\rlap{$^{b}$} & 41.5 / 41.5 & 51.2 / 51.2 & 57.1 / 59.6 & 77.5 / 81.6 \\
with the three-block network, main test & & 28.9 / 28.9 & & & & \\
PM & 7.88 / 11.89 & 17.4 / 26.3 & 34.6 / 52.2 & 60.7 / 91.7 & 106.0 / 160.1 & 160.6 / 242.5 \\
correlated PM & 18.28 / 27.60 & 42.3 / 63.9 & 91.3 / 137.9 & 167.6 / 253.1 & 372.7 / 562.7 & 560.4 / 846.0 \\
Ising-Fast + PM & 45.28 / 47.14 & 53.3 / 57.2 & 61.7 / 69.3 & 79.0 / 92.3 & 101.7 / 123.8 & 136.8 / 173.1 \\
Ising-Accurate + PM & 57.33 / 58.99 & 63.9 / 67.2 & 72.2 / 78.6 & 89.8 / 100.8 & 110.1 / 128.2 & 144.1 / 173.0 \\
Ising-Fast + correlated PM & 47.88 / 51.07 & 61.7 / 69.8 & 81.8 / 99.7 & 119.9 / 153.9 & 177.7 / 238.7 & 256.9 / 354.4 \\
Ising-Accurate + correlated PM & 59.26 / 61.91 & 70.6 / 77.3 & 88.0 / 102.5 & 121.7 / 149.0 & 169.8 / 218.4 & 242.4 / 321.3 \\
our Decoder Switching & 38.03 / 57.41 & 92.1 / 139.0 & 207.5 / 313.3 & 423.7 / 639.7 & 806.2 / 1,217 & 1,591 / 2,402 \\
\midrule
\multicolumn{7}{@{}l}{\emph{BB codes, ms, raw / Grace}} \\
decoder & \multicolumn{2}{c}{BB72, 0.3\pct} & \multicolumn{2}{c}{BB72, 0.2\pct} & \multicolumn{2}{c}{gross, 0.4\pct} \\
\midrule
sector Relay-BP & \multicolumn{2}{c}{2.19 / 3.30} & \multicolumn{2}{c}{1.32 / 1.99} & \multicolumn{2}{c}{84.5 / 127.6} \\
BP-OSD, ldpc & \multicolumn{2}{c}{1.20 / 1.82} & \multicolumn{2}{c}{0.458 / 0.691} & \multicolumn{2}{c}{75.5 / 114.0} \\
full-DEM Relay-BP & \multicolumn{2}{c}{56.5 / 85.4} & \multicolumn{2}{c}{35.0 / 52.8} & \multicolumn{2}{c}{2,524 / 3,811} \\
\sys + sector Relay-BP & \multicolumn{2}{c}{0.305 / 0.415} & \multicolumn{2}{c}{0.108 / 0.117} & \multicolumn{2}{c}{13.57 / 20.35} \\
\sys + full-DEM Relay-BP & \multicolumn{2}{c}{3.25 / 4.86} & \multicolumn{2}{c}{0.310 / 0.422} & \multicolumn{2}{c}{283 / 428} \\
\bottomrule
\end{tabular}
\end{adjustbox}

\vspace{2pt}
\parbox{\textwidth}{\scriptsize GPU parts, Ising's plans and BP-OSD from CUDA-Q (0.891, 0.540 and 13.70\ms{} on the BB
codes) are the same on both bases, and the fast mode has no CPU part. $^{b}$The one-block composition of the second test at
$d=9$, whose CPU part was converted with that session's factor, 1.5423.}
\end{table*}

\head{The speed margins.} The fastest baseline decoder is PM at $d=7$--13, Ising-Fast + PM at $d=15$ and
Ising-Accurate + PM at $d=17$. PM takes 0.05 and 2.63\us{} less than \sys + correlated PM at $d=7$ and, with the three-block
network, at $d=9$. The one-block composition at $d=9$ takes 5.06\us{} less than PM. From $d=11$ on, the fastest baseline
decoder takes 10.7--91.4\us{} more than \sys + correlated PM.

{\clubpenalty=10000 %
\head{The margins at $d=17$.} The re-test converts every CPU time at $d=17$ with the same fitted factor. There,
\sys + correlated PM and correlated PM alone take 81.6 and 846.0\us. On NVIDIA's own circuit for correlated PM at $d=17$, one of the
fit's reference points, the fitted factor gives 1,114\us, above the 860.71\us{} that NVIDIA measured on a Grace
Neoverse-V2 CPU. At $d=9$ it gives 66.8\us{} against NVIDIA's 67.59\us. The routed CPU term
of Table~\ref{tab:time} is 12.0\us{} at $d=17$. Ising-Accurate + PM, the fastest baseline decoder
there, takes 173.0\us, 91.4\us{} more than \sys + correlated PM. All eight speed comparisons at $d=17$ hold.\par}

\begin{table}[tbp]
\centering
\caption{Where the composite time goes. The network term takes most of the time on the surface
code, and routed fallback work dominates all but one BB composition.}
\label{tab:time}
\footnotesize
\setlength{\tabcolsep}{2.7pt}
\begin{adjustbox}{max width=\linewidth}
\begin{tabular}{@{}l r r r r r r@{}}
\toprule
setting & routed & network & \parbox[b]{1.02cm}{\centering routed\\fallback} & handoff & total &
\parbox[b]{0.8cm}{\centering CPU\\share} \\
\midrule
\multicolumn{7}{@{}l}{\emph{Surface code, \textmu s}} \\
$d=7$ & 2.05\pct & 8.4 & 1.0 & 2.5$^{m}$ & 11.9 & 9\pct \\
$d=9$ & 6.07\pct & 14.2 & 5.9 & 1.1$^{a}$ & 21.2 & 28\pct \\
$d=9$, main test & 0 & 27.8 & 0 & 1.1$^{a}$ & 28.9 & 0 \\
$d=11$ & 0 & 40.4 & 0 & 1.1$^{a}$ & 41.5 & 0 \\
$d=13$ & 0 & 50.1 & 0 & 1.1$^{a}$ & 51.2 & 0 \\
$d=15$ & 0.98\pct & 51.2 & 7.2 & 1.1$^{a}$ & 59.6 &
12\pct \\
$d=17$ & 1.02\pct & 68.5 & 12.0 & 1.1$^{a}$ & 81.6 &
15\pct \\
\midrule
\multicolumn{7}{@{}l}{\emph{BB codes, ms}} \\
BB72 fast, 0.2\pct & 0 & 0.088 & 0 & 0.001 & 0.089 & 0 \\
BB72 fast, 0.3\pct & 0 & 0.090 & 0 & 0.001 & 0.091 & 0 \\
BB72 bal., 0.2\pct & 0.247\pct & 0.088 & 0.027 & 0.001 & 0.117 &
23\pct \\
BB72 bal., 0.3\pct & 1.94\pct & 0.088 & 0.325 & 0.001 & 0.415 &
78\pct \\
BB72 acc., 0.2\pct & 0.247\pct & 0.088 & 0.333 & 0.001 & 0.422 &
79\pct \\
BB72 acc., 0.3\pct & 1.94\pct & 0.088 & 4.775 & 0.001 & 4.864 &
98\pct \\
gross balanced & 0.96\pct & 0.254 & 20.097 & 0.001 & 20.352 &
99\pct \\
gross accurate & 2.00\pct & 0.254 & 427.3 & 0.001 & 427.5 &
99.9\pct \\
\bottomrule
\end{tabular}
\end{adjustbox}

\vspace{2pt}
\parbox{\columnwidth}{\scriptsize Surface rows in \textmu s and BB rows in ms. Every routed-fallback term is a
workstation time converted to the Grace basis (Appendix~\ref{app:timing}). The routed fraction is that of
the timing shots. The $d=9$ row is the one-block composition of the second test, timed in its own session. The main test's
three-block network follows it. The fast mode at $p=0.3\pct$ uses the first BB72 network, and every other BB72 row
the second. Totals use unrounded terms. $^{m}$Measured. $^{a}$The 1.1\us{} allowance, above the
estimate.}
\end{table}

\begin{table}[tbp]
\centering
\caption{Decoder settings and timing.}
\label{tab:config}
\footnotesize
\setlength{\tabcolsep}{0pt}
\begin{adjustbox}{max width=\linewidth}
\begin{tabular}{@{}l@{\hspace{9pt}}>{\raggedright\arraybackslash}p{3.3cm}@{\hspace{7pt}}l@{}}
\toprule
decoder & software and settings & timed on \\
\midrule
PM, correlated PM & PyMatching 2.4.0 & CPU$^{a}$ \\ \addlinespace[1.5pt]
Ising + PM$^{b}$ & Ising-Decoding 0.1.0, TensorRT 10.14 & H100, CPU$^{a}$ \\ \addlinespace[1.5pt]
BP-OSD (surface) & CUDA-Q QEC 0.8.0, OSD-CS order 10, 10 (min-LLR) or 60 BP iterations & \begin{tabular}[t]{@{}l@{}}H100,\\host CPU$^{e}$\end{tabular} \\ \addlinespace[1.5pt]
Relay-BP$^{c}$ & relay-bp 0.2.2, fp32 & CPU$^{a}$ \\ \addlinespace[1.5pt]
BP-OSD (BB), ldpc & ldpc 2.4.1 BpOsdDecoder, min-sum, 80 iterations, OSD-CS order 60 & CPU$^{a}$ \\ \addlinespace[1.5pt]
BP-OSD (BB), CUDA-Q & CUDA-Q QEC 0.8.0, same settings & \begin{tabular}[t]{@{}l@{}}H100,\\host CPU$^{e}$\end{tabular} \\ \addlinespace[1.5pt]
Cascade (ours, BB72) & 14 blocks of width 512, BF16, compiled into CUDA graphs at batch 1 and 4,096 & \begin{tabular}[t]{@{}l@{}}RTX PRO 6000,\\H100\end{tabular} \\ \addlinespace[1.5pt]
Cascade (ours, $d=13$) & 13 blocks of width 256, BF16, compiled at batch 1 and 16 & \begin{tabular}[t]{@{}l@{}}RTX PRO 6000,\\H100\end{tabular} \\
\addlinespace[1.5pt]
Relay-BP (chain) & relay-bp 0.2.2 on the full DEM, 16 worker processes & host CPU$^{d}$ \\ \addlinespace[1.5pt]
corr.\ PM, $d=13$ & PyMatching 2.4.0, 1 thread per decode or 16 workers & host CPU$^{d}$ \\
\bottomrule
\end{tabular}
\end{adjustbox}

\vspace{2pt}
\parbox{\columnwidth}{\scriptsize $^{a}$One core of the Linux workstation, converted to the Grace basis. $^{b}$Ising-Fast
and Ising-Accurate are NVIDIA's Ising-Decoder-SurfaceCode-1-Fast and -Accurate. Accuracy comes from NVIDIA's official
PyTorch numerics and time from each model's fastest TensorRT plan (FP8) in the speed re-test, a
combination that favours Ising. $^{c}$A memory strength sets how much of a bit's belief carries over from the previous
iteration. Strength 0.15 for 80 iterations, then up to 300 relay legs of 60 iterations with strengths drawn from
$[-0.23,0.62]$, stopping after 5 solutions, the gross-code values of the example notebook in the Relay-BP repository
(relay-bp 0.2.2). The min-LLR setting of BP-OSD
orders bits for OSD by the smallest log-likelihood ratio (LLR) each reached, and the other surface-code setting runs 60
BP iterations. $^{d}$The card's host CPU, an AMD EPYC 9555 with the RTX PRO 6000 and an Intel Xeon with the H100,
timed end to end and not converted. $^{e}$One core of the H100 machine's host CPU, timed host to host and not
converted.}
\end{table}

\head{Timing of the Cascade comparisons.} The speed re-test timed every Cascade comparison end to end and host to host
on two cards, one full RTX PRO 6000 with its node's AMD EPYC 9555 and one H100 80GB SXM with its own Intel Xeon Platinum 8480+ host.
Both cards ran the same processes with the same arguments on the same timing shots. Four jobs per card ran the processes of the original
timing loops in their original order, three fresh processes for each decoder, with Cascade alone in the same job and
loop. Each composition's value is the slowest accepted process of its three, and Cascade alone's the fastest of its three. The
registered analysis took the slowest process on both sides. Cascade's fastest process changes no outcome on either card
and lowers no speed-up by more than 3.2\pct. Single mode decodes
20,000 shots one at a time after 200 untimed ones and reports the mean and the 99.9th percentile of the latency.
Batched mode decodes 500,000 shots with the networks at batch 4,096 and Relay-BP in 16 worker processes and reports the
wall time per shot. Routed shots queue on the GPU, and Cascade runs whenever 4,096 of them wait and once more, padded,
at the end of the run.

\head{Mean times.} Every speed-up that we report divides the fastest of the baseline's processes by the slowest of ours.
The mean of a decoder's processes lies between its fastest and its slowest, so mean times on both sides can only raise a
speed-up. Table~\ref{tab:means} shows this for the comparisons of Table~\ref{tab:cascade}. There, the three processes
of each decoder agree closely, and the means raise the speed-ups by at most 2.9\pct. The classical tests follow the same
rule, with our times the slowest of five GPU processes (three for the gross-code network) and three CPU processes and each baseline's the fastest of its
processes. Appendix~\ref{app:trt} reports the means of the TensorRT re-test.

\begin{table}[tbp]
\centering
\caption{Speed-ups of Table~\ref{tab:cascade} as reported and, in parentheses, from the mean of the three processes on
both sides. A mean never lowers a speed-up.}
\label{tab:means}
\footnotesize
\setlength{\tabcolsep}{4pt}
\begin{adjustbox}{max width=\linewidth}
\begin{tabular}{@{}l r r r r@{}}
\toprule
& \multicolumn{2}{c}{RTX PRO 6000} & \multicolumn{2}{c}{H100} \\
\cmidrule(lr){2-3}\cmidrule(lr){4-5}
setting & single & batched & single & batched \\
\midrule
BB72, $p=0.2\pct$ & 10.23 (10.46) & 8.85 (8.86) & 7.80 (7.84) & 4.31 (4.32) \\
BB72, $p=0.3\pct$ & 5.81 (5.81) & 5.15 (5.15) & 4.69 (4.71) & 3.17 (3.20) \\
$d=13$, $p=0.6\pct$ & 3.63 (3.65) & 4.47 (4.47) & 3.33 (3.37) & 4.35 (4.38) \\
$d=13$, $p=0.35\pct$ & 5.53 (5.69) & 7.66 (7.67) & 5.17 (5.21) & 7.18 (7.22) \\
\bottomrule
\end{tabular}
\end{adjustbox}
\end{table}

\head{Clock records and acceptance.} A sampler recorded each card's SM clock, its throttle reasons and its power limit
every 100\,ms in every timed window. The re-test rejected every window with outside interference, such as another
process on the card, a hardware slowdown or a lowered power limit. A fresh process then replaced the rejected one, up
to three attempts per process. A workload's own power cap counts as part of the card's speed. On the RTX PRO 6000, 0 of
153 windows were rejected, and 44 accepted processes had a clock sample below 97\pct{} of
the maximum, at lowest 93.8\pct. On the H100, these counts were 2 of 155
and 77, at lowest 68.2\pct, and 78 windows ran at the card's own power cap.
Every comparison was established on both cards.

\head{The compositions on the multi-shot engine.} On the multi-shot engine, the follow-up's compositions took 137.5\us{}
per single shot at $p=0.2\pct$ and 276.1\us{} at 0.3\pct{} on the RTX PRO 6000, and 160.8 and
285.7\us{} on the H100. The first test's composition with the three-block network took 0.510\us{} per
round at $p=0.2\pct$ and 0.998\us{} at 0.3\pct{} on the RTX PRO 6000, and 0.595 and
0.890\us{} on the H100. In the same job, this is 1.74 and 1.38 times less than on the BB engine on the
RTX PRO 6000 and 1.81 and 1.55 times less on the H100, and 23.5, 12.0,
12.2 and 8.2 times less than Cascade alone, at the accuracy of Table~\ref{tab:cascade-first}.

\head{The routed fraction sets the composition's throughput.} In batches every slot of Cascade's graph costs the same, so
$\kappa$ in Eq.~(\ref{eq:speedup}) is close to 1, and the six-block network costs about 0.04 of Cascade's
time per shot on the RTX PRO 6000 and 0.09 on the H100. For 2\pct{} and 10\pct{} routed,
Eq.~(\ref{eq:speedup}) then gives speed-ups of about 17 and 7 times on the RTX PRO 6000,
against the measured 16.3 and 7.0, and of about 9 and 5 times on
the H100, against 8.2 and 4.9. On the RTX PRO 6000, whose component times the follow-up's
predictions use, these predictions matched its measured pipeline times within 4.9\pct. The batched costs of
Section~\ref{sec:where} exclude the waits of batch formation and of Cascade's queue, which at 2\pct{} routed fills after
about 204,800 incoming shots, so we keep them apart from the single-shot results of Section~\ref{sec:rq1}.

\head{The first test's networks on the H100.} On the H100, our Cascade's batched time is about 44\us{}
per shot, and the multi-shot engine runs a three-block network in 2.82\us{} per shot, against
1.87\us{} on the RTX PRO 6000. The batched speed-up of the first test's composition over Cascade alone is
8.2--12.2 times there. The first test's $p=0.3\pct$ chain was slower than batched Cascade alone.

{\clubpenalty=10000 %
\head{Timing of the comparison at $d=13$.} One job per card timed the composition, the chain, \sys + correlated PM and
Cascade alone at both error rates in both modes, three fresh processes each in the original loop's order. Batched mode
runs the network at batch 4,096, Cascade at batch 16 and correlated PM in 16 worker processes on the card's host CPU.
The batched cost per syndrome round divides the time per shot by the 13 rounds of a shot. On the H100, the kernel
caches were compiled on the card before the timed block, from passes over shots that are not timing shots.\par}

\head{Batched timing of the matching and Relay-BP compositions.} A third stage of the speed re-test timed the
compositions, \sys alone, PM, correlated PM and both Relay-BP decoders of Figures~\ref{fig:surface} and~\ref{fig:bb}
in batches of 4,096 under the same registration rules~\cite{convlines_code}. The network term is the host-to-host time per shot on one H100, from shots in pinned host
memory, with the routing and the packing of routed shots inside the timed loop. It used the multi-shot engine on BB72,
the Triton engine on the surface code and the BB engine's kernels on the gross code, whose decisions equalled the
deployed engines' on validation shots. The CPU term is the wall time of one or eight single-threaded processes, each
on its own core of the timed die, decoding the routed shots with the decoder's batch call, divided by all shots. Each
baseline decoded the timing shots in the same harness. Full-DEM Relay-BP decoded the first 4,000 at each BB72 error
rate and the first 40 on the gross code, and sector Relay-BP the first 1,000 on the gross code. The timing sets held 200,000 shots per
surface-code distance, 100,000 per BB72 error rate and 2,000 on the gross code. Our side takes the slowest of five GPU
processes (three on the gross code) and of three CPU runs, and each baseline the fastest of three. All 89 GPU processes
and 24 CPU windows were accepted at their first attempt. No RTX PRO 6000 was available before the timing, so this
stage timed the H100 only. On the Grace basis, \sys + correlated PM took 0.187, 0.698 and 0.822\us{} per round at
$d=7$, 9 and 11 with one CPU process and 0.0856, 0.187 and 0.822\us{} with eight, against 3.22, 6.45 and 10.8\us{}
for correlated PM alone with one. The release lists every cost on both bases.

\section{Our Cascades and Their Comparisons}
\label{app:cascade}

This appendix describes our BB72 and $d=13$ Cascade implementations and comparisons, including the first registered test, exploratory search and follow-up. Appendix~\ref{app:stats} reports their statistics.

\begin{table*}[tp]
\centering
\caption{The first registered test on BB72, with three-block networks in front of our Cascade, timed on two cards.
Neither composition beat Cascade alone in accuracy.}
\label{tab:cascade-first}
\footnotesize
\setlength{\tabcolsep}{4pt}
\begin{adjustbox}{max width=\linewidth}
\begin{tabular}{@{}l r r l r r r r@{}}
\toprule
& & & & \multicolumn{2}{c}{RTX PRO 6000} & \multicolumn{2}{c}{H100} \\
\cmidrule(lr){5-6}\cmidrule(lr){7-8}
decoder & routed (\%) & failures & accuracy & single & batched & single & batched \\
\midrule
\multicolumn{8}{@{}l}{\emph{$p=0.2\pct$, 5,000,000 shots}} \\
Cascade alone & & 3,858 & & 1,750.7 & 12.00 & 1,479.6 & 7.34 \\
\sys + Cascade & 1 & 3,896 & n.s. & 107.1 & 0.89 & 123.5 & 1.14 \\
speed-up & & & & 16.35 & 13.52 & 11.98 & 6.46 \\
\addlinespace[3pt]
\multicolumn{8}{@{}l}{\emph{$p=0.3\pct$, 1,000,000 shots}} \\
Cascade alone & & 4,459 & & 1,748.8 & 12.01 & 1,479.8 & 7.34 \\
\sys + Cascade & 5 & 4,550 & worse & 175.0 & 1.38 & 189.6 & 1.35 \\
speed-up & & & & 9.99 & 8.72 & 7.80 & 5.42 \\
\bottomrule
\end{tabular}
\end{adjustbox}

\vspace{2pt}
\parbox{\textwidth}{\scriptsize The fitted noise model. Both rows of a setting read the same fresh shots, and one count
serves both modes. The accuracy column, times and speed-ups are defined as in Table~\ref{tab:cascade}. One read of the
shots decided each accuracy comparison for both modes, and a check on separate shots found that the timed paths of
both modes decide as that read does.}
\end{table*}

\head{Our Cascade on BB72.} We re-implemented Cascade on our BB72 circuits with 14 blocks of width 512 and trained one copy per
noise setting. Cascade's paper does not fully specify its noise model, so we used two settings, our uniform circuit-level
noise and a variant with extra depolarizing noise after resets and before measurements and a noisy final measurement.
The variant reproduces within their errors the Tesseract, BP-OSD and Relay-BP failure rates that Cascade's paper
reports, a fit that rests on 17--20 Tesseract failures. In this fitted setting, pooled over all our reads of our
Cascade alone, the published logical error rates lie 3.7, 2.0 and 5.2\pct{} above ours at $p=0.1$, 0.2 and 0.3\pct,
inside our 95\pct{} interval at 0.1\pct{} only, and 19\pct{} below ours at 0.4\pct. The fit leaves open whether our network shares the published network's error structure. Cascade runs in
BF16, and the H200 times of its paper, whose precision it does not state, enter none of our speed-ups. Its paper reports no single-shot latency on BB72,
because it amortizes its BB-code timing over batches. The comparisons timed each composition on each card as the slowest
of three processes and Cascade alone, in the same~job, as the fastest of its three~(Appendix~\ref{app:timing}).

\head{The first registered test.} Every shot of the comparison is fresh, and no test set of the classical tests was
read. The first test followed a plan written and reviewed before its development and confirmation shots were drawn.
Development shots fixed the configuration of every decoder and its thresholds. An accuracy read is a pass that counts
each decoder's failures on a set of shots. In the fitted setting, the three-block
compositions routed 1\pct{} of the shots to Cascade at $p=0.2\pct$ and 5\pct{} at 0.3\pct. A check then tested that the timed paths
decide as the accuracy reads do, and it passed every row. One read of the confirmation shots decided each accuracy
comparison. A change to the plan replaced the selection rule after the development curves had been seen and before any
confirmation shot was read. The new rule takes, among the speed-feasible candidates, the one with the lowest predicted
single-shot latency within $F+\sqrt{F}$ of the fewest development failures $F$. As the plan's secondary statement, the excess block failure rate of the three-block composition stayed below the frozen
margin at both points (Appendix~\ref{app:stats}). In the uniform setting the composition showed no significant difference at
0.2\pct{} and was worse at 0.3\pct, and the chain was worse at 0.2\pct{} and better at 0.3\pct. No tested decoder there
was better than Cascade at
both points. Compositions with full-DEM Relay-BP as the fallback failed more often than Cascade alone at both points in both
settings.

\head{The first test's results.} As Table~\ref{tab:cascade-first} shows, the three-block \sys-BB networks
(483,357 parameters) in front of Cascade in the fitted setting showed no significant difference from Cascade at
$p=0.2\pct$ and failed more often at 0.3\pct, 4,550 times against 4,459 in 1,000,000 shots. On the shots that it kept,
the network broke 155 of Cascade's correct decisions and repaired 64. The composition was still faster than
Cascade, by 9.9--16.3 times per single shot and 8.7--13.5 times
in batches on the RTX PRO 6000 and by 7.8--11.9 and 5.4--6.4
times on the H100. On the 500,000 development shots at $p=0.3\pct$,
the three-block network alone fails 4,931 times.

\head{The Relay-BP chain.} \looseness=1The chain of the first test sent Cascade's least confident routed shots on to full-DEM
Relay-BP, 0.1\pct{} of all shots at $p=0.2\pct$ and 0.25\pct{} at 0.3\pct{} in the fitted setting. Relay-BP draws the
memory strengths of each decode from its decoder's random generator, so the seed and the decoder's state decide its
answer. The accuracy reads built a new Relay-BP decoder for every routed shot, seeded with the shot's index in the read,
so each decision depends on the shot and its fixed seed, independent of earlier decodes. The timed chain kept one
Relay-BP decoder per process across shots, seeded once with the library's default seed, as a deployment would run it.
These are two stochastic implementations of the chain. In the accuracy reads, which routed 2\pct{} and 5\pct{} of the
shots to Cascade, the chain was better than Cascade at both points, with 3,596 failures against 3,858 and 3,957 against
4,459. This accuracy result holds for the first implementation. The timing results come from the second, which took
217.1\us{} per single shot and 1.96\us{} per syndrome round in batches at $p=0.2\pct$ and
702.1\us{} and 10.98\us{} at 0.3\pct{} on the RTX PRO 6000, and 313.5,
3.27, 1,178.7 and 19.87\us{} on the H100. On the RTX PRO 6000, the chain was faster
than Cascade alone on average at both points in both modes. On the H100, it was faster in single mode and at
$p=0.2\pct$ in batches, and slower at 0.3\pct{} in batches, where Cascade alone took 7.34\us{} per round.
Its tails were slower on both cards. At $p=0.3\pct$ its 99.9th-percentile single-shot latency was 188.0\ms{}
against 1.77\ms{} for Cascade alone on the RTX PRO 6000, and 355.7 against 1.61\ms{}
on~the~H100.

{\clubpenalty=10000 %
\head{The exploratory search.} After the first test's results were known, we evaluated 52 configurations on its
fresh shots, without adjusting for multiplicity. At $p=0.3\pct$ and 2--10\pct{} routed, the three-block network broke
3.5 to 11 times as many kept shots as the six-block network, which also differs from it in training
(Appendix~\ref{app:train}). With the six-block network, the composition that routed 2\pct{} at $p=0.2\pct$ failed 3,822 times
against Cascade's 3,858 (McNemar $p=2.1\times10^{-7}$). The composition that routed 10\pct{} at 0.3\pct{} showed no
significant difference (4,452 against 4,459, $p=0.12$). These compositions defined the registered follow-up hypotheses.\par}

\head{The registered follow-up.} The follow-up froze its plan, its configurations and its code in one commit before any
read job, and it drew its shots from a seed range reserved for it. The follow-up read Cascade on its eager accuracy path, which runs uncompiled,
and on the two compiled paths that the timings run, at batch one and at batch 4,096. Its hypotheses came from the exploratory look, after the
first test's results were known. The first test had picked both of these configurations before its selection rule
changed. The follow-up kept their thresholds unchanged. One Holm procedure at $\alpha=0.05$ judged four comparisons,
each error rate on Cascade's accuracy path and on the compiled paths, and all four were ``better''. The re-test read
the same frozen compositions, thresholds and analysis on fresh shots of the same sizes, and all four comparisons were again
``better''. On the timed computation of the re-test, the composition failed less often than Cascade by $8.00\times10^{-6}$ per
shot at $p=0.2\pct$ (95\pct{} interval $[6.17, 9.87]\times10^{-6}$, single path) and by $1.41\times10^{-5}$ at
0.3\pct{} ($[1.17, 1.64]\times10^{-5}$), relative reductions of about 1.1\pct{} and 0.31\pct{} on both compiled
paths. Routing cannot gain more than Cascade's failures on the kept shots, 4.3\pct{} of them at 0.2\pct{} and
3.1\pct{} at 0.3\pct.

\head{Our Cascade on the surface code.} We re-implemented Cascade's surface-code network at $d=13$ from its paper. It
has 13 blocks, one per unit of distance as the paper prescribes, with zero-padded $3\times3\times3$ convolutions inside
bottleneck blocks and a read-out that averages over the observable's support. It trains with the paper's optimizers,
schedule and batch of 3,328 shots. The paper does not state the width of its surface-code networks, so we took 256, the width of
its surface-code roofline and ablation examples, for 2.53 million parameters. It calls width 128 deployment-sized, and a
wider network may be more accurate and slower. The
network trained on the circuit of our registered test at $d=13$, with the paper's three-stage schedule of the error
rate up to $p=0.7\pct$. It stopped after 81,130 steps, about $2.7\times10^8$ shots, at the first check past its budget
of 80,000 steps, because its failures on a monitor block had fallen by less than 2\pct{} at both monitored error rates
over the last $3\times10^7$ shots. This circuit has reset and measurement flips at $2p/3$ and a
noiseless last round and readout, and it is the only noise model tested at $d=13$.

{\clubpenalty=10000 %
\head{The comparison at $d=13$.} \looseness=1The comparison tested the composition, a chain that sends Cascade's least confident shots on
to correlated PM, \sys + correlated PM as registered, and correlated PM alone, each against Cascade alone. The
network ran on the release's engine with fp16 inputs and fp32 accumulation, and Cascade in BF16. One rule fixed in the
plan chose the thresholds on development shots at each error rate, taking the candidate with the fewest development
failures among those whose predicted single-shot time stayed below 0.9 times Cascade's. A commit on our server locked
the thresholds and the analysis before the confirmation shots were drawn, 1,000,000 at $p=0.6\pct$ and 30,000,000 at
0.35\pct, and each computation read them once in each mode. A bitwise check confirmed that the timed pipelines decide
every shot as these reads do. The re-test drew fresh confirmation shots of the same sizes and read them with the
thresholds and the analysis frozen, and we report its counts. Correlated PM alone failed 26\pct{} more often than
Cascade alone at $p=0.6\pct$ and 30\pct{} more at 0.35\pct, and \sys + correlated PM failed more often than
Cascade alone at both error rates, 5,341 times against 4,936 and 2,537 against 2,354. The composition's 19-shot gain at
$p=0.6\pct$ equals 40 repairs minus 21 regressions on the kept shots, and its 29-shot gain at 0.35\pct{} equals 29
repairs with no regression. The chain routed 40\pct{} of the shots to Cascade and 0.49\pct{} on to correlated PM at
$p=0.6\pct$. It failed 4,351 times in single mode and 4,353 in batches, 12\pct{} less often than Cascade alone and
better in both modes, in 240.4\us{} per single shot and 5.62\us{} per round on the RTX PRO
6000 and 279.4 and 5.10\us{} on the H100. At 0.35\pct{} it routed 60\pct{} and 0.1\pct{} and
failed 2,632 and 2,628 times, more often than Cascade alone in both modes, in 310.8\us{} per single shot
and 7.75\us{} per round on the RTX PRO 6000 and 372.7 and 6.97\us{} on the H100. The chain's thresholds always send at least
0.1\pct{} of the shots to correlated PM, so better routing at 0.35\pct{} is possible.\par}

\head{The joint decision.} The joint decision of Section~\ref{sec:joint} changes only the routed shots where the
network's and Cascade's decisions differ, about 0.08 and 0.56\pct{} of the BB72 shots and 0.62 and 0.01\pct{} of the
$d=13$ shots. We fitted $w$ by the fewest failures, over a logarithmic grid on 5,500,000 and 8,500,000 fresh BB72 shots
($w=6.31$ and 3.76) and over every breakpoint on 2,000,000 and 15,000,000 fresh $d=13$ shots ($w=7.87$ and 13.84), and
kept it for the test shots of Table~\ref{tab:cascade}. On held-out shots of the same draws, it recovered 59--61\pct{} of
what a perfect choice between the two answers would recover on BB72 and 2--4\pct{} at $d=13$. On the routed shots where
the decisions differ at $d=13$, the network was wrong three times as often as Cascade. On BB72, the class score also
combines the two decoders' bits, and about a third of the shots it repairs return a class that neither decoder chose.
The timed BB72 pipelines rerun the network's vote on these shots to keep all 4,096 class scores, in fp32 with ties to
the lowest class. One Holm procedure at $\alpha=0.05$ over its eight exact McNemar tests against Cascade alone rejected all eight. The largest p-value, 0.033, is single mode at $d=13$ and $p=0.6\pct$, with 289 shots failing only with Cascade alone and 239 only with the joint decision.

\head{Timing the joint decision.} A later session timed the joint decision by the rules of the speed re-test on fresh
shots. Each composition process was followed by one with the joint decision on the same shots, and Cascade alone was
timed again. The RTX PRO 6000 was a node of the type in Table~\ref{tab:cascade}. The H100 was another card, on which Cascade alone took 27--28\pct{} longer per single BB72 shot than in the speed re-test, so the
joint decision's single-mode speed-ups on BB72 there exceed the composition's in Table~\ref{tab:cascade}. The joint
decision took as long as the composition within 2.4\pct{} in 15 of the 16 timings. In the 16th, single mode at
$p=0.3\pct$ on BB72 on the RTX PRO 6000, its processes ran at a 5\pct{} lower GPU clock and took 6.7\pct{} longer. A separate run at equal clock found 0.1--0.3\pct{} at $p=0.2\pct$ and 0.2\pct{} at 0.3\pct.

\section{Our Cascades on TensorRT}
\label{app:trt}

A faster engine for Cascade could shrink the speed-ups of Table~\ref{tab:cascade}. To test this, we compiled both of
our Cascades with NVIDIA TensorRT~10.16 and timed every Cascade comparison again on both cards, in a second registered
speed test. Table~\ref{tab:trt} lists its results.

\head{Engines.} \looseness=1 We exported each Cascade to ONNX and built BF16 TensorRT engines at batch size 1 and at the batch size
of the batched mode, 4,096 on BB72 and 16 at $d=13$. Each engine runs inside a CUDA graph. The engines decide as the
compiled path does. On fresh shots, the two paths took the same decision on 99.955--100\pct{} of the shots, and their
failure counts differed by at most two in every setting, none significantly (exact McNemar test).

\head{Protocol.} \looseness=1 The test ran the processes and arguments of the speed re-test (Appendix~\ref{app:timing}) on fresh
timing shots. It timed Cascade alone on both engines, in three fresh processes each. For each comparison on each card,
the engine with the lower Cascade-alone time served Cascade alone and the compositions. A composition's time is the slowest of its three
processes and Cascade alone's the fastest of its three. The freeze, the seeds and the analysis code were public before
the first timed process~\cite{convlines_code}.

\head{Results.} \looseness=1 TensorRT ran Cascade alone 3.1 to 4.7 times faster per single shot on BB72 and 1.3 to 1.4 times
faster at $d=13$. In batches, it ran the $d=13$ Cascade 1.8 and 1.1 times faster on the two cards. On BB72 at batch
4,096, it took 1.6 and 2.0 times as long as the compiled path, so the batched BB72 comparisons kept the compiled path.
Every composition still took less time than Cascade alone on its faster engine. Per single shot, the compositions took 2.53 to 4.20
times less time on the RTX PRO 6000 and 2.38 to 4.33 times less on the H100. In batches, they took 3.15 to 16.36 and
2.42 to 8.16 times less. With mean times, these four ranges become 2.60--4.22, 2.40--4.40, 3.15--16.38 and 3.21--8.26.
The largest gain from the means, on the H100 at $p=0.3\pct$ in batches, comes from one process of the first test's composition that ran twice as long as the other two.

\begin{table*}[tp]
\centering
\caption{The registered TensorRT re-test. In each comparison, Cascade runs on its faster engine (TRT for
TensorRT, comp.\ for the compiled path), alone and inside the compositions. Every composition takes less time than Cascade alone.}
\label{tab:trt}
\footnotesize
\setlength{\tabcolsep}{4pt}
\begin{adjustbox}{max width=\linewidth}
\begin{tabular}{@{}l l r r r r l r r r r@{}}
\toprule
& \multicolumn{5}{c}{RTX PRO 6000} & \multicolumn{5}{c}{H100} \\
\cmidrule(lr){2-6}\cmidrule(lr){7-11}
setting & engine & Cascade alone & \sys + Cascade & speed-up & mean & engine & Cascade alone & \sys +
Cascade & speed-up & mean \\
\midrule
\multicolumn{11}{@{}l}{\emph{BB72, six-block \sys-BB on the BB engine}} \\
$p=0.2\pct$, single & TRT & 369.9 & 146.3 & 2.53 & 2.60 & TRT & 479.3 & 170.2 & 2.82 & 2.94 \\
$p=0.2\pct$, batched & comp. & 12.01 & 1.365 & 8.80 & 8.82 & comp. & 7.20 & 1.702 & 4.23 & 4.25 \\
$p=0.3\pct$, single & -- & -- & -- & -- & -- & TRT & 478.7 & 201.5 & 2.38 & 2.40 \\
$p=0.3\pct$, batched & comp. & 12.01 & 2.339 & 5.13 & 5.14 & comp. & 7.20 & 2.287 & 3.15 & 3.21 \\
\addlinespace[3pt]
\multicolumn{11}{@{}l}{\emph{BB72, six-block \sys-BB on the multi-shot engine}} \\
$p=0.2\pct$, single & TRT & 369.9 & 109.8 & 3.37 & 3.40 & TRT & 479.3 & 134.3 & 3.57 & 3.67 \\
$p=0.2\pct$, batched & comp. & 12.01 & 0.734 & 16.36 & 16.38 & comp. & 7.20 & 0.882 & 8.16 & 8.26 \\
$p=0.3\pct$, single & -- & -- & -- & -- & -- & TRT & 478.7 & 173.2 & 2.76 & 2.82 \\
$p=0.3\pct$, batched & comp. & 12.01 & 1.712 & 7.01 & 7.02 & comp. & 7.20 & 1.445 & 4.98 & 5.05 \\
\addlinespace[3pt]
\multicolumn{11}{@{}l}{\emph{BB72, the first test\textquoteright s three-block \sys-BB}} \\
$p=0.2\pct$, single & TRT & 363.0 & 94.2 & 3.85 & 3.87 & TRT & 479.0 & 110.6 & 4.33 & 4.40 \\
$p=0.2\pct$, batched & comp. & 12.02 & 0.888 & 13.53 & 13.54 & comp. & 7.19 & 1.147 & 6.28 & 6.64 \\
$p=0.3\pct$, single & -- & -- & -- & -- & -- & TRT & 482.1 & 133.2 & 3.62 & 3.69 \\
$p=0.3\pct$, batched & comp. & 12.02 & 1.377 & 8.73 & 8.73 & comp. & 7.21 & 2.985 & 2.42 & 3.77 \\
\addlinespace[3pt]
\multicolumn{11}{@{}l}{\emph{Surface code, $d=13$}} \\
$p=0.6\pct$, single & TRT & 274.5 & 93.5 & 2.94 & 2.96 & TRT & 324.8 & 115.4 & 2.81 & 2.91 \\
$p=0.6\pct$, batched & TRT & 5.82 & 1.846 & 3.15 & 3.15 & TRT & 7.97 & 1.944 & 4.10 & 4.13 \\
$p=0.35\pct$, single & TRT & 285.9 & 68.1 & 4.20 & 4.22 & TRT & 338.3 & 83.9 & 4.03 & 4.07 \\
$p=0.35\pct$, batched & TRT & 5.82 & 1.317 & 4.42 & 4.42 & TRT & 7.98 & 1.239 & 6.44 & 6.47 \\
\bottomrule
\end{tabular}
\end{adjustbox}

\vspace{2pt}
\parbox{\textwidth}{\scriptsize Times are the mean latency per single shot in \textmu s and the batched cost per
syndrome round in \textmu s, Cascade alone's from the fastest of its three processes and that of \sys + Cascade from the
slowest of three. Mean gives the speed-up from the mean of the three processes on both sides. The first test's
network is the three-block network of Table~\ref{tab:cascade-first}. A dash marks a comparison that the frozen
analysis withholds, because some of Cascade alone's three TensorRT processes were not established within three attempts.}
\end{table*}

\section{The Output-Head Comparison}
\label{app:head}

\begin{figure}[tbp]
\centering
\includegraphics[width=\columnwidth]{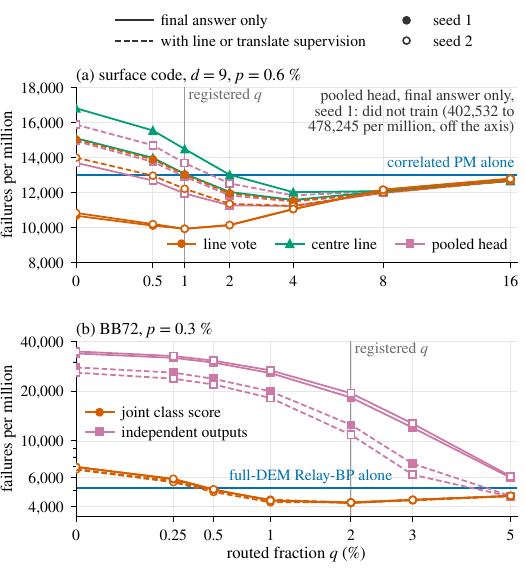}
\caption{Failures per million test shots of each composition against the routed fraction $q$, for every arm and seed of
Table~\ref{tab:head}, with the fallback alone in blue. Colour and marker name the output head, the line style its
supervision and the marker fill its seed, filled for seed 1 and hollow for seed 2. The $q$ axis is square-root scaled, and the failure axis of (b) is logarithmic. The vertical lines mark
the registered $q$.}
\label{fig:headcurves}
\end{figure}

This appendix gives the arms and their matched conditions, the calibration shots, the check for untrained runs, the
rule that decided the second seed, the fallbacks and devices and further results. Table~\ref{tab:headcmp} gives every
comparison of Section~\ref{sec:rq2}, and Figure~\ref{fig:headcurves} the failures of every composition as $q$ grows.

\begin{table*}[tp]
\centering
\caption{Every comparison of the output-head study in its re-test, by seed, with the outcomes that Table~\ref{tab:head}
summarizes.}
\label{tab:headcmp}
\footnotesize
\setlength{\tabcolsep}{4pt}
\begin{adjustbox}{max width=\linewidth}
\begin{tabular}{@{}l l r l l r l@{}}
\toprule
X vs Y & factor & $q$ (\%) & seed 1, X$-$Y [95\pct] & seed 2, X$-$Y [95\pct] & adjusted p & outcome \\
\midrule
\multicolumn{7}{@{}l}{\emph{Surface code, $d=9$, $p=0.6\pct$, 1,000,000 test shots}} \\
d vs b & head & 0 & $+$467,584 [$+$466,585, $+$468,582] & $+$2,862 [$+$2,686, $+$3,039] & $3.1\times10^{-225}$ & worse \\
d vs b & head & 1 & $+$463,573 [$+$462,576, $+$464,569] & $+$2,012 [$+$1,863, $+$2,162] & $7.8\times10^{-156}$ & worse \\
c vs b & head & 0 & $+$6,137 [$+$5,922, $+$6,354] & $+$4,266 [$+$4,069, $+$4,464] & $<10^{-300}$ & worse \\
c vs b & head & 1 & $+$4,549 [$+$4,362, $+$4,737] & $+$3,137 [$+$2,968, $+$3,307] & $4.3\times10^{-297}$ & worse \\
e vs a & head & 0 & $-$94 [$-$291, $+$103] & $+$1,878 [$+$1,680, $+$2,077] & 1 & n.c. \\
e vs a & head & 1 & $-$124 [$-$293, $+$45] & $+$1,465 [$+$1,294, $+$1,636] & 1 & n.c. \\
b vs a & supervision & 0 & $-$4,357 [$-$4,551, $-$4,164] & $-$3,159 [$-$3,339, $-$2,980] & $2.2\times10^{-264}$ & better \\
b vs a & supervision & 1 & $-$3,094 [$-$3,262, $-$2,927] & $-$2,283 [$-$2,437, $-$2,130] & $3.5\times10^{-190}$ & better \\
e vs d & supervision & 0 & $-$463,321 [$-$464,327, $-$462,315] & $+$2,175 [$+$1,976, $+$2,375] & 1 & n.c. \\
e vs d & supervision & 1 & $-$460,603 [$-$461,604, $-$459,601] & $+$1,736 [$+$1,564, $+$1,909] & 1 & n.c. \\
c vs a & both & 0 & $+$1,780 [$+$1,556, $+$2,004] & $+$1,107 [$+$904, $+$1,310] & $9.6\times10^{-26}$ & worse \\
c vs a & both & 1 & $+$1,455 [$+$1,260, $+$1,650] & $+$854 [$+$681, $+$1,027] & $3.0\times10^{-21}$ & worse \\
d vs a & both & 0 & $+$463,227 [$+$462,220, $+$464,233] & $-$297 [$-$486, $-$108] & 1 & n.c. \\
d vs a & both & 1 & $+$460,479 [$+$459,477, $+$461,481] & $-$271 [$-$431, $-$111] & 1 & n.c. \\
\multicolumn{7}{@{}l}{\emph{BB72, $p=0.3\pct$, 500,000 test shots}} \\
h vs g & head & 0 & $+$26,774 [$+$26,322, $+$27,232] & $+$27,748 [$+$27,289, $+$28,213] & $<10^{-300}$ & worse \\
h vs g & head & 2 & $+$14,070 [$+$13,731, $+$14,415] & $+$15,162 [$+$14,813, $+$15,517] & $<10^{-300}$ & worse \\
i vs f & head & 0 & $+$21,022 [$+$20,621, $+$21,429] & $+$19,126 [$+$18,741, $+$19,517] & $<10^{-300}$ & worse \\
i vs f & head & 2 & $+$8,258 [$+$7,991, $+$8,531] & $+$6,636 [$+$6,391, $+$6,886] & $<10^{-300}$ & worse \\
g vs f & supervision & 0 & $+$208 [$+$38, $+$379] & $+$240 [$+$69, $+$411] & 0.036 & worse \\
g vs f & supervision & 2 & $-$36 [$-$120, $+$48] & $-$2 [$-$90, $+$86] & 1 & n.c. \\
i vs h & supervision & 0 & $-$5,960 [$-$6,360, $-$5,561] & $-$8,862 [$-$9,260, $-$8,466] & $5.9\times10^{-190}$ & better \\
i vs h & supervision & 2 & $-$5,776 [$-$6,148, $-$5,406] & $-$8,524 [$-$8,898, $-$8,152] & $6.4\times10^{-207}$ & better \\
h vs f & both & 0 & $+$26,982 [$+$26,529, $+$27,441] & $+$27,988 [$+$27,527, $+$28,455] & $<10^{-300}$ & worse \\
h vs f & both & 2 & $+$14,034 [$+$13,696, $+$14,378] & $+$15,160 [$+$14,811, $+$15,515] & $<10^{-300}$ & worse \\
\bottomrule
\end{tabular}
\end{adjustbox}

\vspace{2pt}
\parbox{\textwidth}{\scriptsize The letters name the arms of Table~\ref{tab:head}. The difference is in failures per
million test shots, X minus Y, with Newcombe's 95\pct{} interval. The adjusted p-value is Holm's over the
intersection-union p-values of the 14 surface-code and the 10 BB72 tests. A head comparison holds the supervision
fixed, a supervision comparison holds the head fixed, and the arms of the rows marked both differ in both. Better and
worse need both seeds to reject in the same direction, and n.c.\ marks an inconclusive comparison.}
\end{table*}

\head{Arms and matched conditions.} For each seed, one trunk initialization was drawn, and every arm of that seed
loads it. The heads start from one shared seed, so arms that share a head module start it identically. Every arm of a
seed reads the same training shots in the same order, with the recipe of Appendix~\ref{app:train} for
$2\times10^8$ shots from scratch, in bf16 on H100 GPUs. The line vote with and without the line supervision has 389,395
parameters and 16,394,841 multiply-adds in its output head per shot. The centre line has 388,235 and 12,307,409, and the
pooled surface head 507,115 and 16,388,361. On BB72 the joint class score has 483,357 parameters and 45,524,736
multiply-adds in its output head, and the pooled head with independent outputs 997,905 and 43,573,824. Matched thus
means equal training shots and about equal multiply-adds in the output head, 99.96\pct{} and 96\pct{} of those of
\sys's heads, with 30\pct{} and 106\pct{} more parameters. A pooled head's
training-only read-out is not part of its decision path. Every read uses the exponential moving average of the weights
in fp32 at batch 4,096, and each read's first 2,000 decisions agreed at batch one.

\head{Calibration shots.} After the analysis was frozen, we drew 200,000 calibration shots per code besides the test
shots. Each arm's threshold is the $q$-quantile of its confidence on the calibration shots, and every arm takes the
same fallback answers. Two seeds give only an observed range of training runs.

\head{Untrained runs.} An arm has not trained if its decision loss stays within 1\pct{} of its constant-guess value
after $2\times10^7$ shots. Seed 1's two pooled surface heads failed this check. As the plan requires, both stay in
Table~\ref{tab:head}, and both arms and the seed's vote with the line supervision trained once more with a noise ramp, $p$ rising from
0.1\pct{} to 0.6\pct{} over the first quarter of the same budget. The pooled head with the line supervision trained
later in its first run, and the one without it never did. All three ramp reruns trained. On the test shots, the ramp
reruns of the pooled heads without and with the line supervision failed on 1,222 and 597 more shots per million than
the ramp rerun of the vote with it at $q=0$, and on 1,097 and 449 more at $q=1\pct$. These reruns are descriptive and
outside the Holm family. BB72 runs passed this check.

\head{The rule for the second seed.} After seed 1, a rule on 200,000 validation shots per code decided whether to train
seed 2 and the centre line. At routed fractions of 0, 1, 2 and 5\pct, it compared each of \sys's heads with the pooled
head of the same supervision, with McNemar tests, Holm's procedure and a margin of 3\pct. On BB72, both contrasts
favoured the joint class score at all four fractions. On the surface code, seed 1's pooled heads had not trained, so an
amendment made in training, before any validation record existed, applied the rule to the ramp reruns. There the vote
with the line supervision failed 5.2--8.5\pct{} less often than the pooled head with it, and both codes continued. The
analysis of the test shots was frozen separately, before any calibration or test shot existed.

{\clubpenalty=10000 %
\head{Fallbacks and devices.} Correlated PM decided the surface-code test shots on the circuit's DEM. Full-DEM Relay-BP
decided the BB72 test shots with relay-bp 0.2.2 and a new decoder for every shot, seeded with the shot's index in its
set. The surface-code reads ran on H200 slices and the BB72 reads on RTX PRO 6000 slices, each code on one GPU model.
The routed share of the test shots stayed within 0.1 percentage points of $q$ for every network.\par}

\head{Further results on BB72.} The adjusted p-values of the four head comparisons on BB72 lie below $10^{-300}$. The
pooled head with the translate supervision gets below full-DEM Relay-BP alone only at the largest tested fraction,
$q=5\pct$, at 4,652 and 4,608 failures per million against 5,228, still above the joint class score's compositions at
2\pct. The pooled head without it stays above Relay-BP at 5,992 and 6,116. These contrasts with Relay-BP alone are observed, outside the
tested comparisons. The translate supervision lowered the pooled head's failures alone in both seeds, by 5,960 and
8,862 per million (better, adjusted $p=5.9\times10^{-190}$). Without it, the joint class score alone failed on 208 and
240 more shots per million in the two seeds (worse, adjusted $p=0.036$), and at $q=2\pct$ the comparison is
inconclusive.

{\clubpenalty=10000 %
\head{Further results on the surface code.} The vote trained on the final answer is the best surface-code network of
the study. In an observed comparison outside the tested family, it failed less often than correlated PM alone
(12,989) and than the $d=9$ network of our earlier tests, which was grown from a two-block network. Its compositions at
$q=1\pct$ failed 24\pct{} less often than correlated PM alone. On the same test shots, the $d=9$ network of our earlier
tests fails 12,336 and 10,937 times per million alone and at $q=1\pct$, and the second BB72 network 6,928 and 4,220
times alone and at $q=2\pct$.\par}

\section{Registration and Reproducibility}
\label{app:reg}

Every confirmatory result in this paper comes from a test registered before its test shots existed, except the joint
decision, which was fitted on separate fresh shots and held out.
Table~\ref{tab:prov} lists these tests with their shots, the source of their seeds, the session that timed them and
their status. A freeze commit fixed each test's networks, thresholds, baseline settings, statistics, test-set size and
analysis command. For the tests seeded by the drand randomness beacon~\cite{drand}, the freeze commit reached GitHub
first, and the seeds came from the first drand round at least 600\,s after GitHub's record of the push. Each test set
came from one draw, and the registered pass read each decoder's predictions once. The release~\cite{convlines_code}
keeps the plan, the freeze, the seeds, the run records and the analysis of every test in the table, together with a log
of each run.

\begin{table*}[tp]
\centering
\caption{Registration of the results in this paper. Each row gives a result's test shots, the source of their seeds,
the session that timed it and its status.}
\label{tab:prov}
\footnotesize
\setlength{\tabcolsep}{3.5pt}
\begin{adjustbox}{max width=\linewidth}
\begin{tabular}{@{}>{\raggedright\arraybackslash}p{5.4cm} >{\raggedright\arraybackslash}p{3.8cm}
>{\raggedright\arraybackslash}p{3.0cm} >{\raggedright\arraybackslash}p{2.0cm} >{\raggedright\arraybackslash}p{3.3cm}@{}}
\toprule
result & test shots & seeds & times & status \\
\midrule
Surface code at $d=7$--17 against every baseline and our Decoder Switching (Table~\ref{tab:surface})
& 2,000,000 at $d=7$, 1,000,000 at $d=9$--17 & drand round 6548726 & speed re-test & registered \\ \addlinespace[1pt]
Surface code at $d=9$ with the one-block network (Table~\ref{tab:surface})
& 1,000,000 & drand round 6551422 & own session & registered \\ \addlinespace[1pt]
BB codes (Figure~\ref{fig:bb}, Table~\ref{tab:bb})
& 200,000 and 300,000 on BB72, 150,000 on the gross code & drand round 6547616 & speed re-test & re-test$^{a}$ \\ \addlinespace[1pt]
First Cascade test on BB72 (Table~\ref{tab:cascade-first}) and its Relay-BP chains (Appendix~\ref{app:cascade})
& 5,000,000 and 1,000,000 & seed range of its plan & speed re-test & revised selection rule \\ \addlinespace[1pt]
Six-block compositions, 52 configurations (Appendix~\ref{app:cascade})
& the first Cascade test's shots & --- & none & exploratory \\ \addlinespace[1pt]
Cascade follow-up on BB72 (Table~\ref{tab:cascade})
& 11,000,000 and 17,000,000 & drand round 6547616 & speed re-test & re-test \\ \addlinespace[1pt]
Our Cascade at $d=13$ (Table~\ref{tab:cascade})
& 1,000,000 and 30,000,000 & drand round 6547616 & speed re-test & re-test \\ \addlinespace[1pt]
Joint decision on the routed shots, both codes (Table~\ref{tab:cascade})
& the test shots of Table~\ref{tab:cascade} & as above & own session & held out \\ \addlinespace[1pt]
Multi-shot costs (Section~\ref{sec:batched-throughput})
& 500,000 timing shots & drand round 6549923 & speed re-test & registered, timing only \\ \addlinespace[1pt]
Batched costs of the matching and Relay-BP compositions (Appendix~\ref{app:timing})
& 200,000 per surface-code distance, 100,000 per BB72 rate, 2,000 on the gross code & drand round 6567343 & speed
re-test & registered, timing only \\ \addlinespace[1pt]
Cascade on TensorRT (Appendix~\ref{app:trt})
& fresh timing shots & drand round 6552306 & own session & registered, timing only \\ \addlinespace[1pt]
Output-head comparison (Table~\ref{tab:head})
& 1,000,000 at $d=9$, 500,000 on BB72 & drand round 6547616 & none & re-test, second-seed rule amended in training
(Appendix~\ref{app:head}) \\ \addlinespace[1pt]
Where the compositions gain (Section~\ref{sec:why})
& saved records of earlier surface-code tests and of the Cascade re-test & --- & none & exploratory \\
\bottomrule
\end{tabular}
\end{adjustbox}

\vspace{2pt}
\parbox{\textwidth}{\scriptsize Registered: fixed in a freeze commit before the shots existed. Re-test: the frozen
networks, thresholds and analysis of an earlier registered test, read on fresh shots. Revised: the protocol changed
after registration and before any confirmation shot was read. Held out: fitted on separate fresh shots and fixed
before it was applied to the test shots, not registered. Exploratory: chosen or analysed after results were known.
Every other result in the paper is exploratory or comes from calibration or development shots. The speed re-test
timed the Cascade comparisons on timing shots from drand round 6549923, every other comparison on round 6550668 and the batched costs on round 6567343. $^{a}$The BB72 balanced mode joined the earlier test after its shots were drawn. \sys alone
with the second BB72 network at $p=0.3\pct$ and with the gross-code network is descriptive.}
\end{table*}

\head{Registered criteria.} The registered tests asked that at least one mode, \sys alone or with a fallback, fail less
often than every decoder of a primary group and take less time than each. The primary group holds PM, correlated PM,
the four Ising pipelines and BP-OSD in two settings on surface codes, and sector Relay-BP and the two BP-OSD decoders on
BB codes. On BB codes, the tests also asked for a slower mode that fails less often than full-DEM Relay-BP, the most
accurate decoder in those tests, in less time. On the speed re-test's times, these criteria fail in three cases. PM takes
less time than \sys + correlated PM at $d=7$ and, with the main test's three-block network, at $d=9$. On the gross code,
BP-OSD from CUDA-Q takes less time than both modes that fail less often than sector Relay-BP.

\head{The speed re-test.} The times in this paper come from a registered speed re-test, which timed again every speed
number of our release, ours and the baselines', and added the batched costs of the matching and Relay-BP compositions
(Appendix~\ref{app:timing}). Each of its three stages froze its rules on
GitHub before its timing shots existed. Its results replace the earlier ones regardless of outcome. Two registered tests
timed their results in their own sessions, the second test at $d=9$ and the TensorRT comparison of
Appendix~\ref{app:trt}. A later, unregistered session timed the joint decision (Appendix~\ref{app:cascade}).

\head{The second test at $d=9$.} The speed re-test found PM faster than the main test's three-block network at $d=9$,
where the network's depth sets its time. We therefore trained a two-block network and a one-block network of width 64
on the same training shots in the same order. A rule fixed before their results chose the one-block network, which
routes 6\pct{} of the shots to correlated PM. A registered test then compared this composition with PM and correlated PM
on 1,000,000 fresh shots. The composition had to fail less often than each, by the exact McNemar test at $\alpha=0.025$
with a 97.5\pct{} Newcombe interval below zero, and to take less composite time than the speed re-test's 26.30\us{} for
PM and 63.92\us{} for correlated PM. It failed 11,978 times, against 27,028 for PM ($p<10^{-300}$) and 12,802 for
correlated PM ($p=9.5\times10^{-39}$), in 21.23\us{} per shot. Of this time, the GPU term took 14.21\us{} on an H100,
the routed CPU work 5.93\us{} on the Grace basis and the handoff 1.10\us. The session took the slowest of five
processes for the GPU term and of three for the CPU term, and it converted the CPU term with its own fitted factor,
1.5423. On the same shots, the network alone failed 18,305 times and the three-block network 10,326 times.

\head{The gross-code modes.} We chose both gross-code modes after a look at an earlier registered gross test, whose
300,000 shots we did not reuse. That test ran a 0.5\pct{} balanced mode on a slower fp32 engine and failed 1,177 times
against 1,456 for sector Relay-BP.

\head{Shared fallback decisions.} Eq.~(\ref{eq:identity}) holds when the fallback decides each routed shot in the
composition as it decides that shot alone. Correlated PM is deterministic. In the BB tests and in the Cascade
comparison, each accuracy read decoded a shot at most once with each fallback, and every composition that routed the
shot took that decision. Relay-BP draws the memory strengths of each decode from its decoder's random generator, so a
decoder kept across shots decides a shot according to the shots it decoded before. The accuracy reads of the first
Cascade test therefore seeded a new decoder for every routed shot with the shot's index in the read, and its timed
pipelines kept one decoder per process (Appendix~\ref{app:cascade}).

\section{Analysis of the Discordant Shots}
\label{app:why}

\begin{figure}[tbp]
\centering
\includegraphics[width=\columnwidth]{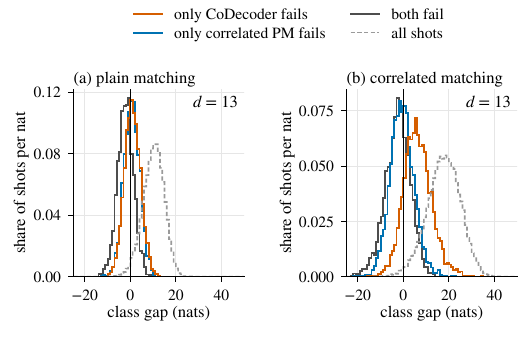}
\caption{Matching's class gap on the registered test shots at $d=13$ (exploratory), the weight of the best correction
in the wrong logical class minus that in the true class, for the shots that only \sys fails (5,416), only
correlated PM fails (3,834) and both fail (2,715), and for every 32nd test shot. The gap is exact for plain matching (a)
and approximate for correlated matching (b), whose axis omits 0.04\pct{} of the shots that both fail.}
\label{fig:gap}
\end{figure}

\begin{figure}[tbp]
\centering
\includegraphics[width=\columnwidth]{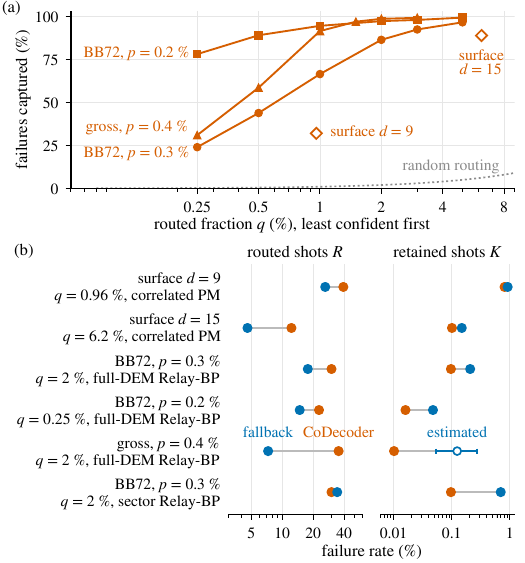}
\caption{Where \sys fails, on shots read before any test. (a) Share of \sys's failures among its least
confident shots, against the routed fraction $q$. The hollow diamonds are fresh development shots of the surface code
(50,000 at $d=9$ and 20,000 at $d=15$), and the curves are calibration shots (200,000 on BB72 and 100,000 on the gross
code). (b) Failure rates of \sys and of the fallback on the routed shots $R$ and on the retained shots $K$ at one
$q$ per case. The gross retained rate of full-DEM Relay-BP is estimated from a 4,100-shot sample, with its 95\pct{}
interval.}
\label{fig:motivation}
\end{figure}

This appendix gives the inputs, the checks and the definitions of the analysis of Section~\ref{sec:why}, and the
descriptions of the shots that Section~\ref{sec:why} summarizes.
Table~\ref{tab:cells} describes its groups of shots, Figure~\ref{fig:gap} shows how close matching comes to a tie on
them, and Figure~\ref{fig:motivation} shows where the network fails on the calibration and development shots of every
code.

\head{Inputs and checks.} The analysis is exploratory, and no registered test has confirmed its descriptions of the
shots. It reads the per-shot records of our earlier registered surface-code tests at $d=13$ and 17, whose networks
learned the line supervision, and of the re-test in front of our Cascade on BB72. The surface-code records hold
the network's $c$, its routing and its decision, and the decisions of correlated PM and PM. We rebuilt every
test shot from its recorded seed with Stim 1.16.0, and each of the 100 chunks at each distance has the hash that the
test's manifest records. PyMatching 2.4.0 on the circuit's DEM reproduces every recorded decision of correlated PM and
of PM, with 0 differences in 1,000,000 shots at each distance. The registered counts come out exactly, 5,627 failures
with 19,405 routed shots at $d=13$ and 2,708 with 79,652 at $d=17$. The re-test's records hold the detector bits,
the observables and the outputs of both networks, and they reproduce its counts. The analysis ran on CPU nodes
alone.

\head{Confidence plot.} Figure~\ref{fig:why} uses equal-count bins on each side of a routing threshold, eight below it and four above it for the surface code or three above it for BB72. The BB72 comparison uses Cascade's eager accuracy path.

\head{Groups and routing.} The network alone decides every shot by the sign of $c$. Crossing its decision with
correlated PM's gives four groups of shots, both right, only the network fails, only correlated PM fails, and both
fail. A shot is routed when $|c|<\tau$, as registered, and kept otherwise. The composition fails on the kept shots that only
the network fails, on the routed shots that only correlated PM fails and on every shot that both fail. The composition's
failures minus correlated PM's therefore equal the kept shots that only the network fails minus the kept shots that
only correlated PM fails, as Eq.~(\ref{eq:identity}) states. A shot's confidence percentile is the rank of $|c|$ among
the 1,000,000 test shots, so the routed shots are the lowest 1.94\pct{} at $d=13$ and 7.97\pct{} at $d=17$. These
percentiles are among the test shots, not the calibration shots that fixed $\tau$, and the routed fractions other than
the registered ones were read after the fact. At $d=13$, correlated PM fails on 17.0, 14.8 and 9.5\pct{} of the least confident 1, 2 and 5\pct{} of test shots, versus 0.65\pct{} overall.

\begin{table}[tbp]
\centering
\caption{Medians of the shot descriptors in our earlier registered surface-code tests. By every measure of the syndrome that
we used, the shots that only the network fails resemble those that only correlated PM fails.}
\label{tab:cells}
\footnotesize
\setlength{\tabcolsep}{2.5pt}
\begin{adjustbox}{max width=\linewidth}
\begin{tabular}{@{}l r r r r@{}}
\toprule
& \parbox[b]{1.25cm}{\centering all shots\\(reference)} & \parbox[b]{1.0cm}{\centering only the\\network\\fails} &
\parbox[b]{1.25cm}{\centering only\\correlated\\PM fails} & \parbox[b]{0.65cm}{\centering both\\fail} \\
\midrule
\multicolumn{5}{@{}l}{\emph{$d=13$}} \\
shots & 1,000,000 & 5,416 & 3,834 & 2,715 \\
routed & 19,405 & 4,060 & 1,556 & 1,326 \\
detection events & 201 & 215 & 215 & 215 \\
time-like pairs per 100 X events & 19.2 & 19.8 & 19.5 & 19.6 \\
space-like pairs per 100 X events & 34.6 & 36.1 & 36.0 & 36.2 \\
space-time pairs per 100 X events & 31.5 & 34.5 & 34.2 & 34.3 \\
rows to the nearer X boundary & 3.50 & 3.53 & 3.51 & 3.51 \\
confidence percentile (\%) & 50.0 & 0.9 & 2.6 & 2.0 \\
plain PM fails (\%) & 2.2 & 40.0 & 42.6 & 73.3 \\
plain class gap (nats) & 10.4 & 0.9 & 0.7 & $-2.0$ \\
correlated class gap (nats) & 18.0 & 5.9 & $-0.7$ & $-2.2$ \\
time-like run of 3 or more (\%) & 5.1 & 8.2 & 6.1 & 6.0 \\
\multicolumn{5}{@{}l}{\emph{$d=17$}} \\
shots & 1,000,000 & 7,380 & 1,811 & 1,132 \\
routed & 79,652 & 7,095 & 1,291 & 938 \\
detection events & 467 & 484 & 486 & 485 \\
time-like pairs per 100 X events & 19.3 & 19.6 & 19.4 & 19.5 \\
space-like pairs per 100 X events & 36.1 & 37.0 & 37.0 & 37.0 \\
space-time pairs per 100 X events & 33.5 & 35.1 & 35.1 & 35.0 \\
rows to the nearer X boundary & 4.50 & 4.52 & 4.51 & 4.50 \\
confidence percentile (\%) & 50.0 & 1.1 & 3.8 & 2.6 \\
plain PM fails (\%) & 1.6 & 27.4 & 39.9 & 72.1 \\
plain class gap (nats) & 11.6 & 2.7 & 1.1 & $-2.1$ \\
correlated class gap (nats) & 22.7 & 10.7 & $-0.2$ & $-2.6$ \\
time-like run of 3 or more (\%) & 9.3 & 12.1 & 9.8 & 12.6 \\
\bottomrule
\end{tabular}
\end{adjustbox}

\vspace{2pt}
\parbox{\columnwidth}{\scriptsize The three disjoint groups are the shots that only the network fails, only correlated
PM fails and both fail. All shots form the reference population, whose column in the last row comes from every 32nd
shot.}
\end{table}

{\clubpenalty=10000 %
\head{Class gap.} To find the least weight of a correction in each logical class, we split the X boundary of the
matching graph and turn its side of the observable's logical line into an extra detector. Decoding with this detector
off and on gives the \mbox{least-weight} correction in each of the two homology classes. The class gap is the weight of the
wrong class minus that of the true class, in nats, and it is negative where the least-weight class is wrong. The gap is
exact for plain matching, and the lower of the two weights equals plain PM's own solution weight on every shot. With
correlations enabled, each forced decode reweights the graph after its own first pass, so the correlated gap is
approximate. Its sign differs from correlated PM's own decision on 1.10\pct{} of all shots at $d=13$ and 0.72\pct{} at
$d=17$, and this appendix uses it only for its broad shape.\par}

\head{Syndrome measures.} Detection events are a shot's fired detectors. Time-like, space-like and space-time pairs are
pairs of fired X-type detectors joined by a DEM edge of that kind, the same site in consecutive rounds, the same round,
or another site in another round. The distance to the boundary is the mean, over a shot's X-type events, of the number
of data rows that a chain must cross to reach the nearer X boundary. The class difference is the set of edges in which
the least-weight plain corrections of the two classes differ, and the analysis records its longest run of time-like
edges at one site. It computes the class difference on every discordant shot and on every 32nd shot as a reference.
Stim samples detectors without recording physical errors. These measures therefore describe syndromes and matching corrections.

{\clubpenalty=10000 %
\head{The two kinds of discordant shot look alike.} By every measure of the syndrome that we used, the shots that only
the network fails resemble those that only correlated PM fails (Table~\ref{tab:cells}). They carry the same number of
detection events, the same mix of time-like, space-like and space-time pairs of events, and events at the same distance
from the boundaries. Both kinds exceed the typical detection-event count by 7\pct{} ($d=13$) and 4\pct{} ($d=17$).\par}

\head{Both kinds are near-ties for matching.} The class gap measures how close matching comes to a wrong decision on a
shot. As Figure~\ref{fig:gap}(a) shows, both kinds of discordant shot are near-ties for plain matching. Their median
class gap is 0.9 and 0.7 nats at $d=13$, against 10.4 nats on a typical shot, and 2.7 and 1.1 nats at $d=17$, against
11.6. The gap lies within 2 nats of zero on 42\pct{} and 43\pct{} of the two kinds at $d=13$, against 4.0\pct{} of
typical shots.

\head{Correlated matching resolves the network's lone failures.} Correlated matching separates the two kinds
(Figure~\ref{fig:gap}(b)). On the shots that only the network fails, its approximate class gap has a median of 5.9 nats
at $d=13$ and 10.7 at $d=17$, and it is positive on 86\pct{} and 92\pct{} of them. On its own lone failures, the median
is $-0.7$ and $-0.2$ nats. On 57\pct{} ($d=13$) and 60\pct{} ($d=17$) of these failures, plain PM decodes the shot, so
the reweighting of correlated matching tips a near-tie the wrong way. No measure that we used sets the network's lone
failures apart by more than a few points. The largest difference, a run of three or more time-like edges at one site
in the class difference (8.2 against 6.1\pct{} at $d=13$), fits its finite temporal reach but covers only about one in fifty shots. Thus, the decoders fail on distinct matching near-ties.

{\clubpenalty=10000 %
\head{Concentration on calibration shots.} As Figure~\ref{fig:motivation}(a) shows, the network's failures concentrate
in its least confident shots on every code. On the gross code, the least confident 2\pct{} of 100,000 calibration shots
hold 707 of the network's 717 failures. On BB72 at $p=0.2\pct$, the least confident 0.25\pct{} hold 114 of 146. On the
surface code at $d=15$, the 6.2\pct{} of fresh development shots below the threshold of our earlier network hold
89\pct{} of the failures.\par}

{\clubpenalty=10000 %
\head{Routed and retained calibration shots.} As Figure~\ref{fig:motivation}(b) shows, every fallback but one beats the
network on the routed calibration shots, as correlated PM does on the routed test shots. On the routed 2\pct{} of the
gross-code calibration shots, the network fails on 707 of 2,000 and full-DEM Relay-BP on 146.
On the retained shots the order reverses, with 10 failures against about 122. The exception is sector Relay-BP on BB72
at $q=2\pct$, which fails more often than the network on the routed shots (Section~\ref{sec:lose}).\par}

\head{BB72.} In front of our Cascade, the analysis reads Cascade's eager accuracy path. Among discordant shots below the routing threshold, the six-block network is the one that fails on 67\pct{} at $p=0.2\pct$ and 75\pct{} at $p=0.3\pct$. Above the threshold, these shares fall to 8\pct{} and 20\pct{}. Only 344 of Cascade's 8,063 failures at $p=0.2\pct$ (4.3\pct) and 2,340 of its 75,957 at $p=0.3\pct$ (3.1\pct) fall on kept shots. On the two compiled paths, the
network repairs 96 and 97 kept shots at $p=0.2\pct$ and 329 and 328 at $p=0.3\pct$, against 90 and 345 on the eager
path, and it breaks 8 and 90, against 8 and 88. Cascade's confidence on a shot is its smallest logit magnitude over the
12 logical qubits. Kept shots that both networks fail, 254 and 1,995, are mostly failures in three or more logical
qubits that Cascade makes with confidence. The periodic code has no boundary to measure.

\head{The repaired failures are small-margin mistakes of Cascade.} On 75 of the 90 repaired shots at $p=0.2\pct$ and
242 of the 345 at 0.3\pct, Cascade gets exactly one of the 12 logical qubits wrong. Its median logit magnitude on that
qubit is 0.84 and 0.74, which gives the right value a probability of about 0.3. The network scores the 12 logical
qubits as one joint class and keeps these shots, with a median margin percentile of 18.5\pct{} and 25.4\pct, well above
its thresholds at 2.00\pct{} and 10.00\pct. Its own 8 and 88 kept failures sit just above the thresholds, at median
margin percentiles of 2.65\pct{} and 13.5\pct.

\section{Implementation, Release and Names}
\label{app:names}

\begin{table*}[tp]
\centering
\caption{Descriptive names in this paper and record names in the release.}
\label{tab:names}
\footnotesize
\setlength{\tabcolsep}{3pt}
\begin{adjustbox}{max width=\linewidth}
\begin{tabular}{@{}>{\raggedright\arraybackslash}p{8.8cm} >{\raggedright\arraybackslash}p{8.0cm}@{}}
\toprule
name in this paper & record name \\
\midrule
\multicolumn{2}{@{}l}{\emph{Tests and their parts}} \\
surface-code test at $d=7$--17 and its calibration shots & SURF-v2 (test set sv2test), sv2cal \\
our Decoder Switching & gap switching (family G) \\
new against earlier surface-code networks & family C \\
re-tests & RETEST \\
earlier surface-code tests at $d=7$--13 and at $d=15$ & study 2 (test set T2), study 3 (T3) \\
earlier registered test at $d=17$ & R17 (test set T6) \\
validation shots at $d=7$--13, 15 and 17 & val\_legacy, V15, V17b \\
BB72 tests and the earlier gross-code test & stage 3 \\
gross-code test at $p=0.4\pct$ & stage 4 \\
addition of the BB72 balanced mode to the earlier test & Amendment S \\
first registered test with our Cascade & phase C \\
change of its selection rule & Amendment 1 \\
registered follow-up on fresh shots & D1 \\
fitted noise model of the Cascade comparison & depspam\_fm \\
analysis of the discordant shots (Section~\ref{sec:why}) & X2 \\
comparison with our Cascade at $d=13$ & W5, surface plan \\
its composition, chain, registered configuration, correlated PM alone and Cascade alone & C5, C1, C4, C3, C2 \\
joint decision: the fits of $w$ on BB72 and at $d=13$ & ARB, ARB2 \\
its failures on the test shots of Table~\ref{tab:cascade}, Cascade's eager and compiled paths & ARB4, ARBCP \\
its timed pipelines and its timing session & ARB3 \\
joint decision with matching's complementary gap (Section~\ref{sec:discussion}) & ARB5 \\
batched timing of the matching and Relay-BP compositions & SPEED-RT stage 3 (STAGE3\_batched) \\
output-head comparison (Section~\ref{sec:rq2}) & A1, readout ablation \\
its arms a to e on the surface code & S-VA, S-V, S-L, S-P, S-PA \\
its arms f to i on BB72 & B-JT, B-J, B-P, B-PA \\
\multicolumn{2}{@{}l}{\emph{Networks and engines}} \\
surface-code networks at $d=7$, 9, 11, 13, 15 and 17 & d7\_b1w64\_sv2, d9\_b1w64\_sh, d11\_b5\_sv2, d13\_b5\_sv2,
d15\_b5\_sv2, d17\_b5\_sv2 \\
the main test's three-block $d=9$ network & d9\_b3\_sv2s1 \\
networks of our earlier surface-code tests & d7\_b1w64\_r740, d9\_b3\_r764, d11\_b5\_r777, d13\_b5\_r775,
d15\_b5\_r782, d17\_b5\_r787 \\
first and second BB72 networks & r11, r12 \\
gross-code network & g2\_D12k113c10 \\
Ising-Fast and Ising-Accurate & M1 and M4 \\
three-block networks of the Cascade comparison, uniform and fitted & S55, S55$'$ \\
six-block network of the Cascade comparison & S55$'$-deep \\
our Cascade on BB72, uniform and fitted & C14, C14$'$ \\
our Cascade at $d=13$ & Cascade-surface, CS13 \\
multi-shot engine & milestone 2 (M2) \\
\bottomrule
\end{tabular}
\end{adjustbox}
\end{table*}

This appendix gives the libraries and the contents of the release, and it maps the names in this paper to the names in
the release.

\head{Implementation and release.} The decoders of the registered tests use PyTorch 2.11, Stim 1.16.0~\cite{stim},
PyMatching 2.4.0~\cite{sparseblossom} and relay-bp 0.2.2~\cite{relaybp}. The multi-shot engine and the Cascade
comparison add their own code and run with Triton 3.6. The release~\cite{convlines_code} holds the decoding and
training code, the weights, the circuits, and the seeds and hashes that rebuild every test set bit for bit. For the
surface-code test, it holds the freeze, the routing policies, the beacon record, the test set's manifest, the timing
records and the analysis, with the registration and the analysis of our Decoder Switching. For the re-tests, it holds
their freeze, beacon, seeds and analysis records. It also holds our re-implementations of Cascade with their weights
and training records, the code and records of the Cascade comparisons, of the joint decision, of the batched timing and of
the output-head comparison, and the records of our earlier registered tests.

{\clubpenalty=10000 %
\head{Names.} Table~\ref{tab:names} maps each descriptive name of this paper to the short internal name under which
the release and the records of the Cascade comparisons, of the joint decision and of the output-head comparison file
it.\par}

\clearpage
\bibliography{refs}

\end{document}